\documentclass[11pt,a4paper]{article}
\pdfoutput=1

\usepackage{jheppub}
\usepackage{amsmath}
\usepackage{amssymb}
\usepackage{graphicx}
\usepackage{url}

\DeclareMathOperator{\re}{Re}

\title{Investigation of the factorization scheme dependence of finite order perturbative QCD calculations}
\author{Karel Kolar}
\affiliation{Institute of Physics, Academy of Sciences of the Czech Republic,\\
Na Slovance 2, 182 21 Prague 8, Czech Republic}
\emailAdd{kolark@fzu.cz}

\abstract{The freedom associated with the definition of parton distribution functions
is analyzed and formulae governing the dependence of parton distribution
functions and hard scattering cross-sections on unphysical quantities
associated with the renormalization and factorization procedure are derived.
The issue of the specification of factorization schemes via the corresponding
higher order splitting functions is discussed in detail. A numerical analysis
of the practical applicability of the so called ZERO factorization scheme,
which could be useful for the construction of consistent NLO Monte Carlo event
generators, is presented.}


\begin{document}

\maketitle

\section{Introduction}

The definition of fundamental objects appearing in perturbative QCD calculations,
such as the color charge and parton distribution functions, depends on some
unphysical quantities. Theoretical predictions for physical quantities are
independent of these unphysical quantities if the appropriate power expansions
are summed to all orders. However, in practice, we are able to perform only
finite order calculations, which provide theoretical predictions depending
on the choice of the numerical values of these unphysical quantities.
The convenient choice of these numerical values thus plays very important
role in finite order calculations. There are three general methods how
to fix the renormalization scale: the Principle of Minimal Sensitivity (PMS)
\cite{pms}, the Effective Charges (EC) \cite{ec} and Brodsky-Lepage-MacKenzie
(BLM) approach \cite{blm}. They all recognize the fact that the existence of well
defined natural physical scale of a given hard process (like $Q^2$ in DIS) does not
by itself imply that the renormalization scale should be identified with it, but
start from different strategies how best to fix it. PMS looks for the region of
local stability of finite order perturbation calculation with respect to
the variation of the renormalization scale, EC (sometimes also called Fastest
Apparent Convergence, FAC approach) prefers the value where all higher order
contributions vanish, and BLM follows closely the recipe used in QED.
The PMS and EC approaches can be applied also for fixing the factorization scale.
All three approaches, together with the conventional approach in which
the renormalization scale is identified with the natural scale
characterizing the hardness of the process and allowed to vary within
some reasonable range \cite{ridolfi}, have been extensively used in phenomenological
analyses of hard scattering processes. See, for example, ref.\ \cite{hahn} for extensive
and detailed comparison of these approaches in QCD analysis of event shapes measured
at LEP. There is also alternative approach to the formulation of perturbative QCD,
which relates directly predictions for different physical quantities
\cite{lubrodsky,maxwell}. However, little attention has so far been
paid to the freedom in the choice of the factorization scheme even though it
is, in principle,  as important as the choice of the factorization scale. All phenomenological
analyses of hard scattering processes carried out so far have been performed in
the $\overline{\rm MS}$ and DIS factorization scheme. This is probably related to
the fact that the freedom in the choice of the factorization scheme is enormous,
even at the NLO.

The aim of this paper is the investigation of the freedom associated
with the definition of parton distribution functions. The immediate motivation
for this study is the potential exploitation of the freedom in the choice of
the factorization scheme for the construction of consistent NLO Monte Carlo event
generators \cite{potter,schorner,webber}.

The paper is organized as follows. The next section contains a review of basic facts,
relations and notation. Section \ref{prtfreedomfactoranal} is devoted to the discussion
of the freedom associated with the factorization procedure in massless perturbative QCD.
The subject of this section is the characterization of the freedom associated with
factorization, the specification of factorization schemes via the corresponding higher
order splitting functions and an overview of general formulae which govern the dependence
of parton distribution functions and hard scattering cross-sections on unphysical quantities
associated with the factorization procedure. Their application at the NLO is described
in Section \ref{prtsituationatnlo}. Since the so called ZERO factorization scheme appears
as the optimal factorization scheme for NLO Monte Carlo event generators, Section
\ref{prtzerofsanalysis} is devoted to the numerical analysis of its practical applicability
at the NLO. The summary and conclusion are presented in Section \ref{prtsumandconcl}. Some
important technical details are the subject of appendices.

\section{Basic facts and notation}
\label{prtfactandnotation}

Within the framework of perturbative QCD, theoretical predictions for physical quantities
are calculated as power expansions in the renormalized coupling parameter of QCD
$a(\mu, {\rm RS})$ ($a = \alpha_{\rm s} / \pi$). In the case of massless perturbative
QCD, which will be the object of our interest, the renormalized coupling parameter
$a(\mu, {\rm RS})$ is the only free parameter that characterizes the theory.\footnote{If
the number of quark flavours is given.} However, the renormalized coupling
parameter $a(\mu, {\rm RS})$ is not a single number, but it is a function of the so called
renormalization scale $\mu$ and the parameters that specify the renormalization scheme RS.
Both the renormalization scale and the renormalization scheme fix the ambiguities
associated with the renormalization procedure, which removes ultra violet singularities
from perturbative calculations.\footnote{The ultra violet singularities are absorbed
into the definition of the renormalized coupling parameter $a(\mu, {\rm RS})$.
The ambiguity of the renormalization procedure arises from the fact that the renormalized
coupling parameter $a(\mu, {\rm RS})$ is not defined uniquely by the requirement
of the absorption of the ultra violet singularities --- an arbitrary finite term can
be absorbed along with every singularity.} The dependence of the renormalized coupling
parameter $a(\mu, {\rm RS})$ on the renormalization scale $\mu$ is determined by
the differential equation
\begin{equation}
  {{\rm d} a(\mu, {\rm RS}) \over {\rm d} \ln\mu} = -b a^2(\mu, {\rm RS})
  \left( 1 + c a(\mu, {\rm RS}) + \sum_{l=2}^{\infty} c_l({\rm RS}) a^l(\mu, {\rm RS}) \right)
  \label{sbbeta}
\end{equation}
where $b = (33 - 2 n_{\rm f})/6$, $c = (153 - 19 n_{\rm f})/(66 - 4 n_{\rm f})$ and $n_{\rm f}$
denotes the number of quark flavours. Whereas the first two coefficients are unique, the
higher order coefficients $c_l(\rm RS)$ are completely arbitrary numbers and, together with
the initial condition of the preceding differential equation, can be used for the
unique specification of the corresponding renormalization scheme RS.

The coefficients of the power expansions that represent theoretical predictions for
physical quantities depend on the renormalization scale and the renormalization scheme
in such a way that if the power expansions are summed to all orders, then the obtained theoretical
predictions are independent of the renormalization scale and the renormalization scheme.
However, the finite order theoretical predictions which we obtain by truncating the corresponding
power expansions depend on these unphysical quantities. The convenient choice of the renormalization
scale and the renormalization scheme thus plays very important role in finite order
calculations\footnote{The ambiguity of the renormalization procedure does not necessarily have
to be seen only as an inconvenience because a convenient choice of the renormalization scale
and the renormalization scheme (which are not fixed but may depend on the process and its
kinematics) can improve the agreement between finite order theoretical predictions
and experiment.} (in practice, we are reliant only on finite order calculations).

To obtain theoretical predictions for processes whose initial state involves hadrons,
we need to know the parton distribution functions $D_{i/H}(x)$, which describe the parton
structure of the relevant hadrons.\footnote{$D_{i/H}(x){\rm d}x$ is the number of partons
of species $i$ inside hadron $H$ each of which carries the fraction of the hadron
momentum that is between $x$ and $x + {\rm d}x$. If it is not necessary to specify to which
hadron parton distribution functions belong, then the corresponding designation of the parton
distribution functions will be dropped in the following text.} As a simple example, consider
deep inelastic lepton-hadron scattering. In this case, the relevant cross-sections can be expressed
in terms of structure functions $F\!\left( x,Q^2\right)$, which, according to the factorization theorem,
are given \cite{politzer} as the convolution integral\footnote{This formula represents a separation of
short distance properties of the theory described by the coefficient functions
$C_i \!\left( x,Q^2,M,{\rm FS},{\rm RS}\right)$ and large distance properties of the theory
described by the parton distribution functions $D_i (x, M, {\rm FS}, {\rm RS})$. The validity
of this formula is not constrained only to perturbative QCD.}
\begin{equation}
  F\!\left( x,Q^2\right) = \sum_i \int_x^1 {{\rm d}y \over y}\, C_i \!\left( {x\over y}, Q^2, M,
  {\rm FS}, {\rm RS} \right) D_i (y, M, {\rm FS}, {\rm RS} )
\end{equation}
where $C_i \!\left( x,Q^2,M,{\rm FS},{\rm RS}\right)$ stands for the corresponding coefficient
functions and $D_i (x, M, {\rm FS}, {\rm RS})$ represents the relevant parton distribution
functions. Both the coefficient functions $C_i \!\left( x,Q^2,M,{\rm FS},{\rm RS}\right)$
and the parton distribution functions $D_i (x, M, {\rm FS}, {\rm RS})$ depend on the factorization
scale $M$, the factorization scheme FS and the renormalization scheme RS, which is used for
the factorization procedure, but the structure function $F\!\left( x,Q^2\right)$ is
independent of these unphysical quantities (at least if all relevant expansions are summed
to all orders), which fix the ambiguity associated with the factorization procedure, which
removes the so called collinear singularities from expressions for physical
quantities.\footnote{The treatment of the factorization procedure in this text is based
on perturbative calculations and therefore the renormalized coupling parameter that is
used for these calculations has to be specified (the factorization procedure thus has to be
preceded by the renormalization procedure). The used renormalized coupling parameter is
specified by the factorization scale $M$ and the renormalization scheme RS. Within
the framework of the factorization procedure, the collinear singularities are absorbed
into the definition of the parton distribution functions. Analogously to the case of
the definition of the renormalized coupling parameter, the (dressed/renormalized) parton
distribution functions are not determined uniquely by the requirement of the absorption
of the collinear singularities. The associated ambiguity is then fixed by the specification
of the factorization scheme FS.}

The parton distribution functions cannot be calculated from perturbative QCD and therefore
must be taken from experimental data. Perturbative QCD determines only their dependence
on unphysical quantities --- the factorization scale, the factorization scheme and
the renormalization scheme used for the factorization procedure. The dependence on
the factorization scale is described by the evolution equations
\begin{equation}
  {{\rm d} D_i(x, M, {\rm FS}, {\rm RS}) \over {\rm d}\ln M} = a(M,{\rm RS})
  \sum_j \int_x^1 {{\rm d} y \over y}\, P_{ij}\!\left({x\over y}, M, {\rm FS}, {\rm RS}
  \right) D_j(y, M, {\rm FS}, {\rm RS})   \label{sbxevolrov}
\end{equation}
where the splitting functions $P_{ij}(x, M, {\rm FS}, {\rm RS})$ can be expanded in powers
of $a(M, {\rm RS})$
\begin{equation}
  P_{ij}(x, M, {\rm FS}, {\rm RS}) = \sum_{k=0}^{\infty} a^k(M, {\rm RS})\, P_{ij}^{(k)}
  (x, {\rm FS}, {\rm RS}) . \label{sbxvetfce}
\end{equation}
Whereas the LO splitting functions $P^{(0)}_{ij}(x)$ are unique (they are independent of
the factorization scheme FS and the renormalization scheme RS), the higher order splitting
functions $P^{(k)}_{ij}(x, {\rm FS}, {\rm RS})$, $k \ge 1$ are completely arbitrary functions
and can be used for labelling factorization schemes.

The coefficient functions $C_i \!\left( x,Q^2,M,{\rm FS},{\rm RS}\right)$ are, at least
in principle, fully calculable within the framework of perturbative QCD and can thus be
expanded in powers of an arbitrary renormalized coupling parameter:
\begin{equation}
  C_i \!\left( x,Q^2,M,{\rm FS},{\rm RS}\right) = \sum_{k=0}^{\infty} a^k(\mu, {\rm RS}_{\rm E})
  \, C_i^{(k)} \!\!\left( x,Q^2, \mu, {\rm RS}_{\rm E}, M,{\rm FS},{\rm RS} \right) . \label{sbxkoef}
\end{equation}
The coefficient functions $C_i \!\left( x,Q^2,M,{\rm FS},{\rm RS}\right)$ are independent of
the renormalization scale $\mu$ and the renormalization scheme ${\rm RS}_{\rm E}$ if
the corresponding expansions are summed to all orders (but still depend on the factorization
scale $M$, the factorization scheme FS and the renormalization scheme RS). The renormalization
scheme ${\rm RS}_{\rm E}$, which is employed for expanding the coefficient functions
$C_i \!\left( x,Q^2,M,{\rm FS},{\rm RS}\right)$, is in principle different from the renormalization
scheme RS, which is used for the factorization procedure. However, in practice both renormalization
schemes are chosen to be identical, which simplifies calculations. Further simplification
of calculations can be achieved by setting $\mu = M$, which is a common and legitimate choice,
but there is a good reason for treating at least the renormalization scale $\mu$ and
the factorization scale $M$ as independent of each other: the renormalization scale emerges
from the renormalization procedure, which deals with ultra violet singularities, which are
related to short distance properties of the theory, while the factorization scale appears
in the factorization procedure, which treats collinear singularities, which are connected
with large distances. Keeping these scales separate can also lead to a better agreement
between finite order theoretical predictions and experimental data.

As a simple example illustrating the freedom in the choice of the factorization scheme,
consider the structure function $F_{2,{\rm NS}}^{\,\rm ep}\!\left( x,Q^2 \right)$
which is defined as
\begin{equation}
  F_{2,{\rm NS}}^{\,\rm ep}\!\left( x,Q^2 \right) = \frac{1}{x} \Bigl(
  F_2^{\,\rm ep}\!\left( x,Q^2 \right) - F_2^{\,\rm en}\!\left( x,Q^2 \right) \Bigr) .
\end{equation}
The structure function $F_{2,{\rm NS}}^{\,\rm ep}\!\left( x,Q^2 \right)$ can be approximately
expressed as\footnote{In the following, we limit ourselves only to such factorization
schemes whose definition preserves the symmetry between quarks and antiquarks and between
quark flavours. For simplicity, the renormalization scheme used for the factorization procedure
is fixed and is the same as that in which the renormalized QCD coupling parameter is defined.
Hence, the dependence on the renormalization scheme is not written out explicitly.}
\begin{equation}
  F_{2,{\rm NS}}^{\,\rm ep}\!\left( x,Q^2 \right) = \int_x^1 \frac{{\rm d}y}{y}\,
  C_{\rm NS}\!\left(\frac{x}{y}, Q^2, M, {\rm FS}\right) q_{\rm NS} (y, M, {\rm FS})
\end{equation}
where (the arguments are suppressed for brevity)
\begin{equation}
  C_{\rm NS}  = 3 \left( C_{2,{\rm u}} - C_{2,{\rm d}} \right)\! , \qquad
  q_{\rm NS}  = \frac{1}{3} \left( D_{{\rm u}/{\rm p}} + D_{\bar{\rm u}/{\rm p}}
  - D_{{\rm d}/{\rm p}} - D_{\bar{\rm d}/{\rm p}} \right)\! .
\end{equation}
The quark non-singlet distribution function $q_{\rm NS}(x, M, {\rm FS})$ satisfies
the evolution equation
\begin{equation}
 {{\rm d} q_{\rm NS} (x, M, {\rm FS}) \over {\rm d}\ln M} = a(M)
 \int_x^1 \frac{{\rm d}y}{y}\, P_{\rm NS}\!\left( \frac{x}{y}, M, {\rm FS} \right)
 q_{\rm NS} (y, M, {\rm FS}) .
\end{equation}
At the NLO approximation, we retain only the first two terms in the expansions
of the coefficient and splitting function:
\begin{align}
  C_{\rm NS}\!\left(x, Q^2, M, {\rm FS}\right) & = \delta(1-x) + a(\mu)\, C^{(1)}_{\rm NS}
  \!\left(x, Q^2, M, {\rm FS}\right)\! , \\[0.6em]
  P_{\rm NS}(x, M, {\rm FS}) & = P^{(0)}_{\rm NS}(x) + a(M)\, P^{(1)}_{\rm NS}(x, {\rm FS}) .
\end{align}
While $P^{(0)}_{\rm NS}(x)$ is unique, both $P^{(1)}_{\rm NS}(x, {\rm FS})$ and $C^{(1)}_{\rm NS}
\!\left(x, Q^2, M, {\rm FS}\right)$ are arbitrary, subjected only to the relation \cite{chyla}
\begin{equation}
  C^{(1)}_{\rm NS}\! \left(x, Q^2, M, {\rm FS}\right) = P^{(0)}_{\rm NS} (x) \ln\frac{Q}{M}
  + \frac{1}{b} P^{(1)}_{\rm NS}(x, {\rm FS}) + \kappa(x)  \label{sbrelnlocoeffandsplit}
\end{equation}
where $\kappa(x)$ is a factorization invariant, i.e.\ independent of both the factorization
scale and the factorization scheme. This relation connects the coefficient function
$C^{(1)}_{\rm NS} \!\left(x, Q^2, M, {\rm FS}\right)$ associated with one-loop perturbative
computations with the splitting function $P^{(1)}_{\rm NS}(x, {\rm FS})$ associated with
two-loop perturbative computations. This connection reflects the fact that the finite
parts associated with one-loop perturbative computations, which cause the ambiguity of
the coefficient function $C^{(1)}_{\rm NS}$, also appear in two-loop perturbative computation
of $P^{(1)}_{\rm NS}$. However, relation (\ref{sbrelnlocoeffandsplit}) does not imply that
the splitting function $P^{(1)}_{\rm NS}$ can be determined from a one-loop perturbative
calculation of the coefficient function $C^{(1)}_{\rm NS}$ because determining the factorization
invariant $\kappa(x)$ requires a two-loop perturbative calculation. Relation (\ref{sbrelnlocoeffandsplit})
only expresses the fact that the ambiguity of the coefficient function $C^{(1)}_{\rm NS}$
is correlated with the ambiguity of the splitting function $P^{(1)}_{\rm NS}$. A more detailed
discussion of relation (\ref{sbrelnlocoeffandsplit}) can be found, for instance, in \cite{zerofs}.

The ambiguity associated with the factorization procedure is large, but almost unexploited
in practice \cite{chyla}. The most widely used factorization scheme is the so called
$\overline{\rm MS}$ factorization scheme, which is suitable for theoretical calculations.
In this factorization scheme, both the splitting function $P^{(1)}_{\rm NS} (x, {\rm FS})$
and the coefficient function $C^{(1)}_{\rm NS} \!\left(x, Q^2, M, {\rm FS}\right)$ are nonzero.
Some analyses are also performed in the DIS factorization scheme, which is introduced in
\cite{dis} in order to express the relation between the structure function
$F_2^{\,\rm ep}\!\left(x, Q^2\right)$ and parton distribution functions in the same way as
in the parton model (this can be done for only one, but arbitrary, ratio of $Q$ and $M$),
which means that the coefficient function $C^{(1)}_{\rm NS} \!\left(x, Q^2, M, {\rm FS}\right)$
vanishes for $M = Q$. This implies that all NLO corrections are included
in the NLO splitting function $P^{(1)}_{\rm NS} (x, {\rm FS})$
and thus exponentiated by the evolution equation. A certain opposite to the DIS
factorization scheme is represented by the ZERO factorization scheme in which all NLO
splitting functions vanish and therefore all NLO corrections are contained in the hard
scattering cross-sections (in our case this means in the NLO coefficient function
$C^{(1)}_{\rm NS} \!\left(x, Q^2, M, {\rm FS}\right)$), which implies that the NLO corrections
are completely unexponentiated. The ZERO factorization scheme is mentioned, for instance,
in \cite{zerofs}. In the next paragraph, it is explained why the ZERO factorization
scheme could be useful for constructing consistent NLO Monte Carlo event generators.

At present time many QCD cross-sections at parton level are known at the NLO accuracy
and necessary algorithms for their incorporation in Monte Carlo event generators have
been developed. However, these algorithms attach to them initial state parton showers
only at the LO accuracy because no satisfactory algorithm for generating initial state
parton showers at the NLO accuracy (in the standard $\overline{\rm MS}$ factorization
scheme) has been found so far. Since the initial state parton showers induce the scale
dependence of parton distribution functions, it is inconsistent to attach LO initial
state parton showers to NLO QCD cross-sections, which include NLO parton distribution
functions. This deficiency could be removed by exploiting the ZERO factorization scheme,
in which the NLO initial state parton showers are formally identical to the LO ones.
The main advantage of this approach is the fact that the existing algorithms for
parton showering and for attaching parton showers to NLO cross-sections need not be
changed. The only necessary action is to transform hard scattering cross-sections from
the standard $\overline{\rm MS}$ factorization scheme to the ZERO factorization scheme
and to determine parton distribution functions in the ZERO factorization scheme.

The construction of consistent NLO Monte Carlo event generators in which initial state
parton showers can be taken formally at the LO thus constitutes one of the motivations
for investigating factorization schemes. Hence, the following section will be devoted to
a more detailed discussion of the freedom associated with factorization, with the emphasis
on its quantification.

\section{The freedom associated with factorization in massless QCD}
\label{prtfreedomfactoranal}

\subsection{Basic facts about factorization in massless QCD}

The parton model description of a hard hadronic collision has three basic ingredients:
the parton distribution functions of the colliding hadrons, a set of parton cross-sections
that describes the hard scattering of the partons in the process and some
model for the hadronization of the final state partons into observable hadrons
(the description of hadronization is not necessary if the hadrons in the final
state of the hard hadronic collision are not specified). Parton cross-sections
are, contrary to parton distribution functions and hadronization, fully calculable
within the framework of perturbation theory. However, QCD radiative corrections
to parton cross-sections are typically divergent (even after the renormalization
procedure because the appropriate singularities are related to the behavior of
massless perturbative QCD at large distances) and therefore useless for
the straightforward incorporation in the parton model. Fortunately, according to
the factorization theorem \cite{machacek}, the singularities connected with the partons
in the initial state are process independent and can be extracted from
the parton cross-sections and absorbed into the bare parton distribution
functions $\widehat{D}_i(x)$ of the naive parton model.\footnote{The elimination
of other possible singularities from parton cross-sections, which is related
to the definition of the final state, is independent of the factorization of
the singularities connected with the partons in the initial state.} The convolution
of the singular factors from parton cross-sections with the bare parton
distribution functions, denoted by $\widehat{D}_i(x)$ in the following, then defines
the dressed parton distribution functions $D_i(x,M,{\rm FS},{\rm RS})$, which are finite,
physically measurable, process independent, but ambiguous because arbitrary
finite terms can be added to the singular terms that are absorbed into
the definition of the dressed parton distribution functions. The ambiguity associated
with the definition of the dressed parton distribution functions
$D_i(x,M,{\rm FS},{\rm RS})$ is discussed below.

Within the framework of dimensional regularization in $d = 4 - 2\varepsilon$
space-time dimensions, the relation between dressed and bare parton distribution
functions is given by the convolution
\begin{equation}
  D_i(x, M, {\rm FS}, {\rm RS}) = \sum_j \int_x^1 \frac{{\rm d}y}{y}\,
  A_{ij}\!\left( \frac{x}{y}, M, {\rm FS}, {\rm RS} \right) \widehat{D}_j(y)  \label{scxdefpdf}
\end{equation}
where the functions $A_{ij}(x, M, {\rm FS}, {\rm RS})$, which represent the singular
factors that are absorbed into parton distribution functions, can be expanded
in powers of $a(M,{\rm RS})$:
\begin{equation}
  A_{ij}(x, M, {\rm FS}, {\rm RS}) = \sum_{k=0}^{\infty} a^k(M,{\rm RS})
  A^{(k)}_{ij}(x, {\rm FS}, {\rm RS}), \quad\!\! A_{ij}^{(0)}(x, {\rm FS}, {\rm RS})
  = \delta_{ij}\delta(1-x)   \label{scdefamatrix}
\end{equation}
and the higher order coefficients of the preceding expansion can be expressed in the form
\begin{equation}
  A^{(k)}_{ij}(x, {\rm FS}, {\rm RS}) = \sum_{l=0}^{\infty} \frac{1}{\varepsilon^l}
  A^{(kl)}_{ij}(x, {\rm FS}, {\rm RS}), \quad k \geq 1  \label{scdefakmatrix}
\end{equation}
where the functions $A^{(kl)}_{ij}(x, {\rm FS}, {\rm RS})$ are independent of
$\varepsilon$. The functions $A^{(k0)}_{ij}(x)$, which represent the finite part
of the absorbed factors, can be chosen arbitrarily and their choice defines
the factorization scheme FS. The functions $A^{(kl)}_{ij}(x, {\rm FS}, {\rm RS})$,
$l \geq 1$, which specify the singular part of the absorbed factors, are then
uniquely determined by the properties of the theory and the choice of the factorization
scheme FS (the choice of the functions $A^{(k0)}_{ij}(x)$).

A detailed analysis of the freedom associated with factorization in massless perturbative
QCD is presented in Appendix \ref{prtanalysisoffreedom}. The following two subsections
contain an overview of important results of this analysis. The rest of this paper,
with the exception of Appendices \ref{prtqcdcouplparam} and \ref{prtanalysisoffreedom},
is concerned only with the case of four space-time dimensions ($\varepsilon = 0$). Some
relations are expressed in terms of Mellin moments, which are defined in Appendix
\ref{prtmellintrans}. The last subsection of this section is devoted to a general
discussion of the issue of practical applicability of factorization schemes that are
specified by the corresponding higher order splitting functions.

\subsection{Parton distribution functions}
\label{prtfacfreedompdf}

In this subsection, we present formulae describing the change of the unphysical quantities
on which parton distribution functions depend and some important facts concerning the
freedom associated with the factorization procedure.

Changing the factorization scheme in which parton distribution functions are defined is
described by the formula
\begin{equation}
  \mathbf{D}(n, M, {\rm FS}_1, {\rm RS}) = \mathbf{T}(n, M, {\rm FS}_1, {\rm FS}_2,
  {\rm RS}) \mathbf{D}(n, M, {\rm FS}_2, {\rm RS})
\end{equation}
where the multiplication is matrix multiplication. The parton distribution functions are
represented by a column vector and the transformation matrix $\mathbf{T}(n, M, {\rm FS}_1,
{\rm FS}_2, {\rm RS})$ is a square matrix. The preceding formula determines the change of
the factorization scheme for the fixed factorization scale $M$ and renormalization scheme RS,
which is used for the factorization procedure. The transformation matrix $\mathbf{T}(n, M,
{\rm FS}_1, {\rm FS}_2, {\rm RS})$ can be expanded in powers of $a(M, {\rm RS})$:
\begin{equation}
  \mathbf{T}(n, M, {\rm FS}_1, {\rm FS}_2, {\rm RS}) = \sum_{k=0}^{\infty}
  a^k(M, {\rm RS}) \, \mathbf{T}^{(k)}(n, {\rm FS}_1, {\rm FS}_2)
\end{equation}
where $\mathbf{T}^{(0)}(n, {\rm FS}_1, {\rm FS}_2) = \mathbf{1}$ and the higher order
coefficients $\mathbf{T}^{(k)}(n, {\rm FS}_1, {\rm FS}_2)$ are given as polynomial expressions
in $\mathbf{A}\!^{(l0)}(n, {\rm FS}_1)$ and $\mathbf{A}\!^{(l0)}(n, {\rm FS}_2)$.

For an arbitrary factorization scheme FS and arbitrary renormalization schemes ${\rm RS}_1$
and ${\rm RS}_2$, there exists such a factorization scheme $\mathcal{FS}({\rm RS}_1,
{\rm RS}_2, {\rm FS})$ that
\begin{equation}
  \mathbf{A}(x, M, {\rm FS}, {\rm RS}_1) = \mathbf{A}(x, M,
  \mathcal{FS}({\rm RS}_1, {\rm RS}_2, {\rm FS}), {\rm RS}_2) ,  \label{szchangerseqv}
\end{equation}
independently of $x$ and the factorization scale $M$. The pair of $\{ {\rm FS}, {\rm RS}_1 \} $
thus defines the same singular factors that are absorbed into parton distribution functions as
the pair of $\{ \mathcal{FS}({\rm RS}_1, {\rm RS}_2, {\rm FS}), {\rm RS}_2 \} $, and therefore
both pairs define the same parton distribution functions and hard scattering cross-sections.
This fact can be exploited for changing the renormalization scheme used for the factorization
procedure because it allows to convert the simultaneous change of the factorization scheme and
the renormalization scheme from $\{ {\rm FS}_0, {\rm RS}_0 \}$ to $\{ {\rm FS}, {\rm RS} \}$
to changing the factorization scheme from ${\rm FS}_0$ to $\mathcal{FS}({\rm RS}, {\rm RS}_0,
{\rm FS})$ for the fixed renormalization scheme ${\rm RS}_0$.

The dependence of parton distribution functions on the factorization scale $M$ is described
by the evolution equations
\begin{equation}
  \frac{{\rm d} \mathbf{D}(n, M, {\rm FS}, {\rm RS})}{{\rm d}\ln M} =
  a(M, {\rm RS}) \mathbf{P}(n, M, {\rm FS}, {\rm RS})
  \mathbf{D}(n, M, {\rm FS}, {\rm RS}) .
\end{equation}
These equations are expressed in terms of Mellin moments. Converting them into $x$-space, we
obtain the evolution equations in the form of (\ref{sbxevolrov}). The splitting functions
$\mathbf{P}(n, M, {\rm FS}, {\rm RS})$ can be expanded in powers of $a(M, {\rm RS})$
\begin{equation}
  \mathbf{P}(n, M, {\rm FS}, {\rm RS}) = \sum_{k=0}^{\infty} a^{k}
  (M, {\rm RS})\, \mathbf{P}^{(k)}(n, {\rm FS}, {\rm RS}) .
\end{equation}
Whereas the LO splitting functions $\mathbf{P}^{(0)}(n)$ are independent of the factorization
scheme and the renormalization scheme, the higher order splitting functions $\mathbf{P}^{(k)}
(n, {\rm FS}, {\rm RS})$, $k \geq 1$ can be chosen at will and can be used for the specification
of the appropriate factorization scheme.\footnote{If we specify factorization schemes via the
corresponding higher order splitting functions, then the complete specification of the
factorization scheme requires also the specification of the corresponding renormalization scheme
because the relation between the splitting functions and the functions $A_{ij}^{(k0)}(x)$,
which define the factorization scheme, depends on the renormalization scheme.}

The relation between $\mathbf{T}^{(k)}(n, {\rm FS}_1, {\rm FS}_2)$ for $k \geq 1$ and the
corresponding splitting functions is given by the formula
\begin{multline}
  \left[ \mathbf{T}^{(k)}(n, {\rm FS}_1, {\rm FS}_2), \mathbf{P}^{(0)}(n) \right]
  - kb \mathbf{T}^{(k)}(n, {\rm FS}_1, {\rm FS}_2) =  \mathbf{P}^{(k)}(n, {\rm FS}_1,
  {\rm RS}) - {} \\ {} - \mathbf{P}^{(k)}(n, {\rm FS}_2, {\rm RS})
  + \sum_{l=1}^{k-1} \left\{ \mathbf{P}^{(k-l)}(n, {\rm FS}_1, {\rm RS})
  \mathbf{T}^{(l)}(n, {\rm FS}_1, {\rm FS}_2)  -  {} \right. \\ \left. {} -
  \mathbf{T}^{(l)}(n, {\rm FS}_1, {\rm FS}_2) \mathbf{P}^{(k-l)}(n, {\rm FS}_2, {\rm RS})
  + lbc_{k-l}({\rm RS}) \mathbf{T}^{(l)}(n, {\rm FS}_1, {\rm FS}_2) \right\}  \label{sztcoeffviasplitfce}
\end{multline}
where the definition of the coefficients $c_l({\rm RS})$, which are introduced in formula
(\ref{sbbeta}), is extended by $c_0({\rm RS}) = 1$ and $c_1({\rm RS}) = c$. The preceding formula
(\ref{sztcoeffviasplitfce}) holds for an arbitrary renormalization scheme RS. If we specify
factorization schemes via the corresponding splitting functions, then formula (\ref{sztcoeffviasplitfce})
allows us to determine the appropriate functions $\mathbf{T}^{(k)}(n, {\rm FS}_1, {\rm FS}_2)$,
which are necessary for changing the factorization scheme (formula (\ref{sztcoeffviasplitfce})
forms a set of equations for $\mathbf{T}^{(k)}(n, {\rm FS}_1, {\rm FS}_2)$, which can be solved
iteratively).\footnote{To specify factorization schemes, we can also exploit functions
$\mathbf{T}^{(k)}(n, {\rm FS}_1, {\rm FS}_2)$ --- an arbitrary factorization scheme FS can be
specified by the functions $\mathbf{T}^{(k)}(n, {\rm FS}, {\rm FS}_0)$ where the factorization
scheme ${\rm FS}_0$ is some fixed and familiar factorization scheme. Formula (\ref{sztcoeffviasplitfce})
can then be used in the opposite way to determine the splitting functions in the factorization
scheme FS (provided we know the splitting functions in the factorization scheme ${\rm FS}_0$).}

If higher order splitting functions are used for the specification of factorization schemes,
then the formula for determining the factorization scheme $\mathcal{FS}({\rm RS}_1, {\rm RS}_2,
{\rm FS})$ has the form
\begin{equation}
  \mathbf{P}^{(k)}(x, \mathcal{FS}({\rm RS}_1, {\rm RS}_2, {\rm FS}), {\rm RS}_2)
  = \sum_{l=0}^k h^{(l+1)}_{k-l}({\rm RS}_1, {\rm RS}_2)\, \mathbf{P}^{(l)}
  (x, {\rm FS}, {\rm RS}_1),
\end{equation}
where the coefficients $h^{(k)}_l ({\rm RS}_1, {\rm RS}_2)$ are introduced in Appendix
\ref{prtqcdcouplparameps}.

\subsection{Hard scattering cross-sections}
\label{prtfacfreedomhscs}

This subsection contains an overview of formulae describing the dependence of hard scattering
cross-sections on the unphysical quantities associated with the factorization procedure.

An arbitrary structure function $F\!\left(x, Q^2\right)$ is given as\footnote{The coefficient
functions form a row vector whereas the parton distribution functions are represented by a column
vector, and therefore the multiplication on the right hand side of this equation yields a number
(a matrix $1\times 1$).}
\begin{equation}
  F\!\left(n, Q^2\right) = \mathbf{C}\!\left(n, Q^2, M, {\rm FS}, {\rm RS} \right)
  \mathbf{D}(n, M, {\rm FS}, {\rm RS}) .
\end{equation}
The coefficient functions $\mathbf{C}\!\left(n, Q^2, M, {\rm FS}, {\rm RS} \right)$ can be
expanded in powers of $a(\mu, {\rm RS}_{\rm E})$
\begin{equation}
  \mathbf{C}\!\left(n, Q^2, M, {\rm FS}, {\rm RS} \right) = \sum_{k=0}^{\infty}
  a^k (\mu, {\rm RS}_{\rm E})\, \mathbf{C}^{(k)}\!\!\left(n, Q^2, \mu, {\rm RS}_{\rm E},
  M, {\rm FS}, {\rm RS} \right) \! .
\end{equation}
The dependence of the coefficient functions $\mathbf{C}^{(k)}\!\!\left(n, Q^2, \mu, {\rm RS}_{\rm E},
M, {\rm FS}, {\rm RS} \right)$ on the factorization scale $M$ and the factorization scheme FS
is governed by
\begin{align}
  \frac{{\rm d}\mathbf{C}^{(k)}\!\!\left(n, Q^2, \mu, {\rm RS}, M, {\rm FS}, {\rm RS}
  \right)}{{\rm d}\ln M} & = - \sum_{l=0}^{k-1} \mathbf{C}^{(l)}\!\!\left(n, Q^2,
  \mu, {\rm RS}, M, {\rm FS}, {\rm RS} \right) \times {} \nonumber\\ & {} \times\sum_{m=0}^{k-l-1}
  g^{(m+1)}_{k-l-m-1} (M, \mu, {\rm RS}) \mathbf{P}^{(m)}(n, {\rm FS}, {\rm RS}) ,
  \label{szcoefffuncchangescale} \displaybreak[0]\\
  \mathbf{C}^{(k)}\!\!\left(n, Q^2, \mu, {\rm RS}, M, {\rm FS}, {\rm RS} \right) & =
  \sum_{l=0}^k \mathbf{C}^{(l)}\!\!\left(n, Q^2, \mu, {\rm RS}, M, {\rm FS}_0, {\rm RS} \right)
  \times {} \nonumber\\ & {} \times \sum_{m=0}^{k-l} g^{(m)}_{k-l-m} (M, \mu, {\rm RS})
  \mathbf{T}^{(m)}(n, {\rm FS}_0, {\rm FS}) , \label{szcoefffuncchangescheme}
\end{align}
where the renormalization scheme ${\rm RS}_{\rm E}$, in which the renormalized coupling parameter
used for expanding the coefficient functions is defined, is identical to the renormalization
scheme RS, which is used for the factorization procedure. The coefficients $g^{(k)}_l (\mu_1, \mu_2,
{\rm RS})$ are introduced in Appendix \ref{prtqcdcouplparameps}. The equivalence of the pair of
$\{ {\rm FS}, {\rm RS}_1 \} $ to the pair of $\{ \mathcal{FS}({\rm RS}_1, {\rm RS}_2, {\rm FS}),
{\rm RS}_2 \} $ allows to convert changing the renormalization scheme used for the factorization
procedure to changing the factorization scheme (for the fixed renormalization scheme), which is
described by formula (\ref{szcoefffuncchangescheme}). Formulae (\ref{szcoefffuncchangescale}) and
(\ref{szcoefffuncchangescheme}) together with formula (\ref{abkoefchangeep}), which allows to change
the renormalized coupling parameter used for expanding the coefficient functions, are thus sufficient
for changing all unphysical parameters associated with the renormalization and factorization procedure
(even in the case if the renormalization scheme of the coupling parameter that is employed for
expanding the coefficient functions is different from the renormalization scheme
used for the factorization procedure).

Any inclusive cross-section $\sigma (P)$ (in general differential) depending on observables $P$
and describing a lepton-hadron collision is given as
\begin{equation}
  \sigma (P) = \sum_i \int_0^1 \!{\rm d}x\, \sigma_i (x, P, M, {\rm FS}, {\rm RS})
  D_i (x, M, {\rm FS}, {\rm RS}) .
\end{equation}
The hard scattering cross-section $\sigma_i (x, P, M, {\rm FS}, {\rm RS})$ can be expanded in
powers of $a(\mu, {\rm RS}_{\rm E})$
\begin{equation}
  \sigma_i (x, P, M, {\rm FS}, {\rm RS}) = \sum_{k=0}^{\infty} a^{k+k_0}(\mu, {\rm RS}_{\rm E})
  \, \sigma_i^{(k)} (x, P, \mu, {\rm RS}_{\rm E}, M, {\rm FS}, {\rm RS})
\end{equation}
where $k_0$ is a nonnegative integer. The formulae describing the dependence of the hard
scattering cross-sections $\sigma_i^{(k)} (x, P, \mu, {\rm RS}_{\rm E}, M, {\rm FS}, {\rm RS})$
on the factorization scale $M$ and the factorization scheme FS read
\begin{align}
  \frac{{\rm d}\sigma_i^{(k)}(x, P, \mu, {\rm RS}, M, {\rm FS}, {\rm RS})}{{\rm d}\ln M} & =
  -\sum_j \int_0^1 \!{\rm d}y \left\{ \sum_{l=0}^{k-1} \sigma_j^{(l)} (xy, P, \mu, {\rm RS}
  M, {\rm FS}, {\rm RS}) \times {} \right. \nonumber\\ & \quad\quad\left. {} \times \sum^{k-l-1}_{m=0}
  g^{(m+1)}_{k-l-m-1} (M, \mu, {\rm RS}) P^{(m)}_{ji}(y, {\rm FS}, {\rm RS}) \right\} ,
  \label{szhscsfscaledep} \displaybreak[0]\\
  \sigma_i^{(k)} (x, P, \mu, {\rm RS}, M, {\rm FS}, {\rm RS}) & = \sum_j \int_0^1 \!{\rm d}y
  \left\{ \sum_{l=0}^k \sigma_j^{(l)} (xy, P, \mu, {\rm RS}, M, {\rm FS}_0, {\rm RS})
  \times {} \right. \nonumber\\ & \quad\quad\left. {} \times \sum_{m=0}^{k-l} g^{(m)}_{k-l-m}(M, \mu,
  {\rm RS}) T^{(m)}_{ji}(y, {\rm FS}_0, {\rm FS}) \right\} . \label{szhscsfschemedep}
\end{align}
Note that in the preceding formulae, the renormalization scheme ${\rm RS}_{\rm E}$ is identical to
the renormalization scheme RS. Formulae (\ref{szhscsfscaledep}) and (\ref{szhscsfschemedep})
represent an analogy of formulae (\ref{szcoefffuncchangescale}) and
(\ref{szcoefffuncchangescheme}) and together with formula (\ref{abkoefchangeep})
are sufficient for changing all unphysical quantities associated with the
renormalization and factorization procedure.

In the case of a hadron-hadron collision, any inclusive cross-section $\sigma (P)$
depending on observables $P$ can be expressed as
\begin{multline}
  \sigma (P) = \sum_{ij} \int_0^1 \!\!\int_0^1 \!{\rm d}x_1{\rm d}x_2 \,
  \sigma_{ij}(x_1, x_2, P, M_1, {\rm FS}_1, {\rm RS}_1, M_2, {\rm FS}_2, {\rm RS}_2)
  \times {} \\ {} \times D_{i/H_1}(x_1, M_1, {\rm FS}_1, {\rm RS}_1)
  D_{j/H_2}(x_2, M_2, {\rm FS}_2, {\rm RS}_2) ,
\end{multline}
where the hard scattering cross-section $\sigma_{ij}(x_1, x_2, P, M_1, {\rm FS}_1, {\rm RS}_1,
M_2, {\rm FS}_2, {\rm RS}_2)$ can be expanded in powers of $a(\mu, {\rm RS}_{\rm E})$
\begin{multline}
  \sigma_{ij}(x_1, x_2, P, M_1, {\rm FS}_1, {\rm RS}_1, M_2, {\rm FS}_2, {\rm RS}_2) = {} \\
  {} = \sum_{k=0}^{\infty} a^{k+k_0}(\mu, {\rm RS}_{\rm E}) \, \sigma_{ij}^{(k)}(x_1, x_2, P,
  \mu, {\rm RS}_{\rm E}, M_1, {\rm FS}_1, {\rm RS}_1, M_2, {\rm FS}_2, {\rm RS}_2) .
\end{multline}
The dependence on the factorization scale $M_1$ is determined by the formula
\begin{multline}
  \frac{{\rm d} \sigma_{ij}^{(k)}(x_1, x_2, P, \mu, {\rm RS}_1, M_1, {\rm FS}_1, {\rm RS}_1, M_2,
  {\rm FS}_2, {\rm RS}_2)}{{\rm d}\ln M_1} = {} \displaybreak[0]\\ {} = - \sum_{r} \int_0^1
  \!{\rm d}y \left\{ \sum_{l=0}^{k-1} \sigma_{rj}^{(l)} (x_1 y, x_2, P, \mu, {\rm RS}_1, M_1, {\rm FS}_1,
  {\rm RS}_1, M_2, {\rm FS}_2, {\rm RS}_2) \times {} \right. \\ \left. {} \times
  \sum_{m=0}^{k-l-1} g^{(m+1)}_{k-l-m-1} (M_1, \mu, {\rm RS}_1) P^{(m)}_{ri} (y,
  {\rm FS}_1, {\rm RS}_1) \right\} . \label{szhscshhfscalef}
\end{multline}
The formula for changing the factorization scheme associated with hadron $H_1$ (from ${\rm FS}_1^{(0)}$
to ${\rm FS}_1$) has the form
\begin{multline}
  \sigma_{ij}^{(k)}(x_1, x_2, P, \mu, {\rm RS}_1, M_1, {\rm FS}_1, {\rm RS}_1, M_2, {\rm FS}_2,
  {\rm RS}_2) = {} \displaybreak[0]\\ {} = \sum_{r} \int_0^1 \!{\rm d}y \left\{
  \sum_{l=0}^k \sigma^{(l)}_{rj}(x_1 y, x_2, P, \mu, {\rm RS}_1, M_1, {\rm FS}_1^{(0)}, {\rm RS}_1,
  M_2, {\rm FS}_2, {\rm RS}_2) \times {} \right. \\ \left. {} \times \sum_{m=0}^{k-l}
  g^{(m)}_{k-l-m}(M_1, \mu, {\rm RS}_1) T^{(m)}_{ri} (y, {\rm FS}_1^{(0)}, {\rm FS}_1)
  \right\} .  \label{szhscshhfschemef}
\end{multline}
Note that in the preceding formulae (\ref{szhscshhfscalef}) and (\ref{szhscshhfschemef}),
the renormalization scheme ${\rm RS}_{\rm E}$ is identified with the renormalization scheme
${\rm RS}_1$. Analogous formulae hold for changing the unphysical quantities associated with
hadron $H_2$. The formula governing the dependence on the factorization scale $M_2$ reads
\begin{multline}
  \frac{{\rm d} \sigma_{ij}^{(k)}(x_1, x_2, P, \mu, {\rm RS}_2, M_1, {\rm FS}_1, {\rm RS}_1, M_2,
  {\rm FS}_2, {\rm RS}_2)}{{\rm d}\ln M_2} = {} \displaybreak[0]\\ {} = - \sum_{r} \int_0^1
  \!{\rm d}y \left\{ \sum_{l=0}^{k-1} \sigma_{ir}^{(l)} (x_1, x_2 y, P, \mu, {\rm RS}_2, M_1, {\rm FS}_1,
  {\rm RS}_1, M_2, {\rm FS}_2, {\rm RS}_2) \times {} \right. \\ \left. {} \times
  \sum_{m=0}^{k-l-1} g^{(m+1)}_{k-l-m-1} (M_2, \mu, {\rm RS}_2) P^{(m)}_{rj} (y,
  {\rm FS}_2, {\rm RS}_2) \right\} . \label{szhscshhfscales}
\end{multline}
Changing the factorization scheme associated with hadron $H_2$ (from ${\rm FS}_2^{(0)}$ to
${\rm FS}_2$) is determined by
\begin{multline}
  \sigma_{ij}^{(k)}(x_1, x_2, P, \mu, {\rm RS}_2, M_1, {\rm FS}_1, {\rm RS}_1, M_2, {\rm FS}_2,
  {\rm RS}_2) = {} \displaybreak[0]\\ {} = \sum_{r} \int_0^1 \!{\rm d}y \left\{
  \sum_{l=0}^k \sigma^{(l)}_{ir}(x_1, x_2 y, P, \mu, {\rm RS}_2, M_1, {\rm FS}_1, {\rm RS}_1,
  M_2, {\rm FS}_2^{(0)}, {\rm RS}_2) \times {} \right. \\ \left. {} \times \sum_{m=0}^{k-l}
  g^{(m)}_{k-l-m}(M_2, \mu, {\rm RS}_2) T^{(m)}_{rj} (y, {\rm FS}_2^{(0)}, {\rm FS}_2)
  \right\} .  \label{szhscshhfschemes}
\end{multline}
Note that in the preceding formulae (\ref{szhscshhfscales}) and (\ref{szhscshhfschemes}),
the renormalization scheme ${\rm RS}_{\rm E}$ is identified with the renormalization scheme
${\rm RS}_2$. The above mentioned formulae (\ref{szhscshhfscalef}), (\ref{szhscshhfschemef}),
(\ref{szhscshhfscales}), (\ref{szhscshhfschemes}) together with formula (\ref{abkoefchangeep})
are sufficient for changing all unphysical quantities associated with the renormalization and
factorization procedure.

\subsection{Applicability of factorization schemes specified by splitting functions}
\label{prtapplicabilityfsgen}

As it has already been mentioned, higher order splitting functions can be chosen at will
and can be used for labeling factorization schemes. This subsection will be devoted to
the question of practical applicability of factorization schemes that are specified by
the corresponding splitting functions.

To change the factorization scheme in which parton distribution functions and/or hard
scattering cross-sections are defined, it is necessary to determine the appropriate
functions $T^{(k)}_{ij}(x, {\rm FS}_1, {\rm FS}_2)$. In the case when the factorization
schemes are specified by the corresponding higher order splitting functions, the
necessary functions $T^{(k)}_{ij}(x, {\rm FS}_1, {\rm FS}_2)$ are given as the solution of
the system of equations represented by relation (\ref{sztcoeffviasplitfce}, \ref{sctcoeffviasplitfce}).
The zeros of the denominators in formulae (\ref{absolutioneqnzac})--(\ref{absolutioneqnkon}),
which express the solution of the appropriate equations, can give rise to singularities
in the Mellin moments $T^{(k)}_{ij}(n, {\rm FS}_1, {\rm FS}_2)$. The connection between
the location of the singularitites of Mellin moments $f(n)$ and the low $x$ behaviour of
the original function $f(x)$, which is analysed in Appendix \ref{prtmellintrans}, then
imply that the zeros which are located sufficiently on the right (in the complex plane)
can considerably influence the low $x$ behaviour of the functions $T^{(k)}_{ij}(x,
{\rm FS}_1, {\rm FS}_2)$. It can be proven that for every real number $\xi$, there
exists such a real number $\kappa_0$ that for every $\kappa > \kappa_0$, the denominator
given by (\ref{abdenominatorsoleq}) has some zero point for $\re n > \xi$, and therefore
the influence of the zeros of the denominators on the low $x$ behaviour of
$T^{(k)}_{ij}(x, {\rm FS}_1, {\rm FS}_2)$ cannot be ruled out, at least
at higher orders (for large $k$).

Let us consider some factorization scheme ${\rm FS}_0$ in which the low $x$ behaviour
of the appropriate parton distribution functions is fully determined by "physics" and
is not affected by the choice of the finite parts $A_{ij}^{(k0)}(x)$, which define
the factorization scheme.\footnote{The MS factorization scheme, in which the finite
parts $A_{ij}^{(k0)}(x)$ are set equal to zero, should be an example of such
a factorization scheme.} Singularities in the Mellin moments
$T^{(k)}_{ij}(n, {\rm FS}, {\rm FS}_0)$ induced by the zeros of the denominators may
imply that the parton distribution functions in the factorization scheme FS
have much larger values for low $x$ than those in the factorization scheme
${\rm FS}_0$.\footnote{Singularities in $T^{(k)}_{ij}(n, {\rm FS}, {\rm FS}_0)$ that
are induced by the zeros of the denominators may also cause undesirable behaviour
of hard scattering cross-sections in the factorization scheme FS.}
If this occurs, then in the case of the factorization scheme FS, there must be
an extensive mutual cancellation of those large values in the expressions
for physical quantities which significantly depend on the low $x$ region
because the theoretical predictions for physical quantities have to
be independent of the factorization scheme. The extensive mutual cancellation
can cause problems in numerical calculations, and moreover, it is likely that
the mutual cancellation is incomplete at finite order calculations, which can
result in unreliable theoretical predictions at the low $x$ region. Hence,
the range of the practical applicability of such a factorization scheme can be
significantly restricted even though the corresponding splitting functions
appear at first sight as reasonable.

It is worth noting that if we specify factorization schemes using the appropriate
finite parts $A_{ij}^{(k0)}(x)$, then any unexpected restrictions of their practical
applicability are ruled out because $T^{(k)}_{ij}(x, {\rm FS}, {\rm MS}) = A^{(k0)}_{ij}
(x, {\rm FS})$, which follows from formulae (\ref{scdefakmatrix}), (\ref{screkrelbkoef}),
(\ref{scdeftkoef}) and (\ref{scbformsfs}).

\section{Situation at the next-to-leading order}
\label{prtsituationatnlo}

In the preceding section, we have presented formulae describing the dependence
of parton distribution functions and hard scattering cross-sections on
unphysical quantities associated with the factorization procedure.
This section contains the compendium of these formulae and their consequences
for NLO approximation, which consists in retaining only the first two terms in
perturbative expansions. In this section, we limit ourselves only to the case when
the renormalization scheme used for the factorization procedure is fixed and
is the same as that in which the expansion parameter $a(\mu, {\rm RS})$
is defined. The dependence on the renormalization scheme will thus not
be written out explicitly.

\subsection{Changing the factorization scheme at NLO}

To change the factorization scheme at the NLO, it is necessary to determine
the appropriate functions $T_{ij}^{(1)}(x, {\rm FS}_1, {\rm FS}_2)$.
According to formula (\ref{sztcoeffviasplitfce}, \ref{sctcoeffviasplitfce}),
the Mellin moments $T_{ij}^{(1)}(n, {\rm FS}_1, {\rm FS}_2)$ are given as
the solution of the following equation
\begin{equation}
  \left[ \mathbf{T}^{(1)}(n, {\rm FS}_1, {\rm FS}_2), \mathbf{P}^{(0)}(n) \right]
  - b \mathbf{T}^{(1)}(n, {\rm FS}_1, {\rm FS}_2) = \mathbf{P}^{(1)}(n, {\rm FS}_1)
  - \mathbf{P}^{(1)}(n, {\rm FS}_2) .  \label{sdtfuncdefp}
\end{equation}
The formula for expressing the functions $T_{ij}^{(1)}(x, {\rm FS}_1, {\rm FS}_2)$
in terms of the finite parts $A^{(k0)}_{ij}(x)$ then reads
\begin{equation}
  \mathbf{T}^{(1)}(x, {\rm FS}_1, {\rm FS}_2) = \mathbf{A}\! ^{(10)}(x, {\rm FS}_1)
  - \mathbf{A}\! ^{(10)}(x, {\rm FS}_2) , \label{sdtfuncdefa}
\end{equation}
which follows from relations (\ref{screkrelbkoef}) and (\ref{scdeftkoef}).
In NLO approximation, factorization schemes are thus fully specified by the
corresponding finite parts $A_{ij}^{(10)}(x)$ or NLO splitting functions
$P^{(1)}_{ij}(x)$. If we use the functions $A^{(10)}_{ij}(x)$ or $T^{(1)}_{ij}(x)$
for the specification of factorization schemes,\footnote{A given factorization
scheme FS can be specified by the functions $T_{ij}^{(1)}(x, {\rm FS}, {\rm FS}_0)$
with the factorization scheme ${\rm FS}_0$ to be some fixed and familiar
factorization scheme (e.g.\ the $\overline{\rm MS}$ factorization scheme).} then
equation (\ref{sdtfuncdefp}) allows us to determine the corresponding NLO splitting
functions (provided we know them in some factorization scheme).

\subsection{Parton distribution functions}

The evolution equations in NLO approximation have the form
\begin{equation}
  \frac{{\rm d} \mathbf{D}(n, M, {\rm FS})}{{\rm d}\ln M} = \Bigl( a(M)
  \mathbf{P}^{(0)}(n) + a^2(M) \mathbf{P}^{(1)}(n, {\rm FS}) \Bigr)
  \mathbf{D}(n, M, {\rm FS}) .
\end{equation}
The change of the factorization scheme is then described by the formula
\begin{equation}
  \mathbf{D}(n, M, {\rm FS}) = \Bigl( \mathbf{1} + a(M) \mathbf{T}^{(1)}
  (n, {\rm FS}, {\rm FS}_0) \Bigr) \mathbf{D}(n, M, {\rm FS}_0) ,
\end{equation}
which can be exponentiated to
\begin{equation}
  \mathbf{D}(n, M, {\rm FS}) = \exp\Bigl( a(M) \mathbf{T}^{(1)}
  (n, {\rm FS}, {\rm FS}_0) \Bigr) \mathbf{D}(n, M, {\rm FS}_0) .   \label{sdtranspdfexp}
\end{equation}
An advantage of the exponentiated formula is the fact that the composition
of ${\rm FS}_0 \to {\rm FS}$ and ${\rm FS} \to {\rm FS}_0$ gives the original
distribution functions, which does not hold for the unexponentiated formula.

\subsection{Hard scattering cross-sections}

In NLO approximation, the coefficient functions $\mathbf{C}\!\left(n, Q^2, M, {\rm FS}
\right)$ are given as
\begin{equation}
  \mathbf{C}\!\left( n, Q^2, M, {\rm FS} \right) = \mathbf{C}^{(0)}\!\! \left( n, Q^2
  \right) + a(\mu)\, \mathbf{C}^{(1)}\!\!\left( n, Q^2, M, {\rm FS} \right) \! .
\end{equation}
The LO coefficient functions $\mathbf{C}^{(0)} \!\!\left(n, Q^2 \right)$ are independent
of the unphysical quantities such as $\mu$, $M$ and FS. The dependence of the NLO coefficient
functions $\mathbf{C}^{(1)}\!\! \left( n, Q^2, M, {\rm FS} \right)$ on the unphysical
quantities is determined by
\begin{multline}
  \mathbf{C}^{(1)}\!\!\left( n, Q^2, M, {\rm FS}\right) =
  \mathbf{C}^{(1)}\!\!\left( n, Q^2, M_0, {\rm FS}_0\right) + {} \\ {}
  + \mathbf{C}^{(0)}\!\!\left(n, Q^2 \right) \left( \mathbf{P}^{(0)}(n)
  \ln\frac{M_0}{M} + \mathbf{T}^{(1)}(n, {\rm FS}_0, {\rm FS}) \right) \! ,
\end{multline}
which follows from relations (\ref{szcoefffuncchangescale}, \ref{sccoefffuncchangescale}),
(\ref{szcoefffuncchangescheme}, \ref{sccoefffuncchangescheme}) and (\ref{abkoefchangeep}).

Similarly, the hard scattering cross-section $\sigma_i(x, P, M, {\rm FS})$ is expressed as
\begin{equation}
  \sigma_i (x, P, M, {\rm FS}) = a^{k_0}(\mu) \Bigl( \sigma^{(0)}_i(x, P) + a(\mu)\,
  \sigma_i^{(1)} (x, P, \mu, M, {\rm FS}) \Bigr)
\end{equation}
where $k_0$ is some nonnegative integer. The NLO hard scattering cross-section $\sigma_i
^{(1)}(x, P, \mu, M, {\rm FS})$ satisfies
\begin{multline}
  \sigma_i^{(1)}(x, P, \mu, M, {\rm FS}) = \sigma_i^{(1)} (x, P, \mu_0, M_0, {\rm FS_0})
  + k_0 b \ln\frac{\mu}{\mu_0}\; \sigma_i^{(0)}(x, P) + {} \\ {} + \sum_j \int_0^1
  \!{\rm d}y\, \sigma_j^{(0)}(xy, P) \left( P^{(0)}_{ji}(y) \ln\frac{M_0}{M}
  + T^{(1)}_{ji}(y, {\rm FS}_0, {\rm FS}) \right) \! .
\end{multline}

In the case of hadron-hadron collisions, we then have
\begin{multline}
  \sigma_{ij}(x_1, x_2, P, M_1, {\rm FS}_1, M_2, {\rm FS}_2) = {} \\ {} = a^{k_0}(\mu)
  \Bigl( \sigma^{(0)}_{ij}(x_1, x_2, P) + a(\mu)\, \sigma_{ij}^{(1)}(x_1, x_2, P, \mu,
  M_1, {\rm FS}_1, M_2, {\rm FS}_2) \Bigr)
\end{multline}
with
\begin{multline}
 \sigma_{ij}^{(1)}(x_1, x_2, P, \mu, M_1, {\rm FS}_1, M_2, {\rm FS}_2) = {} \\ {}
 = \sigma_{ij}^{(1)}(x_1, x_2, P, \mu_0, M_1^{(0)}, {\rm FS}_1^{(0)}, M_2^{(0)},
 {\rm FS}_2^{(0)}) + k_0 b\ln\frac{\mu}{\mu_0}\; \sigma^{(0)}_{ij}(x_1, x_2, P)
 + {} \displaybreak[0]\\ {} + \sum_r \int_0^1 \!{\rm d}y \left\{ \sigma^{(0)}_{rj}(x_1 y, x_2, P)
 \left( P^{(0)}_{ri}(y) \ln\frac{M_1^{(0)}}{M_1} + T^{(1)}_{ri}(y, {\rm FS}_1^{(0)},
 {\rm FS}_1) \right) + {} \right. \\ \left. {} + \sigma_{ir}^{(0)}(x_1, x_2 y, P)
 \left( P^{(0)}_{rj}(y) \ln\frac{M_2^{(0)}}{M_2} + T^{(1)}_{rj}(y, {\rm FS}_2^{(0)},
 {\rm FS}_2) \right) \right\} \! .
\end{multline}

\subsection{Applicability of factorization schemes specified by NLO splitting functions}
\label{prtpracappfsnlo}

In Subsection \ref{prtapplicabilityfsgen}, it has been shown that some splitting
functions that appear at first sight as reasonable can correspond to factorization
schemes which have some restrictions on their practical applicability. This
subsection will therefore be devoted to a more detailed analysis of this fact
at the NLO.

A given factorization scheme FS can have some restrictions on its practical
applicability if the Mellin moments $T_{ij}^{(1)}(n, {\rm FS}, \overline{\rm MS})$
have some singularities that are located too much on the right in the complex plane.
If we specify factorization schemes via NLO splitting functions, then such
singularities can unexpectedly arise from the zeros of the denominators in formulae
(\ref{absolutioneqnzac})--(\ref{absolutioneqnkon}), which express the solution
of equation (\ref{sdtfuncdefp}). In the following, we limit ourselves only to the case
when the matrix of the NLO splitting functions $\mathbf{P}^{(1)}(x)$ has the same
structure as that which corresponds to the $\overline{\rm MS}$ factorization scheme,
that is
\begin{alignat}{2}
  P^{(1)}_{q_i q_j}(x) & = P^{(1)}_{\bar{q}_i \bar{q}_j} (x) = \delta_{ij}
  P^{(1)V}_{qq}\! (x) + P^{(1)S}_{qq} (x), & \quad\! P^{(1)}_{q_i G} (x) & =
  P^{(1)}_{\bar{q}_i G} (x) = P^{(1)}_{qG} (x), \nonumber\\
  P^{(1)}_{q_i \bar{q}_j}(x) & = P^{(1)}_{\bar{q}_i q_j} (x) = \delta_{ij}
  P^{(1)V}_{q\bar{q}}\! (x) + P^{(1)S}_{qq} (x), & \quad\! P^{(1)}_{G q_i} (x) & =
  P^{(1)}_{G\bar{q}_i} (x) = P^{(1)}_{Gq} (x) .
\end{alignat}
In this case, the unexpected singularities in the corresponding Mellin moments
$T_{ij}^{(1)}(n, {\rm FS}, \overline{\rm MS})$ can be induced only by the zeros of
\begin{equation}
  b^2 - \left( P^{(0)}_{qq}(n) - P^{(0)}_{GG}(n) \right)^2 - 8n_{\rm f}
  P^{(0)}_{qG} (n) P^{(0)}_{Gq} (n) . \label{sdproblemdenom}
\end{equation}
A detailed analysis of the preceding expression for the number of quark flavours
$n_{\rm f} \in \{3, 4, 5\}$ then shows that it has two simple roots in the
half-plane $\re n > 1$. The approximate numerical values of these roots are:
\begin{align}
  n & \in \{ 1.7329, 4.6306 \} \quad \text{for} \quad n_{\rm f} = 3, \nonumber\\
  n & \in \{ 1.7995, 3.8458 \} \quad \text{for} \quad n_{\rm f} = 4, \nonumber\\
  n & \in \{ 1.9001, 3.1798 \} \quad \text{for} \quad n_{\rm f} = 5 . \label{sdnumvalueofroot}
\end{align}
Analysing formulae (\ref{absolutioneqnzac})--(\ref{absolutioneqnkon}), which
express the solution of equation (\ref{sdtfuncdefp}), we find that the Mellin
moments $T_{ij}^{(1)}(n, {\rm FS}, \overline{\rm MS})$ do not have any pole
at a simple root of (\ref{sdproblemdenom}) denoted by $n_0$ if and only if the NLO
splitting functions $P^{(1)}_{ij}(n)$ corresponding to the factorization scheme FS
satisfy
\begin{align}
  & P^{(0)}_{Gq}(n_0) \left( P^{(0)}_{qq}(n_0) - P^{(0)}_{GG}(n_0) - b \right)
  \left( P^{(1)}_{qG}(n_0) - P^{(1)}_{qG}(n_0, \overline{\rm MS}) \right) + {} \nonumber\\
  +\, & P^{(0)}_{qG}(n_0) \left( P^{(0)}_{qq}(n_0) - P^{(0)}_{GG} (n_0) + b \right)
  \left(P^{(1)}_{Gq}(n_0) - P^{(1)}_{Gq}(n_0, \overline{\rm MS}) \right) - {} \nonumber\displaybreak[0]\\
  -\, & 2 P^{(0)}_{qG}(n_0) P^{(0)}_{Gq}(n_0) \left( P^{(1)V}_{qq} \! (n_0) + P^{(1)V}_{q\bar{q}}
  \! (n_0) + 2n_{\rm f} P^{(1)S}_{qq} (n_0) - P^{(1)}_{GG}(n_0) - {} \right. \nonumber\\
  -\, & P^{(1)V}_{qq} \! (n_0, \overline{\rm MS}) - P^{(1)V}_{q\bar{q}} \! (n_0, \overline{\rm MS})
  - 2n_{\rm f} P^{(1)S}_{qq} (n_0, \overline{\rm MS}) + P^{(1)}_{GG}(n_0,
  \overline{\rm MS}) \Big) = 0 .  \label{sdconofpracapp}
\end{align}

The preceding condition can be expressed in terms of the singlet splitting functions.
At the NLO, the evolution of the quark singlet distribution function $\Sigma(x)$ defined as
\begin{equation}
  \Sigma (x) = \sum_{i=1}^{n_{\rm f}} \bigl( q_i (x) + \bar{q}_i (x) \bigr)
\end{equation}
and the gluon distribution function $G(x)$ is described by the following system
of coupled equations:
\begin{align}
  \frac{{\rm d}\Sigma (n, M)}{{\rm d}\ln M} = \Big( & a(M) P^{(0)}_{QQ}(n)
  + a^2(M) P^{(1)}_{QQ}(n) \Big) \,\Sigma(n, M) + {} \nonumber\\ {} + \Big(
  & a(M) P^{(0)}_{QG}(n) + a^2(M) P^{(1)}_{QG}(n) \Big) \, G(n, M) , \nonumber\displaybreak[0]\\
  \frac{{\rm d}G (n, M)}{{\rm d}\ln M} = \Big( & a(M) P^{(0)}_{GQ}(n)
  + a^2(M) P^{(1)}_{GQ}(n) \Big) \,\Sigma(n, M) + {} \nonumber\\ {} + \Big(
  & a(M) P^{(0)}_{GG}(n) + a^2(M) P^{(1)}_{GG}(n) \Big) \, G(n, M)
\end{align}
where the singlet splitting functions are given by
\begin{alignat}{2}
  P^{(0)}_{QQ}(x) & = P^{(0)}_{qq}(x),  & P^{(1)}_{QQ}(x) & = P^{(1)V}_{qq}\! (x)
  + P^{(1)V}_{q\bar{q}}\! (x) + 2n_{\rm f} P^{(1)S}_{qq} (x), \nonumber\\
  P^{(0)}_{QG} (x) & = 2n_{\rm f} P^{(0)}_{qG} (x), & \;\;\;\quad P^{(1)}_{QG} (x)
  & = 2n_{\rm f} P^{(1)}_{qG} (x), \nonumber\\
  P^{(0)}_{GQ}(x) &= P^{(0)}_{Gq}(x), & P^{(1)}_{GQ}(x) &= P^{(1)}_{Gq}(x) .
\end{alignat}
The condition (\ref{sdconofpracapp}) can then be rewritten in the form
\begin{align}
  & P^{(0)}_{GQ}(n_0) \left( P^{(0)}_{QQ}(n_0) - P^{(0)}_{GG}(n_0) - b \right)
  \left( P^{(1)}_{QG}(n_0) - P^{(1)}_{QG}(n_0, \overline{\rm MS}) \right) + {} \nonumber\\
  +\, & P^{(0)}_{QG}(n_0) \left( P^{(0)}_{QQ}(n_0) - P^{(0)}_{GG} (n_0) + b \right)
  \left(P^{(1)}_{GQ}(n_0) - P^{(1)}_{GQ}(n_0, \overline{\rm MS}) \right) - {} \\
  -\, & 2 P^{(0)}_{QG}(n_0) P^{(0)}_{GQ}(n_0) \left( P^{(1)}_{QQ} (n_0)
  - P^{(1)}_{GG}(n_0) -  P^{(1)}_{QQ} (n_0, \overline{\rm MS}) + P^{(1)}_{GG}(n_0,
  \overline{\rm MS}) \right) = 0 , \nonumber
\end{align}
which means that this condition does not put any restriction on the choice of
the non-singlet NLO splitting functions. Hence, there should be no unexpected
constraints on practical applicability in the non-singlet sector.\footnote{This
is not surprising because the non-singlet case can be analysed in the same way
as the general case in Appendix \ref{prtanalysisoffreedom} with the only difference
that all matrices are of dimension $1 \times 1$. Hence, formula
(\ref{sztcoeffviasplitfce}, \ref{sctcoeffviasplitfce}) does not contain
the commutator, and therefore no unexpected singularities can emerge in
the Mellin moments $T^{(k)}_{ij}(n, {\rm FS}_1, {\rm FS}_2)$.}

\section{Results of numerical analysis at NLO}
\label{prtzerofsanalysis}

At the end of Section \ref{prtfactandnotation}, we have shown that
the ZERO factorization scheme could be useful for phenomenology
and NLO Monte Carlo event generators. The important issue of its
practical applicability at the NLO is the subject of this section.

\subsection{Changing the factorization scheme from $\overline{\rm MS}$ to ZERO}

To convert hard scattering cross-sections and parton distribution functions
from the standard $\overline{\rm MS}$ factorization scheme to the ZERO one,
we need to know the functions $T^{(1)}_{ij}(x, \overline{\rm MS},
{\rm ZERO})$,\footnote{The functions $T^{(1)}_{ij}(x, {\rm ZERO}, \overline{\rm MS})$,
which are required for the conversion of parton distribution functions, are equal to
$-T^{(1)}_{ij}(x, \overline{\rm MS}, {\rm ZERO})$, which follows from formula
(\ref{sdtfuncdefa}).} which will be denoted as $T^{(1)}_{ij}(x)$ in the following.
According to relation (\ref{sdtfuncdefp}), the Mellin moments of these functions are
given as the solution of
\begin{equation}
  \left[ \mathbf{T}^{(1)}(n), \mathbf{P}^{(0)}(n) \right] - b\mathbf{T}^{(1)}(n)
  = \mathbf{P}^{(1)}(n)
\end{equation}
where $P^{(1)}_{ij}(x)$ denotes the $\overline{\rm MS}$ NLO splitting functions
\cite{nlosplitfuncns,nlosplitfuncsng}.
Using formulae (\ref{absolutioneqnzac})--(\ref{absolutioneqnkon}), we get
\begin{align}
  T^{(1)}_{q_i q_j} (n) &= T^{(1)}_{\bar{q}_i \bar{q}_j}(n) = T^{(1)}_3 (n)
  - \frac{1}{b} \Bigl( \delta_{ij} P^{(1)V}_{qq} \! (n) + P^{(1)S}_{qq} (n) \Bigr) ,
  \nonumber\\ T^{(1)}_{q_i \bar{q}_j} (n) &= T^{(1)}_{\bar{q}_i q_j}(n) = T^{(1)}_3 (n)
  - \frac{1}{b} \Bigl( \delta_{ij} P^{(1)V}_{q\bar{q}} \! (n) + P^{(1)S}_{qq} (n) \Bigr) ,
  \nonumber\displaybreak[0]\\ T^{(1)}_{q_i G}(n) &= T^{(1)}_{\bar{q}_i G}(n) = T^{(1)}_1 (n),
  \qquad T^{(1)}_{G q_i}(n) = T^{(1)}_{G\bar{q}_i}(n) = T^{(1)}_2 (n), \nonumber\\
  T^{(1)}_{GG}(n) &= -\frac{1}{b} P^{(1)}_{GG}(n) - 2n_{\rm f} T^{(1)}_3(n)
\end{align}
where the Mellin moments $T^{(1)}_1(n)$, $T^{(1)}_2(n)$ and $T^{(1)}_3(n)$ are given by
(the dependence on $n$ is not written out explicitly in the following formulae)
\begin{align}
  T^{(1)}_1 & = \frac{1}{\nu} \biggl[ P^{(0)}_{qG} \left( P^{(0)}_{qq} - P^{(0)}_{GG} - b
  \right) \left( P^{(1)V}_{qq}\! + P^{(1)V}_{q\bar{q}}\! + 2n_{\rm f} P^{(1)S}_{qq} - P^{(1)}_{GG}
  \right) + {} \biggr. \nonumber\\ & \biggl. \;\;\;\; {} + \left( bP^{(0)}_{qq} - bP^{(0)}_{GG}
  - b^2 + 4n_{\rm f} P^{(0)}_{qG} P^{(0)}_{Gq} \right) P^{(1)}_{qG} + 4n_{\rm f} \left(
  P^{(0)}_{qG} \right) ^2 P^{(1)}_{Gq}\, \biggr] , \displaybreak[0]\\
  T^{(1)}_2 & = \frac{1}{\nu} \biggl[ P^{(0)}_{Gq} \left( P^{(0)}_{qq} - P^{(0)}_{GG} + b
  \right) \left( P^{(1)V}_{qq}\! + P^{(1)V}_{q\bar{q}}\! + 2n_{\rm f} P^{(1)S}_{qq} - P^{(1)}_{GG}
  \right) + {} \biggr. \nonumber\\ & \biggl. \;\;\;\; {} + \left( bP^{(0)}_{GG} - bP^{(0)}_{qq}
  - b^2 + 4n_{\rm f} P^{(0)}_{qG} P^{(0)}_{Gq} \right) P^{(1)}_{Gq} + 4n_{\rm f} \left(
  P^{(0)}_{Gq} \right) ^2 P^{(1)}_{qG}\, \biggr] , \displaybreak[0]\\
  T^{(1)}_3 & = \frac{1}{\nu} \biggl[ P^{(0)}_{qG} \left( P^{(0)}_{qq} - P^{(0)}_{GG} + b
  \right) P^{(1)}_{Gq} + P^{(0)}_{Gq} \left( P^{(0)}_{qq} - P^{(0)}_{GG} - b \right)
  P^{(1)}_{qG} - {} \biggr. \nonumber\\ & \biggl. \;\;\;\; {} - 2P^{(0)}_{qG} P^{(0)}_{Gq}
  \left( P^{(1)V}_{qq}\! + P^{(1)V}_{q\bar{q}}\! + 2n_{\rm f}P^{(1)S}_{qq} - P^{(1)}_{GG}
  \right) \biggr]
\end{align}
and the denominator $\nu$ is expressed as
\begin{equation}
  \nu = b \left( b^2 - \left( P^{(0)}_{qq} - P^{(0)}_{GG} \right) ^2 - 8n_{\rm f}
  P^{(0)}_{qG} P^{(0)}_{Gq} \right) \! .
\end{equation}

To obtain the functions $T^{(1)}_{ij}(x)$ in the $x$-space, it is necessary to determine
the Mellin inversion of $T^{(1)}_1(n)$, $T^{(1)}_2(n)$ and $T^{(1)}_3(n)$, which has to
be performed numerically. This was carried out for three and four (massless) quark
flavours with the result
\begin{equation}
  T^{(1)}_i (x) \approx C_i x^{-\xi} \quad\text{for}\quad x \lesssim 0.1
\end{equation}
where $\xi \doteq 4.63$ for $n_{\rm f} = 3$ and $\xi \doteq 3.85$ for $n_{\rm f} = 4$.
The values of the coefficients $C_i$ are then such that the functions $T^{(1)}_i (x)$
dominate over the NLO splitting functions $P^{(1)}_{ij}(x)$ for $x \lesssim 0.1$, which
means that
\begin{equation}
  T^{(1)}_{ij} (x) \approx C_{ij} x^{-\xi} \quad\text{for}\quad x \lesssim 0.1
\end{equation}
with the same value of $\xi$ as in the case of $T^{(1)}_i (x)$. This low $x$ behaviour
of the functions $T^{(1)}_{ij}(x)$ is in agreement with the fact that the ZERO factorization
scheme does not satisfy the condition (\ref{sdconofpracapp}). Since the rapid growth
of the absolute value of $T^{(1)}_{ij}(x)$ occurs at $x \approx 0.1$, it is likely that
the range of the practical applicability of the ZERO factorization scheme is restricted.

\subsection{The range of practical applicability of the ZERO factorization scheme}

The low $x$ behaviour of the functions $T^{(1)}_{ij}(x, \overline{\rm MS}, {\rm ZERO})$
indicates that the ZERO factorization scheme has some restrictions on its practical
applicability. In this subsection, we will investigate the practical applicability
of the ZERO factorization scheme in the case of the structure function $F_2\!\left(
x, Q^2 \right)$.

A detailed numerical analysis of the ZERO factorization scheme was performed for three
massless quark flavours. Using formula (\ref{sdtranspdfexp}), the ZERO parton distribution
functions at the factorization scale $M_{\rm T} = 100\, {\rm GeV}$ were calculated from
the $\overline{\rm MS}$ ones that corresponded to the MRST 1998 set \cite{mrst}. The ZERO parton
distribution functions at a general factorization scale $M$ were then determined by
the evolution from $M_{\rm T}$ to $M$ in the ZERO factorization scheme.
The theoretical predictions for $F_2\!\left( x, Q^2 \right)$ were obtained by
the numerically calculated Mellin inversion of the Mellin moments $F_2\!\left( n, Q^2
\right)$, which did not include any contributions concerning heavy quark flavours
(only three (effectively) massless quark flavours were taken into account).
Hence, the obtained theoretical predictions for $F_2\!\left( x, Q^2 \right)$
cannot be compared with experimental data, but are applicable for the assesment
of the practical applicability of the ZERO factorization scheme, which is our aim.
\begin{figure}
  \centering
  \includegraphics[width=0.4\textwidth,angle=90]{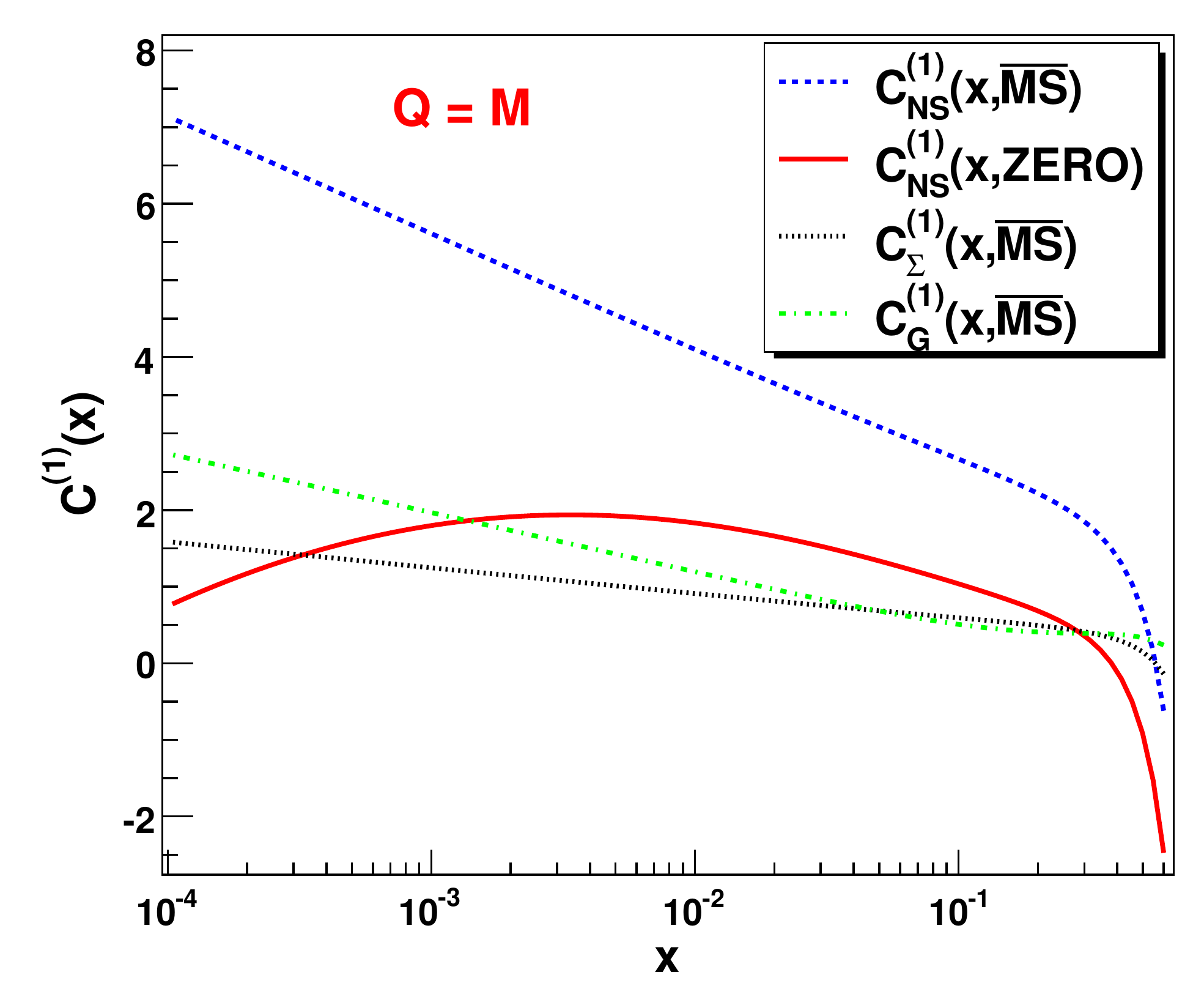}
  \includegraphics[width=0.4\textwidth,angle=90]{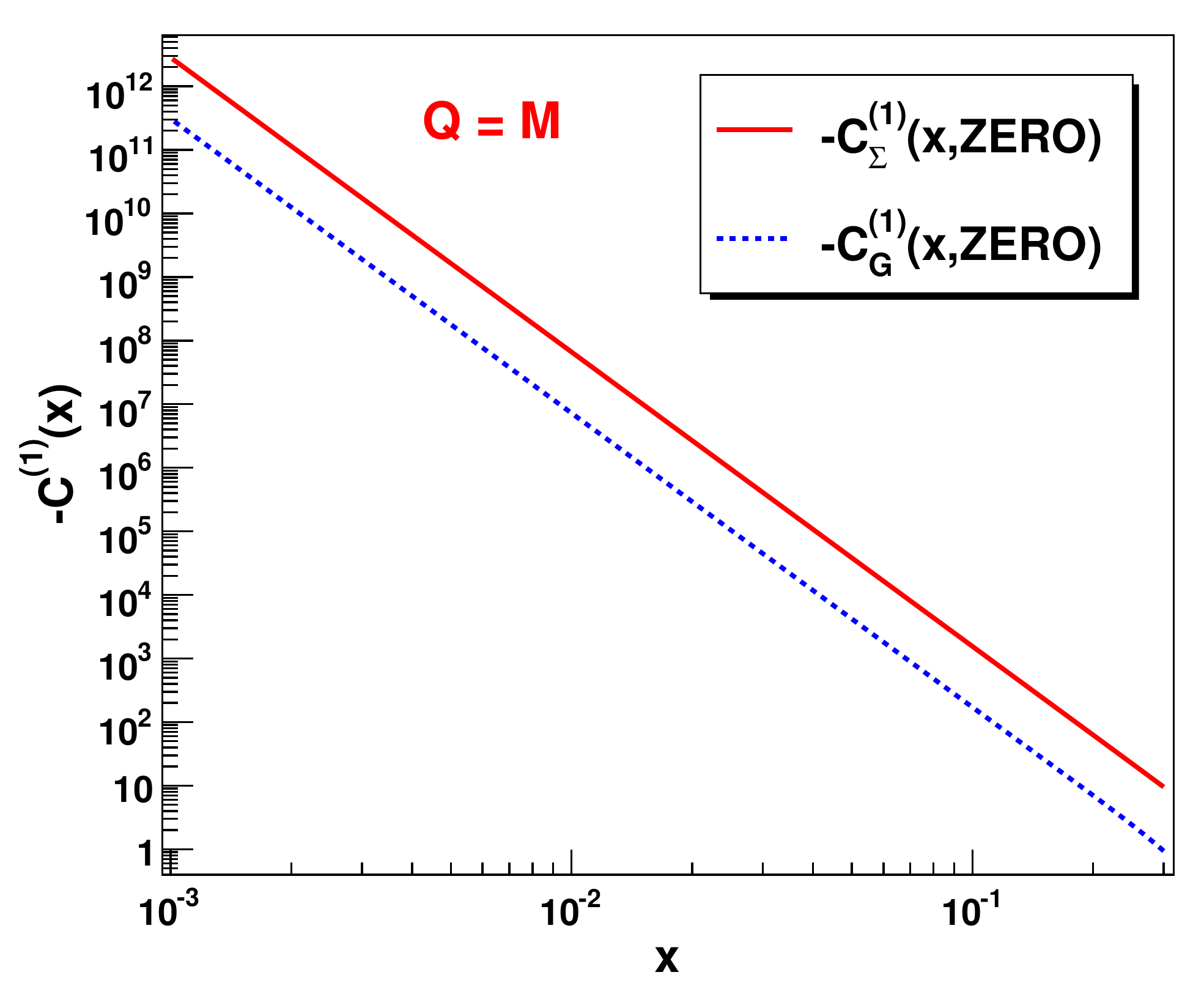}
  \caption{Coefficient functions corresponding to the structure function $F_2\!
  \left( x, Q^2 \right)$ in the $\overline{\rm MS}$ and ZERO factorization scheme
  for $n_{\rm f}=3$. Note that because of the logarithmic scale, the coefficient
  functions in the right graph are plotted with the negative sign.}
  \label{figcoefffunc}
\end{figure}
\begin{figure}
  \centering
  \includegraphics[width=0.4\textwidth,angle=90]{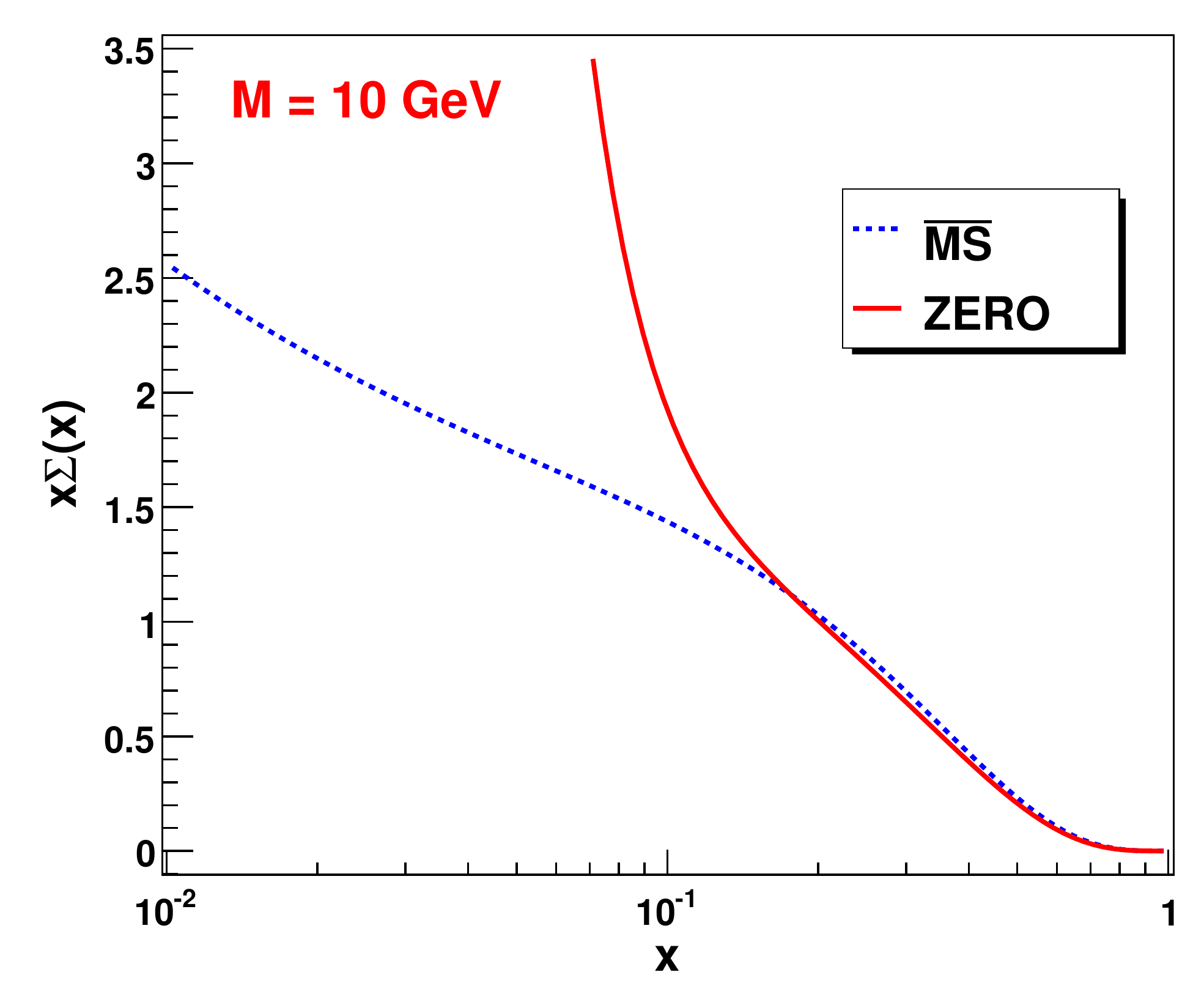}
  \includegraphics[width=0.4\textwidth,angle=90]{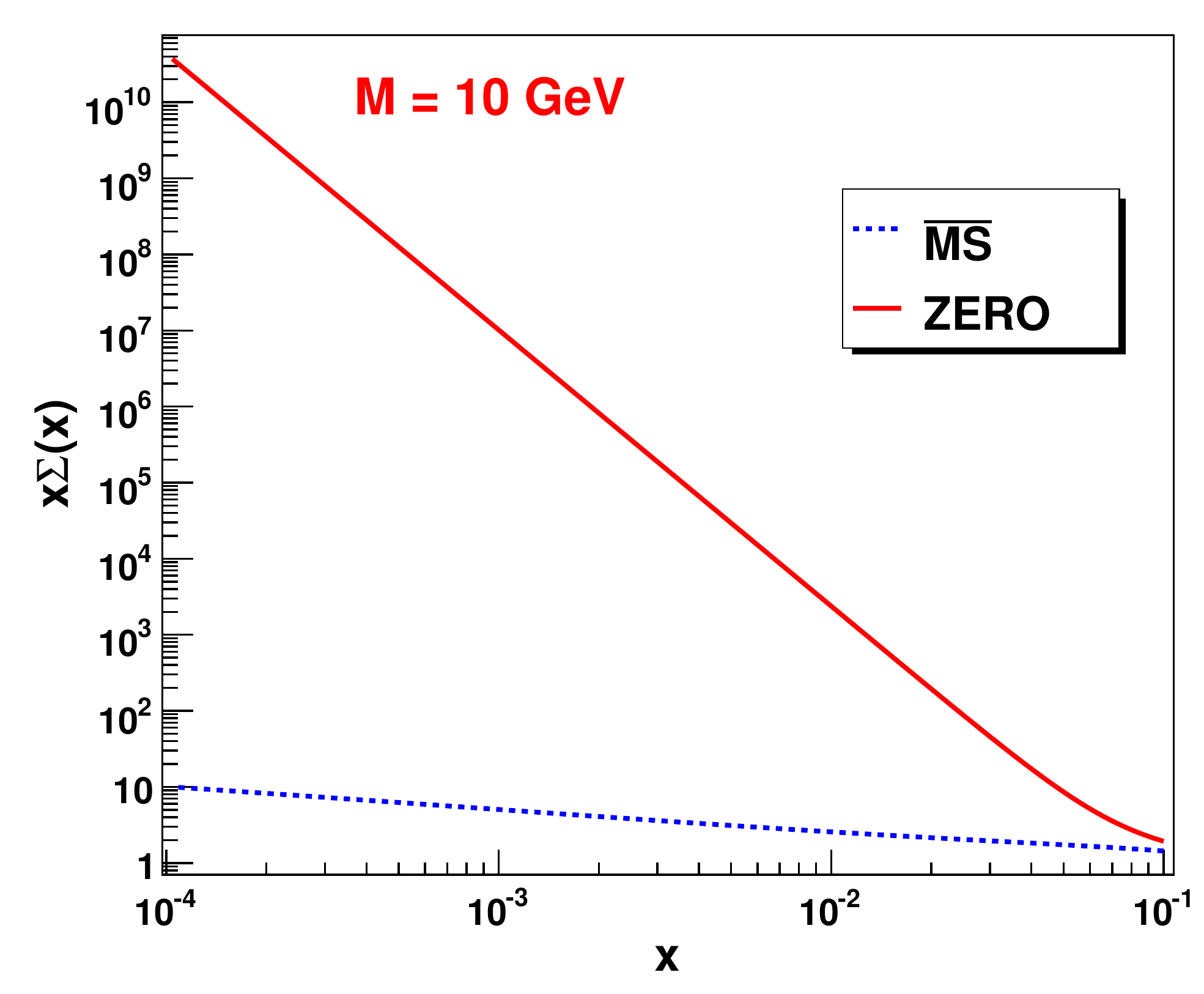}
  \includegraphics[width=0.4\textwidth,angle=90]{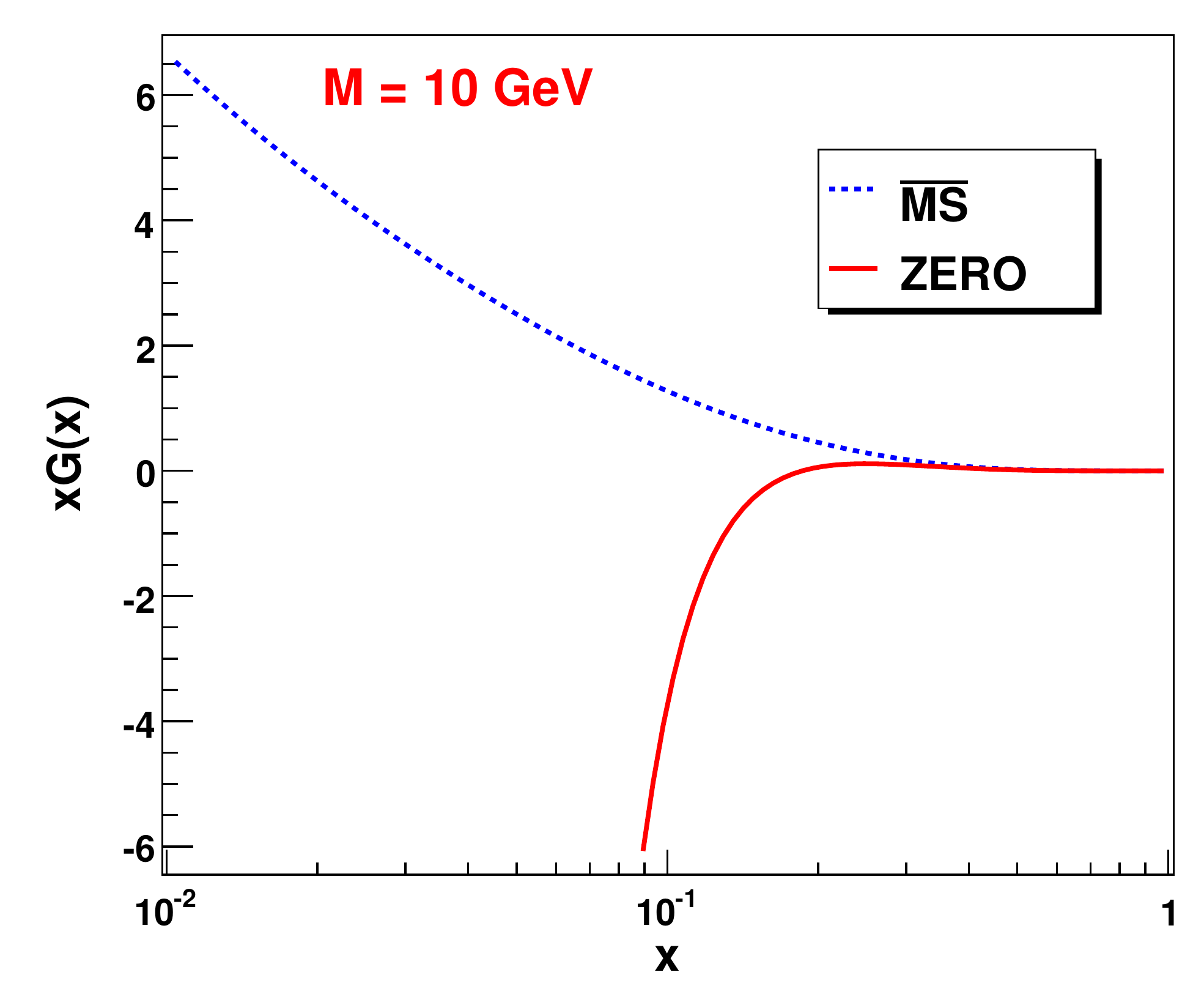}
  \includegraphics[width=0.4\textwidth,angle=90]{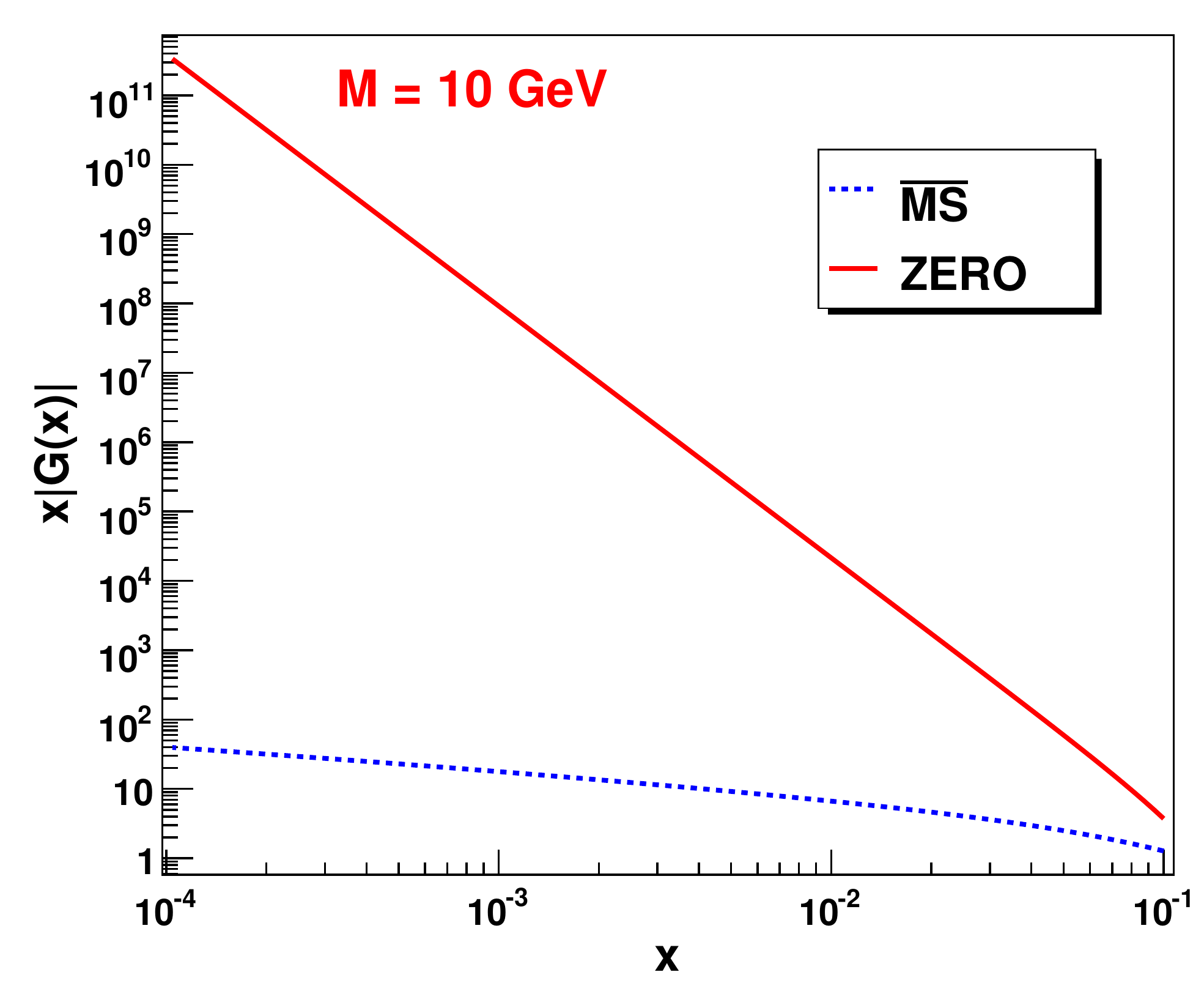}
  \caption{Quark singlet and gluon distribution function in the $\overline{\rm MS}$
  and ZERO factorization scheme for $n_{\rm f}=3$. Note that the gluon distribution
  functions in the lower right graph are plotted in their absolute value ---
  whereas the $\overline{\rm MS}$ distribution is positive, the ZERO distribution
  is negative (see the lower left graph).}
  \label{figpdfthree}
\end{figure}

The right graph in Figure \ref{figcoefffunc} shows that the quark singlet and gluon NLO
coefficient function behave for low $x$ in the same way as the functions $T^{(1)}_{ij}(x,
\overline{\rm MS}, {\rm ZERO})$, and the same is true for the quark singlet and gluon
distribution function, as can be seen from Figure \ref{figpdfthree}. The low $x$ behaviour
of these distribution functions implies that if some physical quantity depends on the
values of these distribution functions in the region $x \lesssim 0.1$, then obtaining
a reasonable theoretical prediction for this quantity requires a considerable mutual
cancellation of large values in the appropriate formula. It is likely that the sufficient
cancellation occurs, if ever, only for some choices of the renormalization and factorization
scale, which means that most of the choices result in unreliable theoretical predictions.
Moreover, the sufficient cancellation causes complications in numerical computations, which
can be hardly soluble.
\begin{figure}
  \centering
  \includegraphics[width=0.4\textwidth,angle=90]{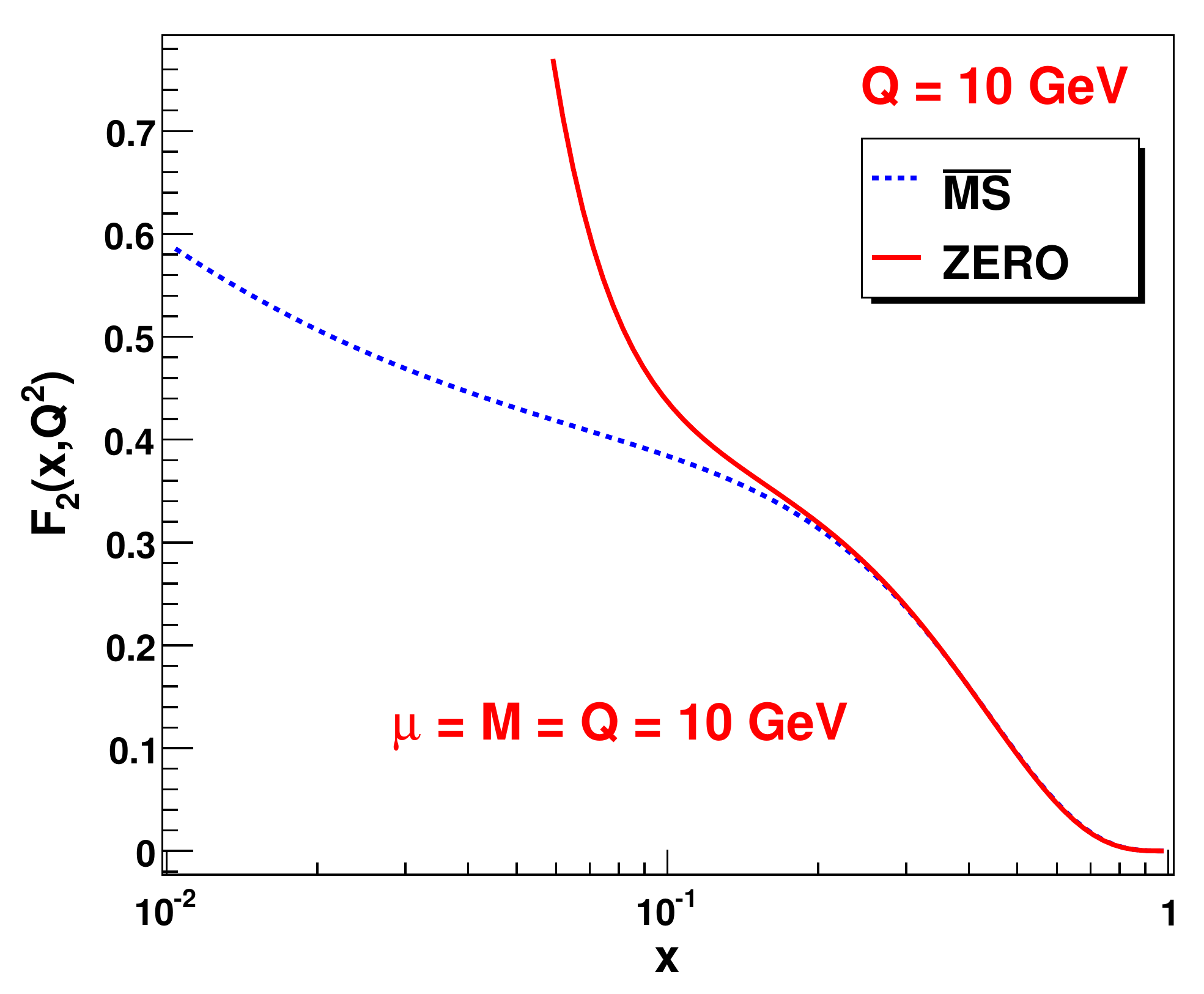}
  \includegraphics[width=0.4\textwidth,angle=90]{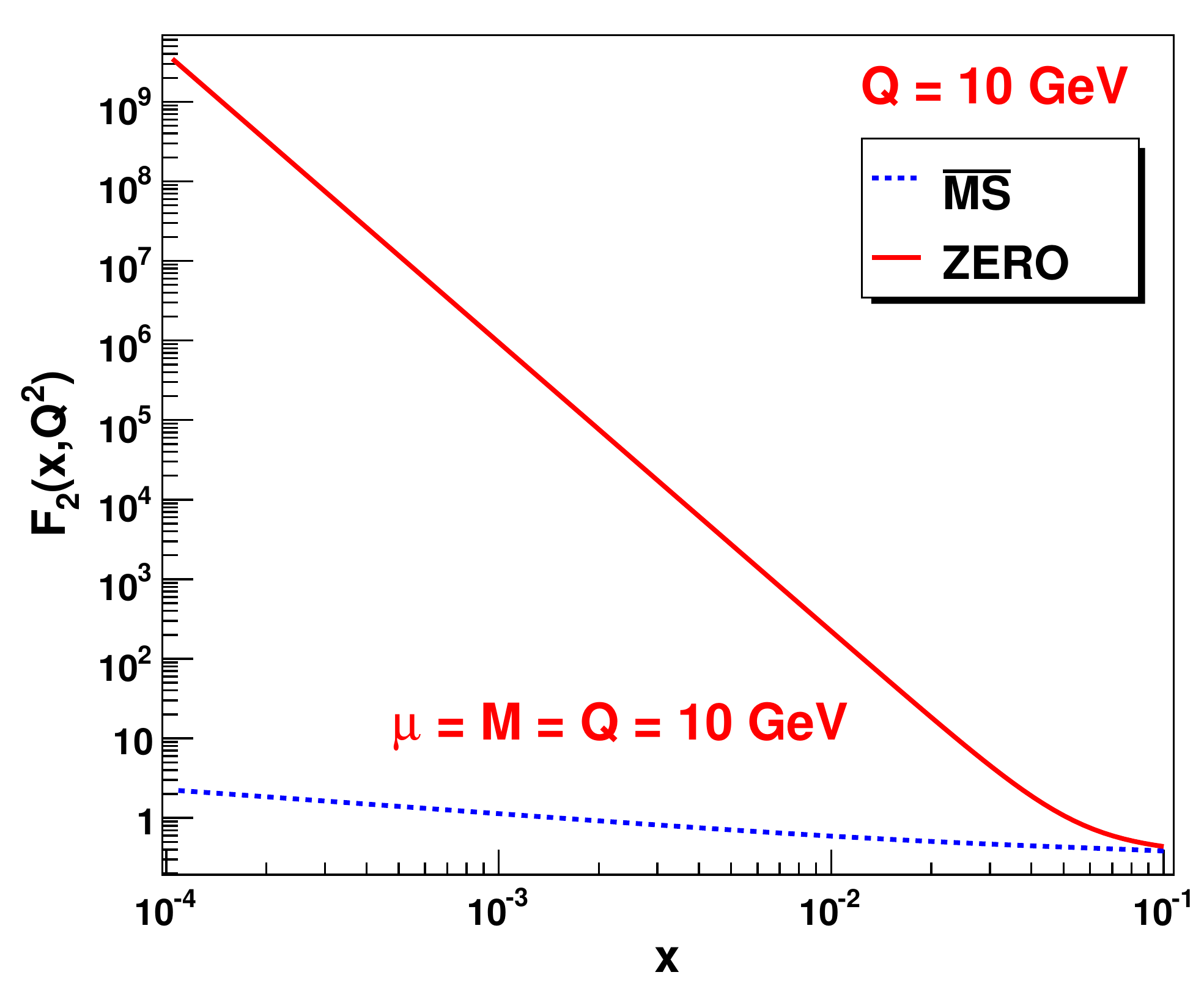}
  \includegraphics[width=0.4\textwidth,angle=90]{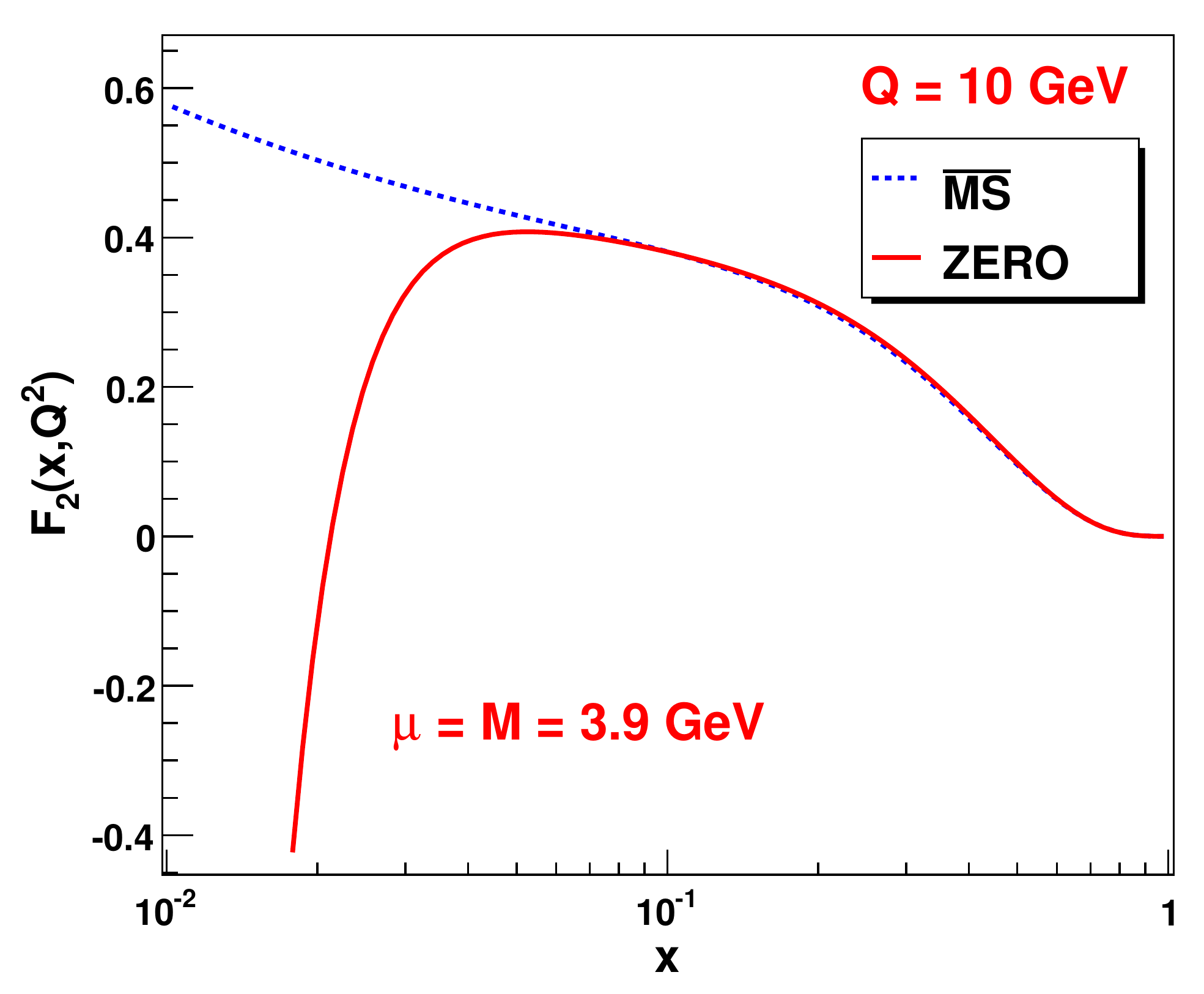}
  \includegraphics[width=0.4\textwidth,angle=90]{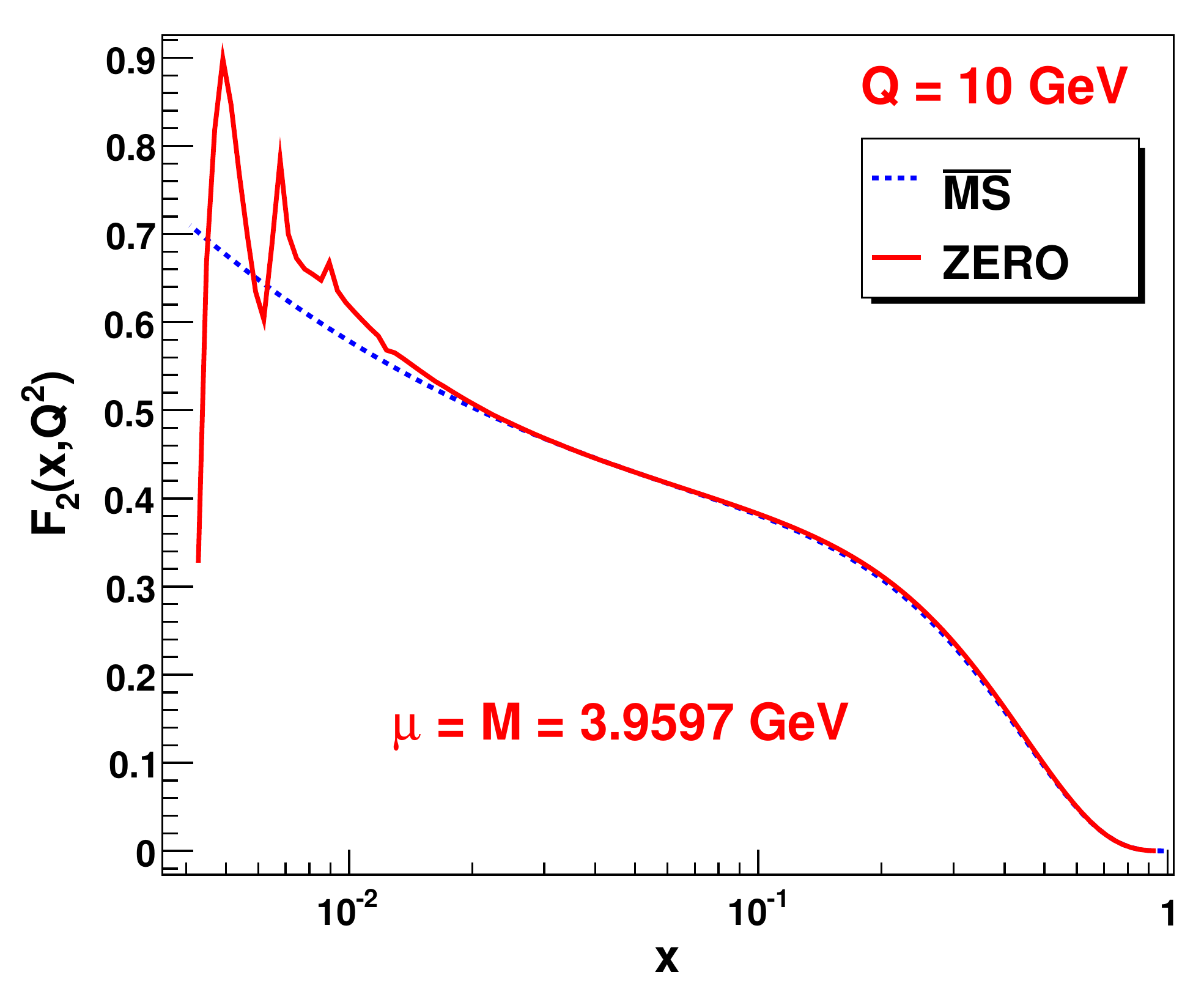}
  \caption{Theoretical predictions for $F_2\! \left( x, Q^2 \right)$ in the
  $\overline{\rm MS}$ and ZERO factorization scheme for $n_{\rm f}=3$.}
  \label{figthpredthree}
\end{figure}
\begin{figure}
  \centering
  \includegraphics[width=0.4\textwidth,angle=90]{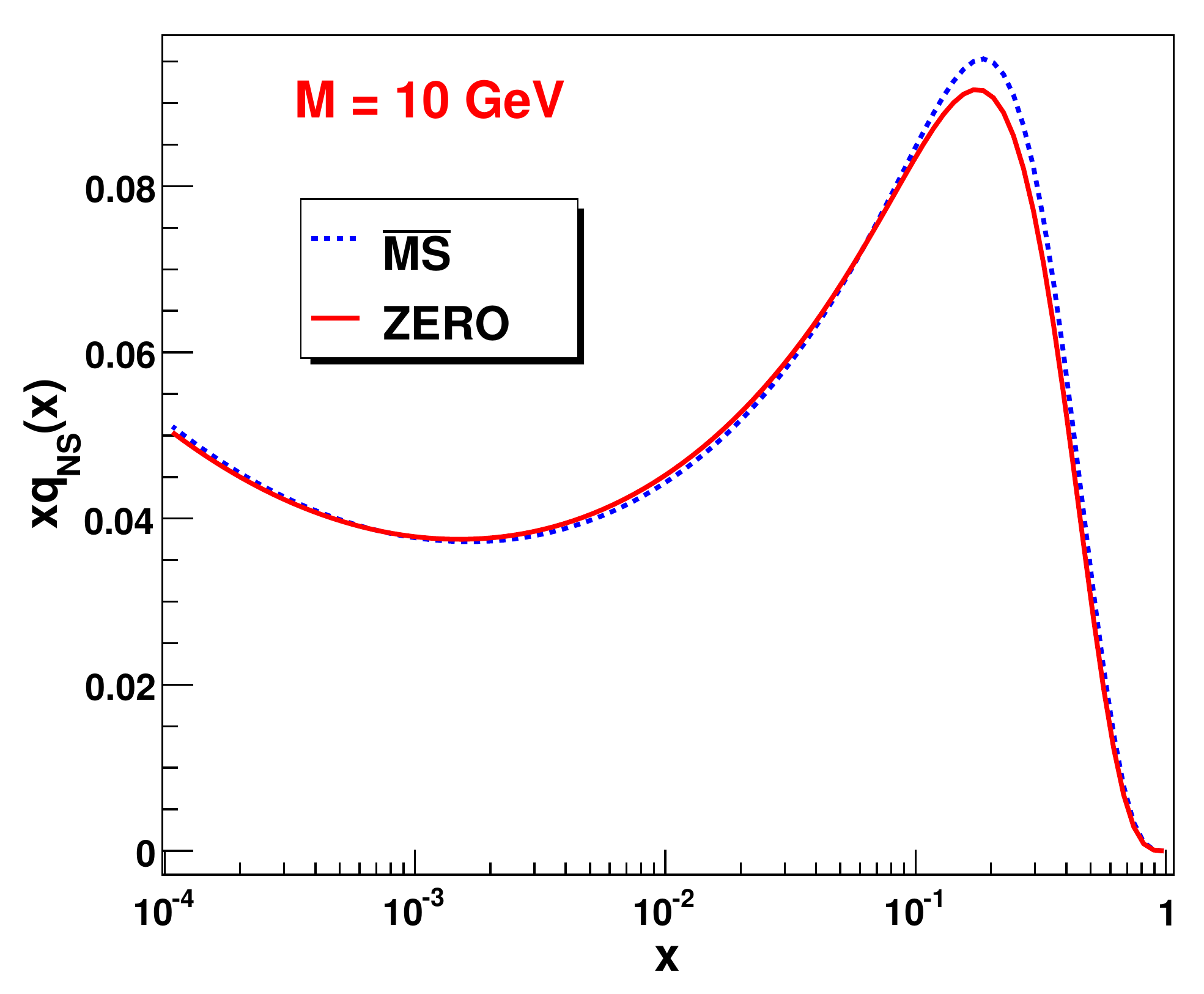}
  \includegraphics[width=0.4\textwidth,angle=90]{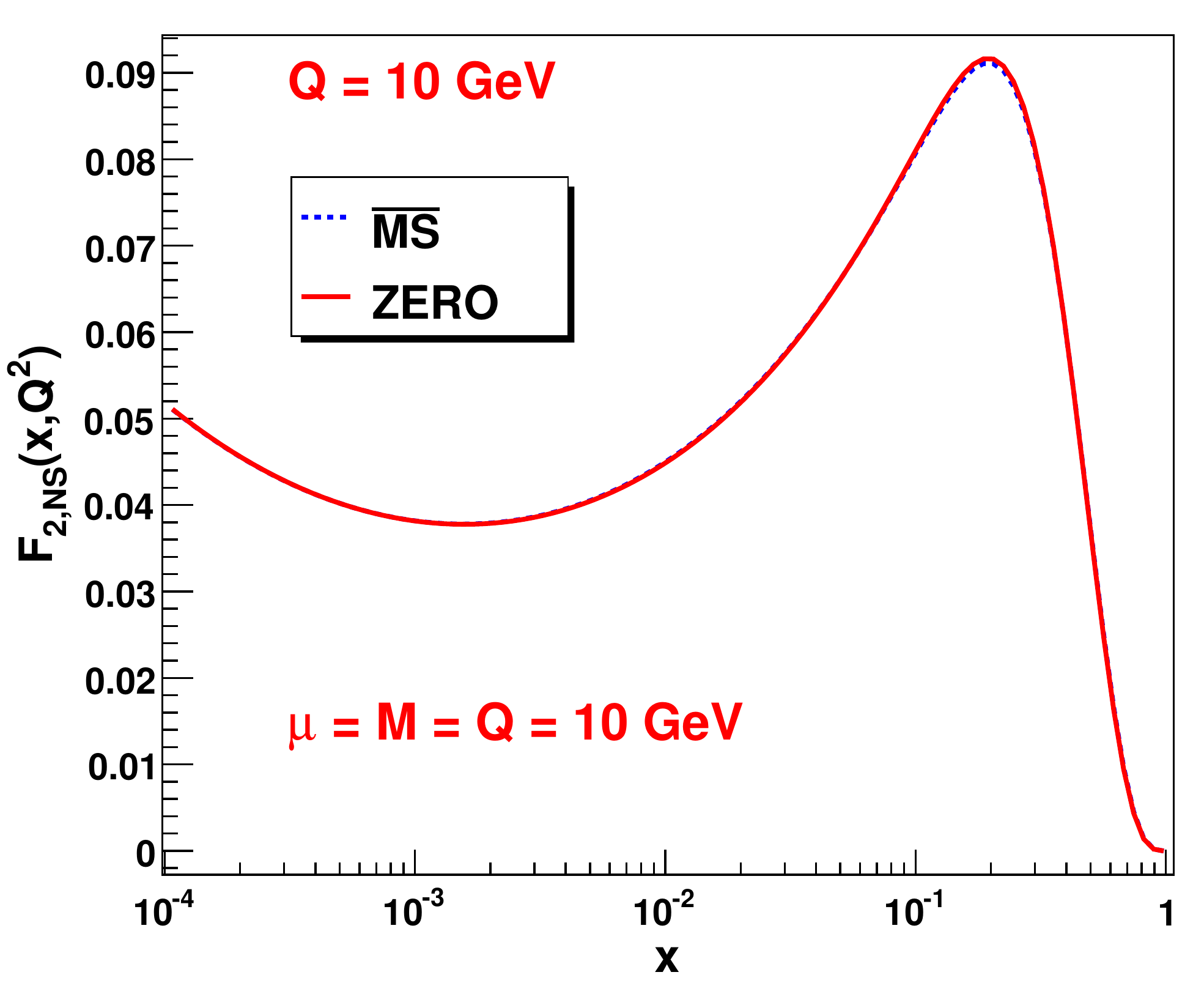}
  \caption{Quark non-singlet distribution function and the non-singlet
  contribution $F_{2,{\rm NS}}\!\left( x, Q^2 \right)$ to $F_2\! \left( x,
  Q^2 \right)$ in the $\overline{\rm MS}$ and ZERO factorization scheme
  for $n_{\rm f} = 3$.}
  \label{fignonsinglet}
\end{figure}

Both undesirable facts manifest themselves in Figure \ref{figthpredthree}. The upper
graphs show a theoretical prediction for $F_2\! \left( x, Q^2 \right)$, which is
unreliable for $x \lesssim 0.1$. In this case, the theoretically predicted
$F_2\! \left( x, Q^2 \right)$ rapidly grows with decreasing $x$ in the region
$x \lesssim 0.1$. The lower left graph indicates that there are some theoretical
predictions for $F_2\! \left( x, Q^2 \right)$ which rapidly fall with decreasing
$x$. Hence, the range of the theoretical predictions for $F_2\! \left( x, Q^2 \right)$
is very large for $x \lesssim 0.1$, and therefore the ZERO factorization scheme
for $n_{\rm f} = 3$ has practically no predictive power in the region $x \lesssim 0.1$.
The range of the theoretical predictions for $F_2\! \left( x, Q^2 \right)$ should
include reasonable values, which means that for a given $x_0$, there exist such
choices of the renormalization and factorization scale that the corresponding
theoretical predictions for the value of $F_2\! \left( x_0, Q^2 \right)$ are
reasonable (but this does not mean that there must be such a choice of the
renormalization and factorization scale that the corresponding theoretical
prediction for $F_2\! \left( x, Q^2 \right)$ is reasonable for all $x$). However,
obtaining the reasonable theoretical predictions in numerical calculations is
difficult, which can be seen in the lower right graph, where the used precision
of the numerical computation is insufficient for $x \lesssim 0.02$. The comparison
of the lower graphs shows that the reasonable theoretical predictions in the
region $x \lesssim 0.1$ are very sensitive to the choice of the renormalization
and factorization scale, which is a consequence of the sensitivity of the extent
of the mutual cancellation of large values to this choice.
\begin{figure}
  \centering
  \includegraphics[width=0.4\textwidth,angle=90]{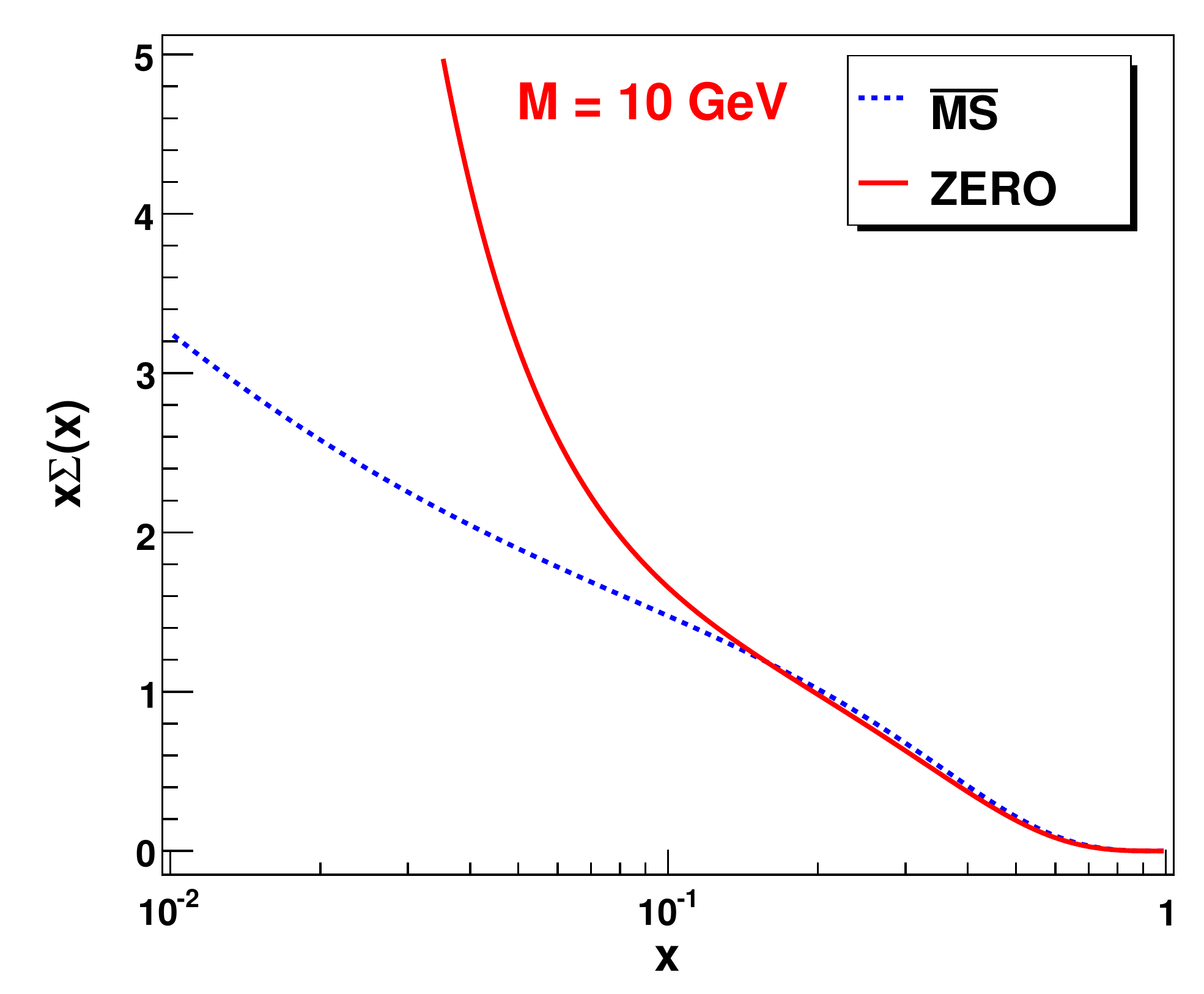}
  \includegraphics[width=0.4\textwidth,angle=90]{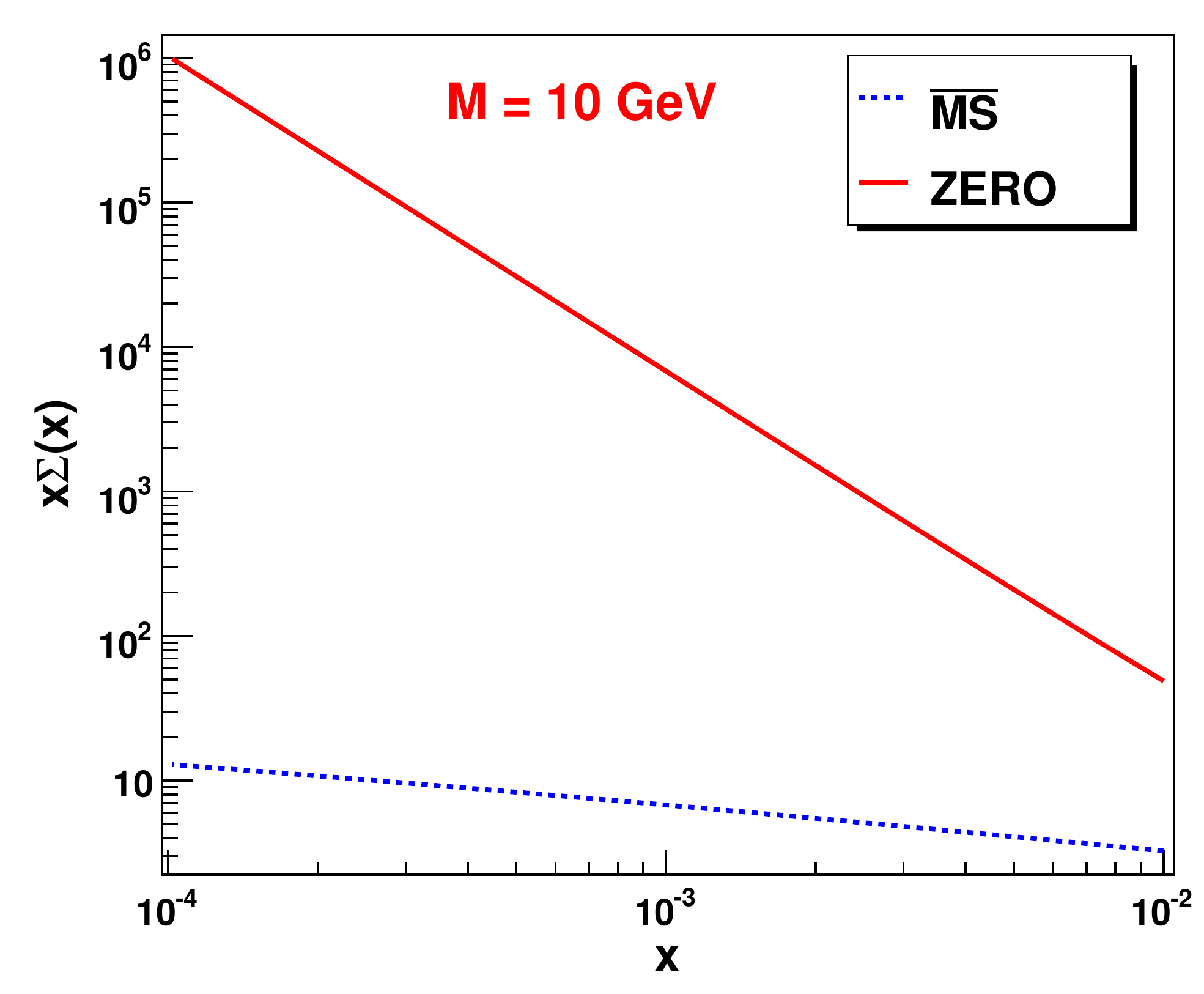}
  \includegraphics[width=0.4\textwidth,angle=90]{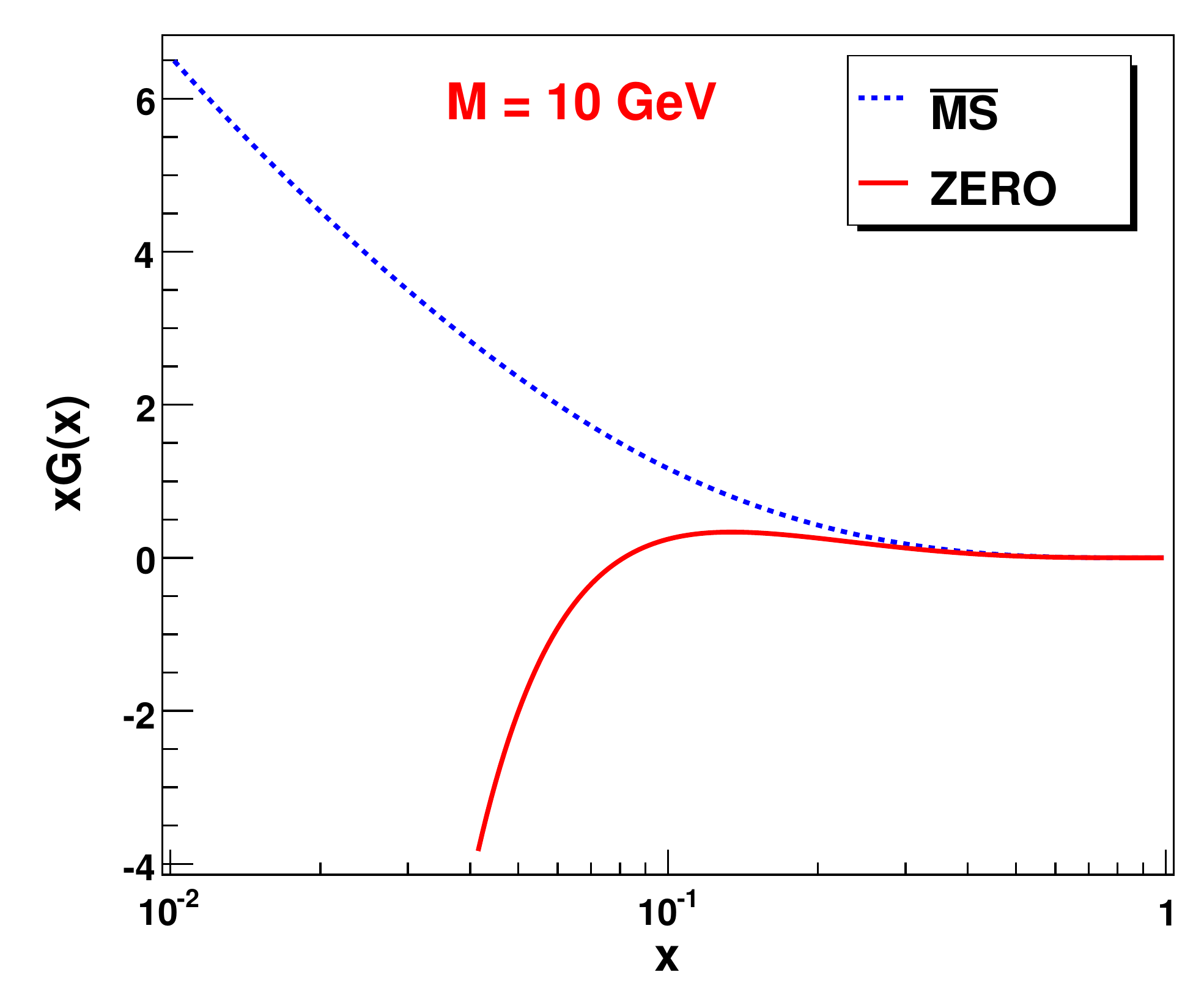}
  \includegraphics[width=0.4\textwidth,angle=90]{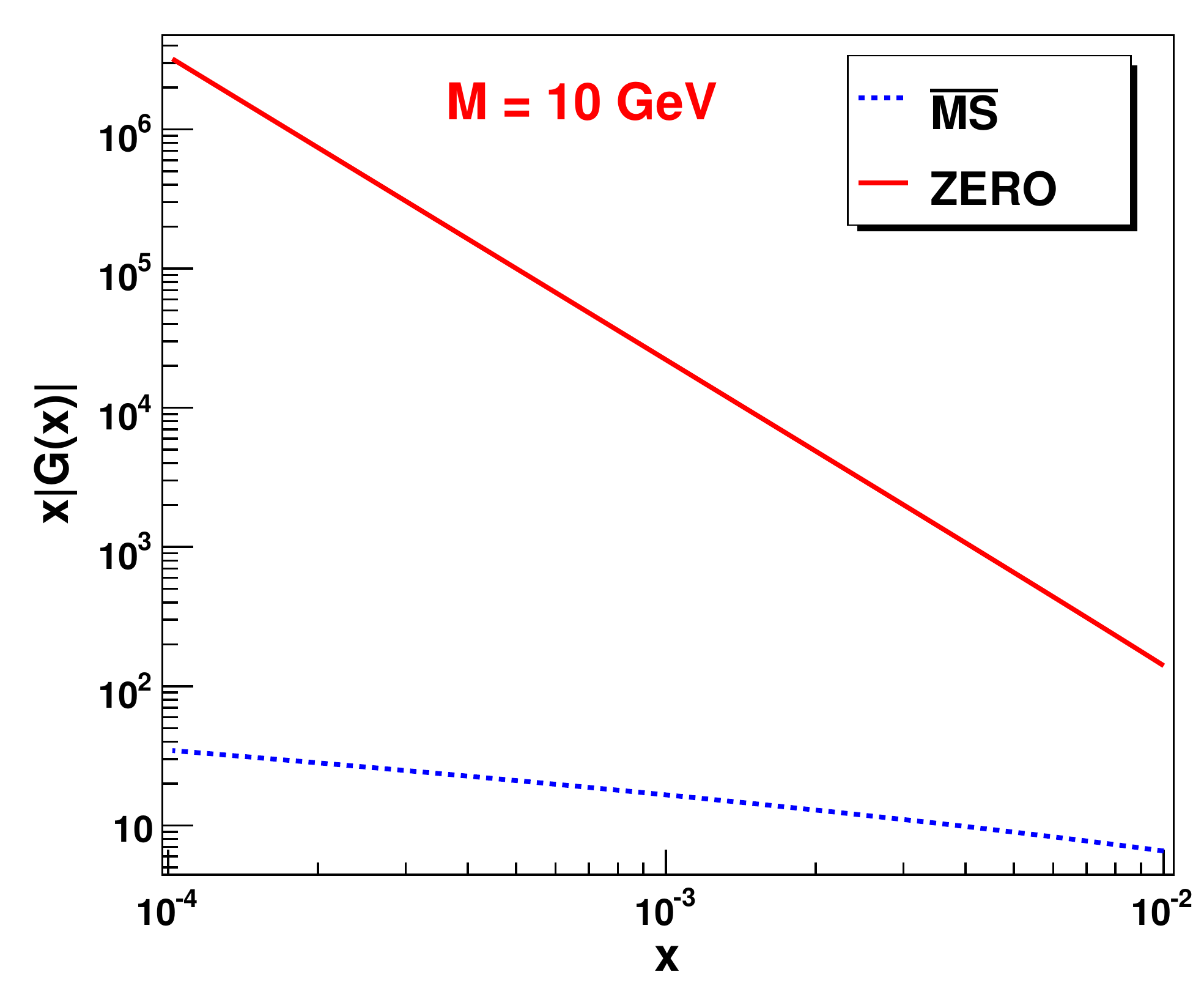}
  \caption{Quark singlet and gluon distribution function in the $\overline{\rm MS}$
  and ZERO factorization scheme for $n_{\rm f}=5$. Note that the gluon distribution
  functions in the lower right graph are plotted in their absolute value ---
  whereas the $\overline{\rm MS}$ distribution is positive, the ZERO distribution
  is negative (see the lower left graph).}
  \label{figpdffive}
\end{figure}
\begin{figure}
  \centering
  \includegraphics[width=0.4\textwidth,angle=90]{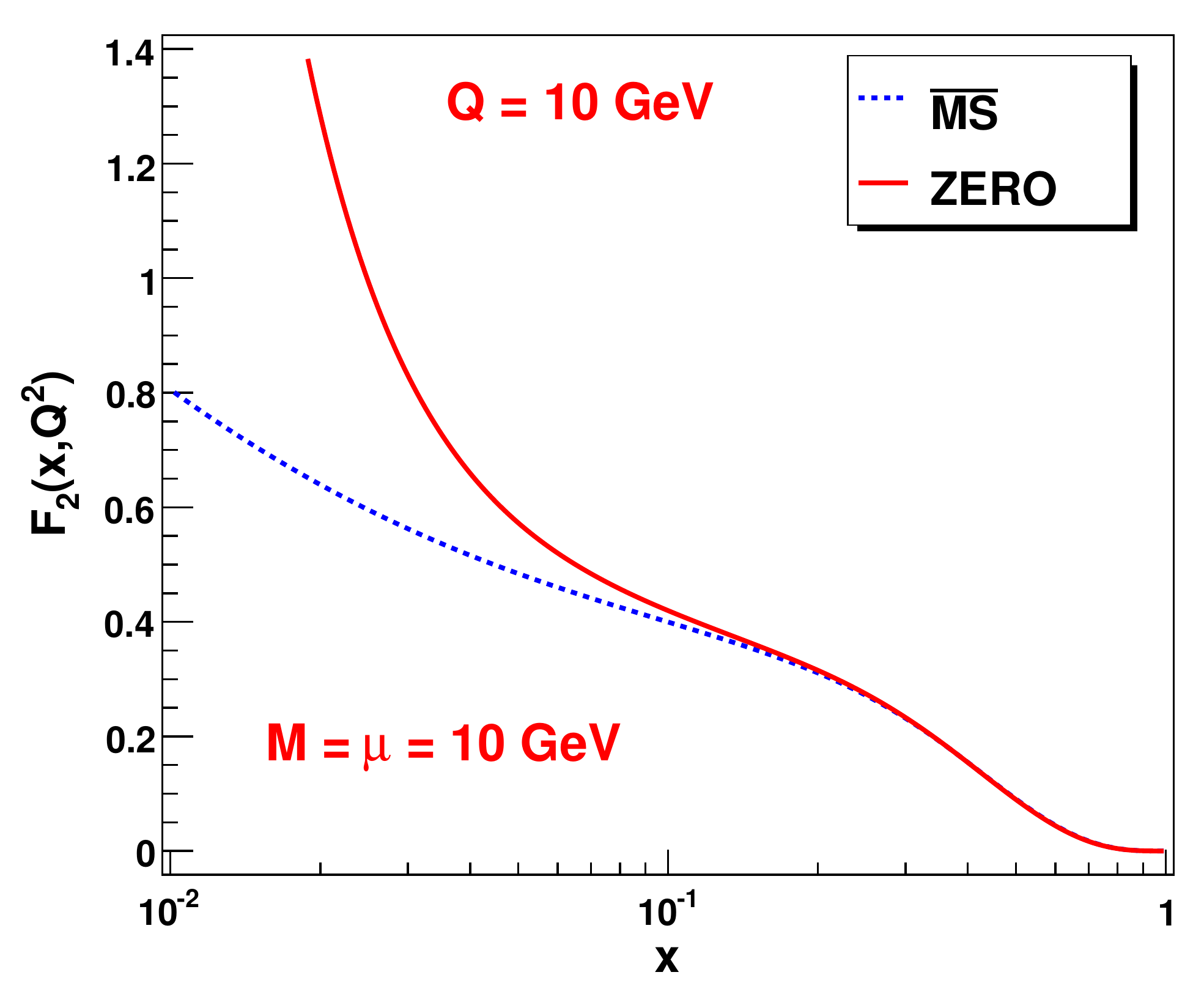}
  \includegraphics[width=0.4\textwidth,angle=90]{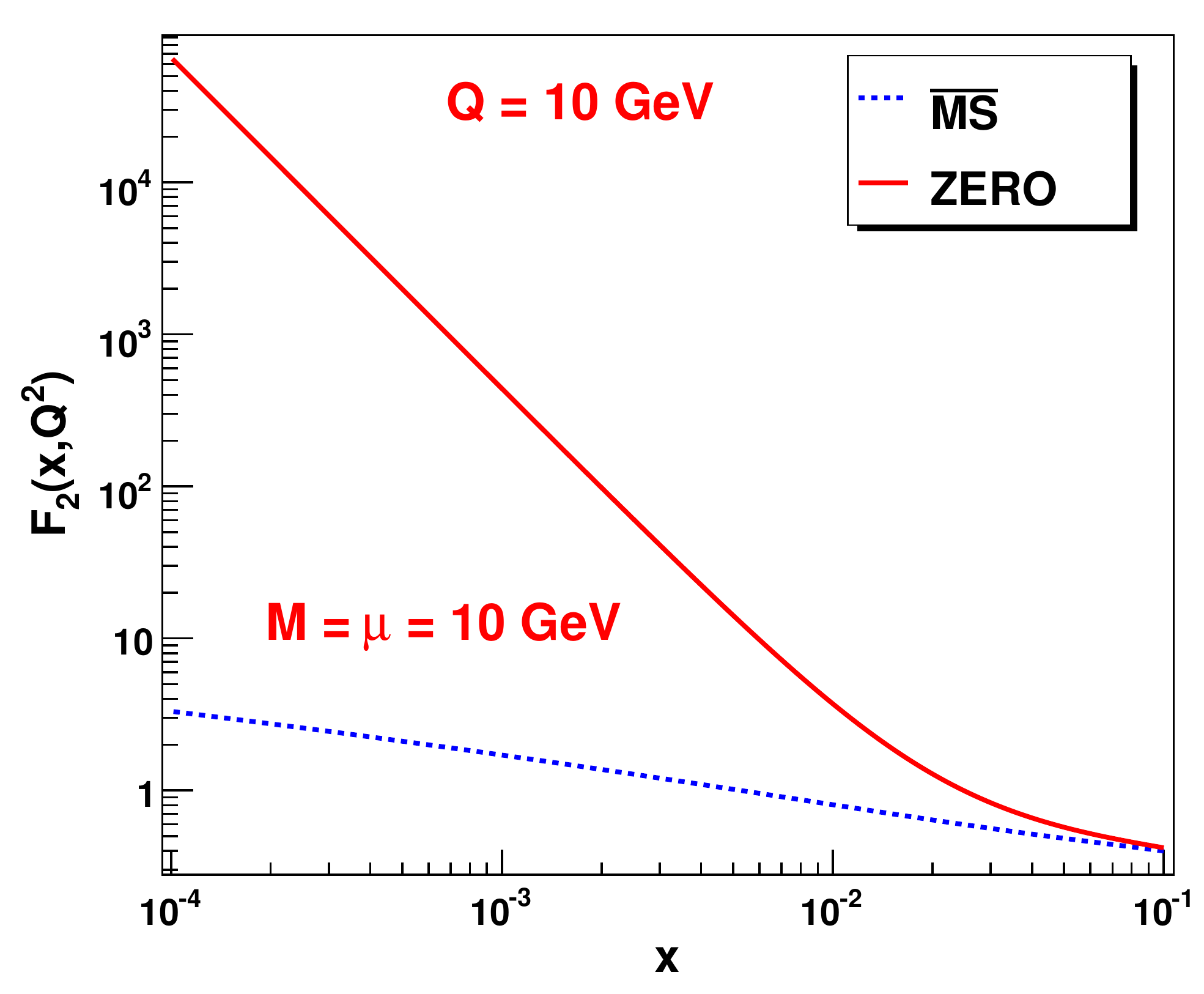}
  \includegraphics[width=0.4\textwidth,angle=90]{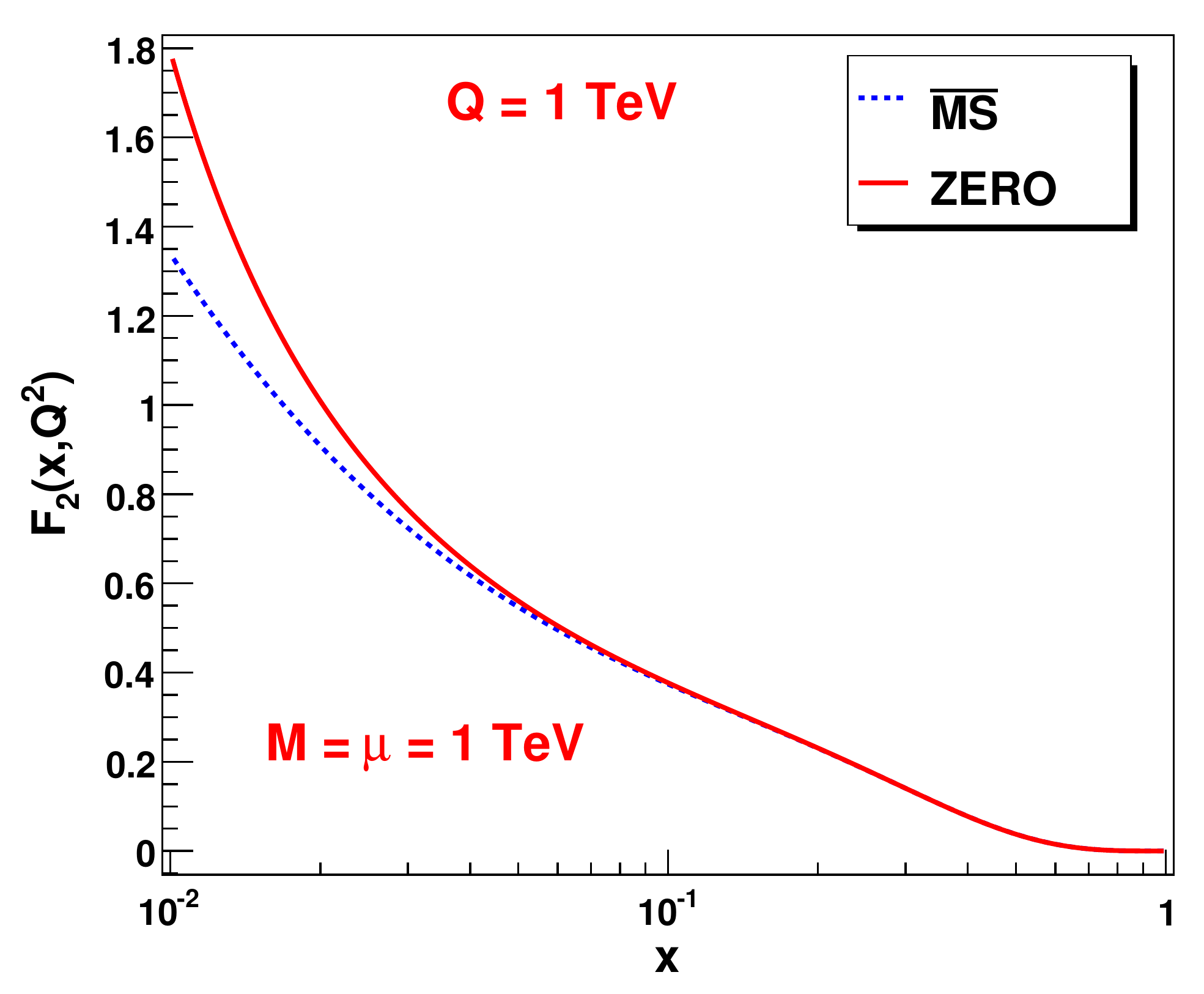}
  \includegraphics[width=0.4\textwidth,angle=90]{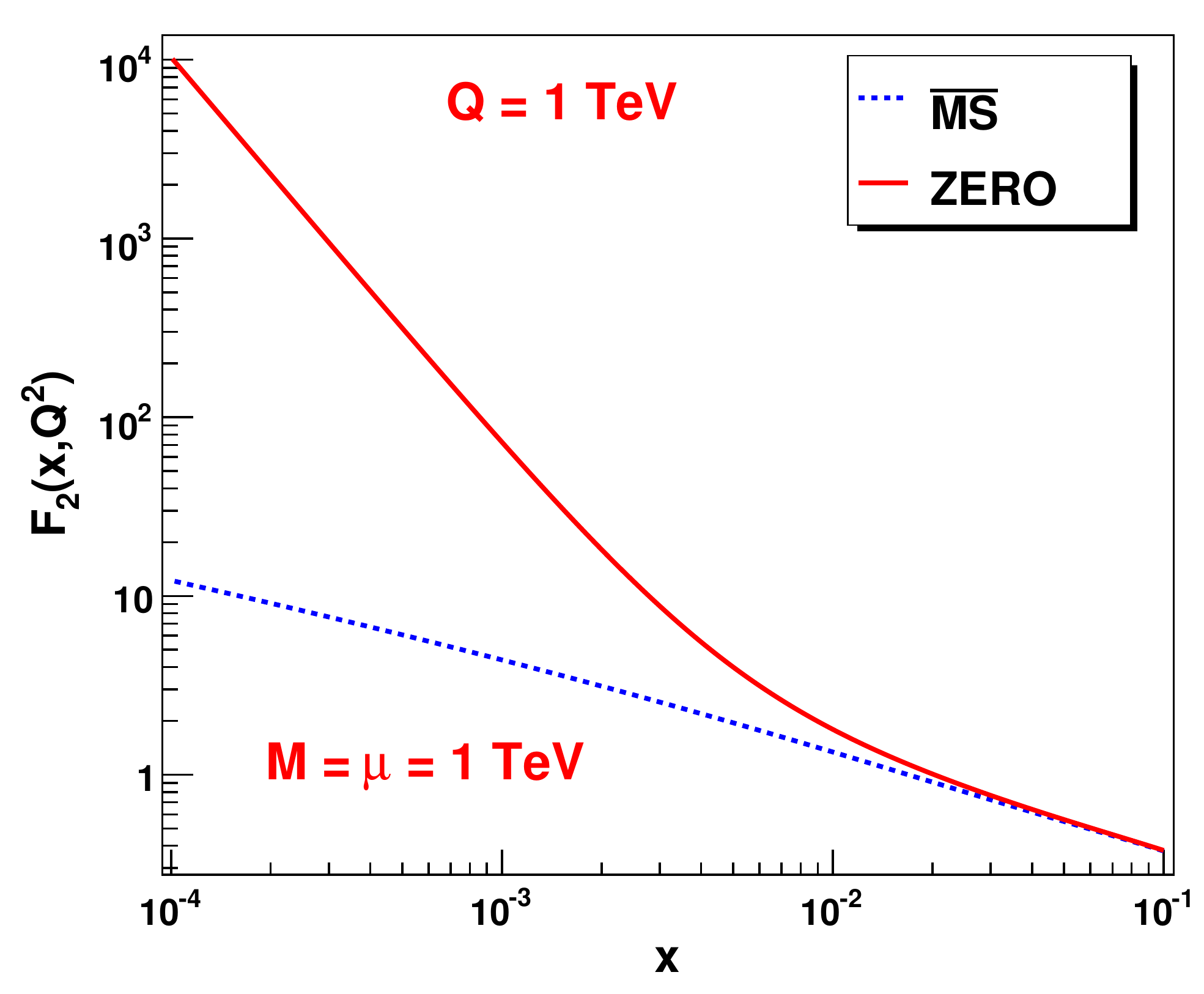}
  \caption{Theoretical predictions for $F_2\! \left( x, Q^2 \right)$ in the
  $\overline{\rm MS}$ and ZERO factorization scheme for $n_{\rm f}=5$.}
  \label{figthpredfive}
\end{figure}

According to Subsection \ref{prtpracappfsnlo}, the ZERO factorization scheme should
have no restrictions on its practical applicability in the non-singlet case, which
is demonstrated in \mbox{Figure \ref{fignonsinglet}}.

The functions $T^{(1)}_{ij}(x, \overline{\rm MS}, {\rm ZERO})$ were not calculated
in the $x$-space for five (massless) quark flavours. Since the ZERO factorization
scheme for $n_{\rm f} = 5$ does not satisfy the condition (\ref{sdconofpracapp}),
the functions $T^{(1)}_{ij}(x, \overline{\rm MS}, {\rm ZERO})$ should behave for
low $x$ in the following way (see (\ref{sdnumvalueofroot})):
\begin{equation}
  T^{(1)}_{ij} (x, \overline{\rm MS}, {\rm ZERO}) \approx C_{ij} x^{-\xi}, \qquad
  \xi \doteq 3.18 , \label{selowxbehavtijfive}
\end{equation}
which indicates that the ZERO factorization scheme has some constraints on its practical
applicability, similarly as in the case of $n_{\rm f} = 3$ and $n_{\rm f} = 4$.
The ZERO parton distribution functions and the theoretical predictions for
$F_2\! \left( x, Q^2 \right)$ were calculated in a similar way\footnote{Seeing
that the Mellin moments of the parton distribution functions and the structure function
$F_2\! \left( x, Q^2 \right)$ tend to zero sufficiently fast for large $n$, the numerical
computation of their Mellin inversion is quite easy, which does not hold for the Mellin
inversion of $T^{(1)}_{ij}(n, \overline{\rm MS}, {\rm ZERO})$.} as in the case of
$n_{\rm f} = 3$ (the only difference was the fact that the value of the factorization
scale $M_{\rm T}$ was $10^7\, {\rm GeV}$ and the $\overline{\rm MS}$ parton distribution
functions corresponded to the MSTW 2008 set \cite{mstw}).
The ZERO parton distribution functions displayed in Figure \ref{figpdffive} behave for
low $x$ in a similar way as in the case of $n_{\rm f} = 3$, which is displayed in Figure
\ref{figpdfthree}. The main difference is the rate of the growth, which is in accordance
with relation (\ref{selowxbehavtijfive}) and is thus slower than in the case of
$n_{\rm f} = 3$. The comparison of Figures \ref{figpdfthree} and \ref{figpdffive} also shows
that in the case of $n_{\rm f} = 5$, the rapid growth occurs at a bit lower value of $x$.
However, the difference is not significant. Hence, it is not surprising that
the ZERO factorization scheme for $n_{\rm f} = 5$ gives unreliable theoretical predictions
for $F_2\! \left( x, Q^2 \right)$ in the region of low $x$, as can be seen in Figure
\ref{figthpredfive}. The comparison of the upper and lower graphs in Figure \ref{figthpredfive}
shows that the region of the practical applicability increases slowly with increasing energy,
which also holds for $n_{\rm f} = 3$.

For the number of massless quark flavours from three to five, which represents the
relevant values for QCD phenomenology, the ZERO factorization scheme is fully
applicable for the physical quantities that do not depend on the values of the quark
singlet and gluon distribution function in the region $x \lesssim x_0$ where $x_0 = 0.1$. However,
in the case of the physical quantities which depend on the values of these distribution functions
in the region $x \lesssim x_0$, the ZERO factorization scheme gives unreliable theoretical
predictions and has little predictive power. Even for these physical quantities, obtaining
reasonable theoretical predictions is possible for some special choices of the renormalization and
factorization scale, but requires a considerable mutual cancellation of large values, which causes
difficulties in numerical calculations. In some cases, the value of the bound $x_0$ can
be lowered, however, it is practically impossible to lower the value of $x_0$ in such a way
that the ZERO factorization scheme would be applicable in the full range that is used
in QCD phenomenology because the extent of the undesirable features rapidly grows with
decreasing $x_0$.

As it has already been mentioned, the ZERO factorization scheme puts all NLO corrections
into hard scattering cross-sections and is therefore a certain opposite to the DIS factorization
scheme, in which all NLO corrections (to $F_2\!\left( x, Q^2 \right)$) are included in the
corresponding NLO splitting functions. The restricted practical applicability of the ZERO
factorization scheme is thus surprising because the DIS factorization scheme can be applied
without any restrictions.

There are other reasons why the restricted practical applicability of the ZERO factorization
scheme is unexpected and not obvious. As it has already been shown, the ZERO factorization
scheme has no restrictions of its practical applicability in the non-singlet case. In addition,
the condition of practical applicability (eq.\ (\ref{sdconofpracapp})) is nontrivial and its
consequences depend on the number of massless quark flavours (e.g.\ the growth of the ZERO
quark singlet distribution function in the low $x$ region is a result of the fact that the
ZERO factorization scheme does not satisfy the condition of practical applicability and the
rate of this growth depends nontrivially on the number of massless quark flavours (see
Figures \ref{figpdfthree} and \ref{figpdffive})).

\section{Summary and conclusion}
\label{prtsumandconcl}

In this text, we have analysed the freedom associated with factorization in
massless perturbative QCD, with the emphasis on its quantification.
We have derived the formulae that are useful for changing the unphysical
quantities associated with the renormalization and factorization
procedure and we have shown that factorization schemes can be specified
using the higher order splitting functions, which can be chosen at will.
This allows us to introduce the so called ZERO factorization scheme,
which is defined by setting the higher order splitting functions equal to zero,
which means that the ZERO factorization scheme represents a certain
opposite to the more familiar DIS factorization scheme. The potential
exploitation of the ZERO factorization scheme for the construction of
consistent NLO Monte Carlo event generators in which initial state parton
showers can be taken formally at the LO has provided the main motivation for
this study. The related questions have been discussed in more detail
at the end of Section \ref{prtfactandnotation}.

The discussion of the practical applicability of general factorization schemes
has shown that the factorization schemes specified by splitting functions
can have unexpected restrictions on its practical applicability. Even if
splitting functions appear at first sight as reasonable, they can specify
such a factorization scheme that the values of the appropriate parton
distribution functions are significantly larger in the low $x$
region than in the case of the standard $\overline{\rm MS}$
factorization scheme. The cancellation of these large values, which has to
occur in the expressions for physical quantities, can then result in unreliable
theoretical predictions, little predictive power and difficulties in numerical
calculations, which restricts the practical applicability of such a factorization
scheme.

The question of the practical applicability of general factorization schemes
has been studied in more detail at the NLO.
We have found the condition (eq.\ (\ref{sdconofpracapp})) that must
be satisfied by NLO splitting functions in order that any unexpected restrictions
on the practical applicability of the corresponding factorization scheme are ruled out.
Unfortunately, this condition is not satisfied for the ZERO factorization scheme.
The numerical analysis of the ZERO factorization scheme at the NLO has then shown
that its application is reliable for the physical quantities which do not depend
on the values of the quark singlet and gluon distribution function in the region
$x \lesssim 0.1$. However, if we apply the ZERO factorization scheme for the other
physical quantities (i.e.\ those which depend on the values of these distribution
functions in the region $x \lesssim 0.1$), then the undesirable features mentioned
in the preceding paragraph can occur. The restricted practical applicability
of the ZERO factorization scheme has been illustrated on the structure function
$F_2\!\left( x, Q^2 \right)$ in the low $x$ region. However, even for the physical
quantities for which the ZERO factorization scheme is unreliable, there should
exist some special choices of the renormalization and factorization scale that
result in reasonable theoretical predictions, but the exploitation of this fact
in Monte Carlo event generators is practically impossible because obtaining
reasonable theoretical predictions in this case requires a considerable mutual
cancellation of large values.

Although the ZERO factorization scheme has some restrictions of its practical
applicability, its phenomenological exploitation still makes sense because
it can be applied for the description of the production of heavy objects,
which is important in searches for new physics.

\appendix

\section{Mellin transform and related formulae}
\label{prtmellintrans}

The Mellin transform of a function $f(x)$ defined on the interval $(0,1)$ is given as
\begin{equation}
  f(n) = \int_0^1 x^{n-1} f(x)\, {\rm d}x   \label{abdefmeltrans}
\end{equation}
where $n$ is in general a complex number. The preceding integral, if exists at least for
some real value of $n$, defines a function of
complex variable $n$ that is holomorphic at least in some right half plane. Using
analytical continuation, the Mellin moments $f(n)$ (i.e.\ the Mellin transform of the function
$f(x)$)\footnote{In this text, the same symbol is used for both the function and its
Mellin moments. The distinction between them is made by the argument --- using the argument
$n$ always refers to the Mellin moments.} can be defined even for some values of $n$ for which
the integral does not exist (however, this does not mean that the Mellin moments can
be defined for all $n$). The Mellin moments $f(n)$ uniquely determine the original function
$f(x)$, and therefore the Mellin transform can be inverted. The Mellin transform and its
inversion are linear. The formula for the inverse Mellin transform reads
\begin{equation}
  f(x) = \frac{1}{2\pi{\rm i}} \int_{\xi - {\rm i}\infty}^{\xi + {\rm i}\infty}
  f(n) x^{-n} {\rm d}n
\end{equation}
where $\xi$ is a real parameter chosen in such a way that the integration contour
is located to the right of all singularities of $f(n)$ in the plane of complex $n$.

The location of the singularities of the Mellin moments $f(n)$ is related to the behaviour
of the function $f(x)$ in the vicinity of $x=0$. If the Mellin moments $f(n)$ have some
singularity for $\re n > \xi$ and if there exists a nonnegative integer $k$ such that
the function $g(x) = (1-x)^k f(x)$ is bounded on some interval $[\varepsilon, 1]$, then
there is no real number $C$ such that
\begin{equation}
  |f(x)| \leq C \!\left( \frac{1}{x} \right) ^{\!\xi} \,\text{ for }\,
  x \in (0,\varepsilon)
\end{equation}
because if the preceding relation held for some real number $C$, then the function $g(x)$
would have to satisfy
\begin{equation}
  |g(x)| \leq K \!\left( \frac{1}{x} \right) ^{\!\xi}
\end{equation}
for some real number $K$ and for all $x$ from the interval $(0,1]$. This constraint would
then imply the following bound
\begin{equation}
  |g(n)| \leq \int_0^1 |x^{n-1}| |g(x)| {\rm d}x \leq K\! \int_0^1 x^{\re n - 1 - \xi} {\rm d}x ,
\end{equation}
which would mean that the Mellin moments $g(n)$ could not have any singularity for $\re n > \xi$,
but this would be in contradiction with the assumption that the Mellin moments $f(n)$ have some
singularity for $\re n > \xi$ because
\begin{equation}
  g(n) = \int_0^1 x^{n-1} \sum_{j=0}^k (-1)^j \binom{k}{j} x^j f(x)\, {\rm d}x =
  \sum_{j=0}^k (-1)^j \binom{k}{j} f(n+j) .
\end{equation}

Some relations concerning factorization (e.g.\ the relation between dressed and bare parton
distribution functions) are expressed as a convolution given by
\begin{align}
  \left( f\ast g \right) \!(x) &= \int_0^1 \!\int_0^1 \!{\rm d}y{\rm d}z\, f(y) g(z) \delta
  (yz - x) = \label{abdefconvul}\\ &= \int_x^1 \frac{{\rm d}y}{y}\, f\!\left( \frac{x}{y} \right) \! g(y) =
  \int_x^1 \frac{{\rm d}y}{y}\, f(y) \:\! g\!\left( \frac{x}{y} \right) \! .
\end{align}
From formula (\ref{abdefconvul}), we easily get
\begin{equation}
  \left( f\ast g \right) \!(n) =  f(n) g(n),  \label{abmeltransconvul}
\end{equation}
which means that the Mellin transform converts the convolution (\ref{abdefconvul}) to ordinary
multiplication of Mellin moments.

Using formula (\ref{abdefconvul}), we find
\begin{align}
  \int_0^1 \!{\rm d}x\, f(x) \left( g\ast h \right) \!(x) & = \int_0^1 \!\int_0^1 \!{\rm d}y
  {\rm d}z \, f(yz) g(y) h(z) = \nonumber\\ &= \int_0^1 \!{\rm d}x \left( f\otimes g \right)
  \!(x) h(x) = \int_0^1 \!{\rm d}x \left( f\otimes h \right) \!(x) g(x) \label{abconeqncrosssec}
\end{align}
where the symbol $\otimes$ is defined as
\begin{equation}
  \left( f\otimes g \right) \!(x) = \int_0^1 \!{\rm d}y\, f(xy)g(y)   \label{abdefotimes}
\end{equation}
and satisfies
\begin{align}
  \left[ \left( f\otimes g \right) \otimes h\right] \!(x) & = \int_0^1 \!{\rm d}z
  \left(f \otimes g\right) \! (xz) h(z) = \int_0^1\!\int_0^1
  \!{\rm d}y{\rm d}z\, f(xyz) g(y) h(z) = \nonumber\\
  & = \int_0^1\!\int_0^1 \!{\rm d}y{\rm d}z\, g(y)h(z) \int_0^1 \!{\rm d}\xi\, f(x\xi)
  \delta (\xi - yz) = \nonumber\displaybreak[0]\\
  & = \int_0^1 \!{\rm d}\xi\, f(x\xi) \int_0^1 \!\int_0^1 \!{\rm d}y{\rm d}z\, g(y) h(z)
  \delta(\xi - yz) = \nonumber\\
  & = \int_0^1 \!{\rm d}\xi\, f(x\xi) \left(g \ast h\right) \!(\xi) =
  \left[f \otimes \left( g \ast h\right) \right] \!(x) . \label{abconrelcrosssec}
\end{align}
Formulae (\ref{abconeqncrosssec}) and (\ref{abconrelcrosssec}) are useful
for manipulations with expressions for cross-sections.

Using relations (\ref{abdefmeltrans}), (\ref{abdefconvul}) and (\ref{abdefotimes}),
we easily get the following properties of the function $f(x) = \delta (1 - x)$:
\begin{equation}
  f(n) = 1, \qquad f \ast g = g \ast f = g, \qquad g \otimes f = g  \label{abdeltafcejeident}
\end{equation}
where $g$ is an arbitrary function.

\section{Some important formulae for the QCD coupling parameter $a(\mu, {\rm RS})$}
\label{prtqcdcouplparam}

\subsection{Changing expansion parameter}

Within the framework of perturbative QCD, the QCD coupling parameter $a(\mu, {\rm RS})$
is a function of the renormalization scale $\mu$ and the renormalization scheme RS.
Hence, it is useful to be familiar with relations between coupling parameters
corresponding to different arguments.

Consider an expansion parameter $a(P)$ depending on a set of parameters denoted by $P$.
Let us assume that the relation between expansion parameters corresponding to
different sets of parameters $P_1$ and $P_2$ has the form of power series:
\begin{equation}
  a(P_1) = \sum_{l=0}^{\infty} c^{(1)}_l (P_1, P_2)\, a^{l+1}(P_2).  \label{abchangerel}
\end{equation}
This relation can then be generalized as follows:
\begin{equation}
  a^k(P_1) = \sum_{l=0}^{\infty} c^{(k)}_l (P_1, P_2)\, a^{l+k}(P_2) \label{abgenrel}
\end{equation}
where $k$ is a nonnegative integer and
\begin{align}
  c^{(0)}_l (P_1, P_2) &= \delta_{l0}, \label{abkoefcz}   \\
  c^{(k+1)}_l (P_1, P_2) & = \sum_{m=0}^l c^{(k)}_{l-m} (P_1, P_2) \,
  c^{(1)}_m (P_1, P_2) \quad\text{for}\quad k \geq 1 .  \label{abkoefck}
\end{align}
Formula (\ref{abgenrel}) is useful for changing an expansion parameter of power
expansions. Consider some quantity $F$ which can be expressed as an expansion in
powers of $a(P_0)$:
\begin{equation}
  F = \sum_{k=0}^{\infty} F^{(k)}(P_0)\, a^{k+k_0}(P_0) .
\end{equation}
Using formula (\ref{abgenrel}), we can then obtain an expansion of $F$ in powers
of an arbitrary expansion parameter $a(P)$:
\begin{equation}
  F = \sum_{k=0}^{\infty} F^{(k)}(P)\, a^{k+k_0}(P)
\end{equation}
where
\begin{equation}
  F^{(k)}(P) = \sum_{l=0}^k F^{(l)}(P_0)\, c^{(l+k_0)}_{k-l}(P_0,P). \label{abkoefchangeep}
\end{equation}

\subsection{The QCD coupling parameter in $d = 4-2\varepsilon$ space-time dimensions}
\label{prtqcdcouplparameps}

Within the framework of dimensional regularization in $d = 4 - 2\varepsilon$ space-time
dimensions, equation (\ref{sbbeta}), which describes the dependence of the QCD
coupling parameter $a(\mu, {\rm RS})$ on the renormalization scale $\mu$, has the following form
\begin{equation}
  \frac{{\rm d}a(\mu, {\rm RS})}{{\rm d}\ln\mu} = \sum_{l=0}^{\infty} \beta_l ({\rm RS})\,
  a^{l+1}(\mu, {\rm RS})  \label{abbeta}
\end{equation}
where $\beta_0({\rm RS}) = -2\varepsilon$ and all higher order coefficients $\beta_l({\rm RS})$,
$l \geq 1$ are linear functions of $\varepsilon$ and depend on the renormalization scheme RS
(in the case of the coefficients $\beta_1({\rm RS})$ and $\beta_2({\rm RS})$, only the part
proportional to $\varepsilon$ is dependent on the renormalization scheme RS).

The renormalization scale of the QCD coupling parameter can be changed using the formula
\begin{equation}
  a(\mu, {\rm RS}) = \sum_{l=0}^{\infty} g^{(1)}_l (\mu, \mu_0, {\rm RS})\, a^{l+1} (\mu_0,
  {\rm RS}),   \label{abchangescale}
\end{equation}
which can be derived from (\ref{abbeta}). The coefficients\footnote{These coefficients can also
be written as $g^{(1)}_l (\frac{\mu}{\mu_0}, {\rm RS})$ because they depend only on the ratio
of $\mu$ and $\mu_0$.} $g^{(1)}_l (\mu, \mu_0, {\rm RS})$ depend on $\varepsilon$ and
are finite in the limit $\varepsilon \to 0$. The change of the renormalization scheme of
the QCD coupling parameter is then described by the formula\footnote{Note that
the coefficients $h^{(1)}_l ({\rm RS}_1, {\rm RS}_2)$ do not depend on $\mu$.}
\begin{equation}
  a(\mu, {\rm RS}_1) = \sum_{l=0}^{\infty} h^{(1)}_l ({\rm RS}_1, {\rm RS}_2) \,
  a^{l+1} (\mu, {\rm RS}_2), \quad h^{(1)}_0 ({\rm RS}_1, {\rm RS}_2) = 1 .  \label{abchangescheme}
\end{equation}
Contrary to the coefficients $g^{(1)}_l (\mu, \mu_0, {\rm RS})$, the coefficients
$h^{(1)}_l ({\rm RS}_1, {\rm RS}_2)$ are independent of $\varepsilon$. Both formulae
(\ref{abchangescale}) and (\ref{abchangescheme}) have the form of (\ref{abchangerel})
and can thus be generalized to
\begin{align}
  a^k(\mu, {\rm RS}) & = \sum_{l=0}^{\infty} g^{(k)}_l (\mu, \mu_0, {\rm RS}) \,
  a^{l+k} (\mu_0, {\rm RS}), \\
  a^k(\mu, {\rm RS}_1) & = \sum_{l=0}^{\infty} h^{(k)}_l ({\rm RS}_1, {\rm RS}_2) \,
  a^{l+k} (\mu, {\rm RS}_2),   \label{abgenchangescheme}
\end{align}
where the coefficients $g^{(k)}_l (\mu, \mu_0, {\rm RS})$ and
$h^{(k)}_l ({\rm RS}_1, {\rm RS}_2)$ are determined by relations (\ref{abkoefcz})
and (\ref{abkoefck}).

\section{Solution of equations (\ref{schigherpfsdependence}) and (\ref{sztcoeffviasplitfce}, \ref{sctcoeffviasplitfce})}
\label{prtsolutionmatrixeq}

This appendix contains formulae for the solution of the matrix equation
\begin{equation}
  \left[ \mathbf{X}(n), \mathbf{P}^{(0)} (n) \right] -
  \kappa \mathbf{X}(n) = \mathbf{Y}(n), \quad \kappa\neq 0  \label{abdefmatrixeqn}
\end{equation}
with respect to $\mathbf{X}(n)$. The matrix $\mathbf{P}^{(0)}(n)$ represents
the LO splitting functions and the matrix $\mathbf{Y}(n)$ is given.
All three matrices in the preceding equation are square
matrices with indices corresponding to parton species: $q_i$, $\bar{q}_i$ and
$G$, and therefore their dimension is $2n_{\rm f}+1$ where $n_{\rm f}$ denotes
the number of quark flavours. For the following, it is useful to introduce
a unified labeling for quarks and antiquarks: $Q_i = q_i$, $Q_{i+n_{\rm f}} = \bar{q}_i$.
The structure of the matrix $\mathbf{P}^{(0)}(n)$ can then be expressed as follows:
\begin{equation}
  P^{(0)}_{Q_{i}Q_{j}}(n) = P^{(0)}_{qq}(n) \delta_{ij}, \quad
  P^{(0)}_{Q_{i}G}(n) = P^{(0)}_{qG}(n), \quad
  P^{(0)}_{GQ_{i}}(n) = P^{(0)}_{Gq}(n), \quad
  P^{(0)}_{GG}(n).
\end{equation}

For better readability of the formulae expressing the solution of equation
(\ref{abdefmatrixeqn}), it is convenient not to write out explicitly the
dependence on $n$. The formulae then read
\begin{align}
  X_{Q_{i}Q_{j}} = {} &\overline{X} - \frac{1}{\kappa} Y_{Q_{i}Q_{j}} +
  \frac{P^{(0)}_{qG} P^{(0)}_{Gq} \left( \sum_{k=1}^{2n_{\rm f}} Y_{Q_{i}Q_{k}}
  - Y_{qq} \right) + \kappa P^{(0)}_{Gq} \left( Y_{Q_{i}G} - Y_{qG} \right)}
  {\kappa \left( \kappa P^{(0)}_{GG} - \kappa P^{(0)}_{qq} - \kappa^2
  + 2n_{\rm f} P^{(0)}_{qG} P^{(0)}_{Gq} \right)} + {} \nonumber\displaybreak[0]\\
  & {} + \frac{P^{(0)}_{qG} P^{(0)}_{Gq} \left( \sum_{k=1}^{2n_{\rm f}} Y_{Q_{k}Q_{j}}
  - Y_{qq} \right) - \kappa P^{(0)}_{qG} \left( Y_{GQ_{j}} - Y_{Gq} \right)}
  {\kappa \left( \kappa P^{(0)}_{qq} - \kappa P^{(0)}_{GG} - \kappa^2
  + 2n_{\rm f} P^{(0)}_{qG} P^{(0)}_{Gq} \right)} , \label{absolutioneqnzac}\displaybreak[1]\\
  X_{Q_{i}G} = {} & X_{qG} + \frac{\kappa \left( Y_{Q_{i}G} - Y_{qG} \right)
  + P^{(0)}_{qG} \left( \sum_{k=1}^{2n_{\rm f}} Y_{Q_{i}Q_{k}} - Y_{qq} \right)}
  {\kappa P^{(0)}_{GG} - \kappa P^{(0)}_{qq} - \kappa^2 + 2n_{\rm f}
  P^{(0)}_{qG} P^{(0)}_{Gq}} , \displaybreak[1]\\
  X_{GQ_{i}} = {} & X_{Gq} + \frac{\kappa \left( Y_{GQ_{i}} - Y_{Gq} \right)
  - P^{(0)}_{Gq} \left( \sum_{k=1}^{2n_{\rm f}} Y_{Q_{k}Q_{i}} - Y_{qq} \right)}
  {\kappa P^{(0)}_{qq} - \kappa P^{(0)}_{GG} - \kappa^2 + 2n_{\rm f}
  P^{(0)}_{qG} P^{(0)}_{Gq}} , \displaybreak[0]\\
  X_{GG} = {}& -\frac{1}{\kappa} Y_{GG} - 2n_{\rm f} \overline{X}
\end{align}
where
\begin{align}
  \overline{X} & = \frac{1}{\nu} \left[ P^{(0)}_{Gq} \left( P^{(0)}_{qq} - P^{(0)}_{GG}
   - \kappa \right) Y_{qG} + P^{(0)}_{qG} \left( P^{(0)}_{qq} - P^{(0)}_{GG} + \kappa
  \right) Y_{Gq} - {} \right. \nonumber\\
  & \quad \left. {} - 2 P^{(0)}_{qG} P^{(0)}_{Gq} \left( Y_{qq} - Y_{GG} \right)
  \right] \! , \displaybreak[0]\\
  X_{qG} & = \frac{1}{\nu} \left[ \left( P^{(0)}_{qq} - P^{(0)}_{GG} - \kappa \right)
  \left( \kappa Y_{qG} + P^{(0)}_{qG} \left( Y_{qq} - Y_{GG} \right) \right) + {} \right.
  \nonumber\\ & \quad \left. {} + 4n_{\rm f} P^{(0)}_{qG} \left( P^{(0)}_{Gq} Y_{qG}
  + P^{(0)}_{qG} Y_{Gq} \right) \right] \! , \displaybreak[0]\\
  X_{Gq} & = \frac{1}{\nu} \left[ \left( P^{(0)}_{GG} - P^{(0)}_{qq} - \kappa \right)
  \left( \kappa Y_{Gq} - P^{(0)}_{Gq} \left( Y_{qq} - Y_{GG} \right) \right) + {} \right.
  \nonumber\\ & \quad \left. {} + 4n_{\rm f} P^{(0)}_{Gq} \left( P^{(0)}_{Gq} Y_{qG}
  + P^{(0)}_{qG} Y_{Gq} \right) \right] \label{absolutioneqnkon}
\end{align}
and the denominator $\nu$ is given as
\begin{equation}
  \nu  = \kappa \left( \kappa^2 - \left( P^{(0)}_{qq} - P^{(0)}_{GG} \right)^2
  - 8n_{\rm f} P^{(0)}_{qG} P^{(0)}_{Gq} \right) \! .  \label{abdenominatorsoleq}
\end{equation}
The symbols $Y_{qq}$, $Y_{qG}$ and $Y_{Gq}$ are defined as follows:
\begin{equation}
  Y_{qq} \equiv \frac{1}{2n_{\rm f}} \sum_{k,l=1}^{2n_{\rm f}} Y_{Q_{k}Q_{l}} , \quad
  Y_{qG} \equiv \frac{1}{2n_{\rm f}} \sum_{k=1}^{2n_{\rm f}} Y_{Q_{k}G} , \quad
  Y_{Gq} \equiv \frac{1}{2n_{\rm f}} \sum_{k=1}^{2n_{\rm f}} Y_{GQ_{k}} .
\end{equation}
If all denominators in expressions (\ref{absolutioneqnzac})--(\ref{absolutioneqnkon})
are nonzero, then the solution of equation (\ref{abdefmatrixeqn}) exists and is
unique for an arbitrary right hand side $\mathbf{Y}$.

It can be shown that for every nonzero $\kappa$, there exists some right half plane
in which all denominators in expressions (\ref{absolutioneqnzac})--(\ref{absolutioneqnkon})
are nonzero. Hence, if the right hand side $\mathbf{Y}(n)$ of (\ref{abdefmatrixeqn})
is holomorphic in some right half plane, then there exists some right half plane
in which the solution $\mathbf{X}(n)$ of equation (\ref{abdefmatrixeqn}) is
unique and holomorphic.

\section{Analysis of the freedom associated with factorization in massless QCD}
\label{prtanalysisoffreedom}

This appendix contains a detailed analysis of the freedom associated with the factorization
procedure in massless perturbative QCD. Important results of this appendix are summarized
in Subsections \ref{prtfacfreedompdf} and \ref{prtfacfreedomhscs}.

To investigate the ambiguity associated with the definition of dressed parton distribution
functions, it is useful to express the relevant formulae in terms of Mellin moments, which
are defined in Appendix \ref{prtmellintrans}. Converting relation (\ref{scxdefpdf}) into
the space of Mellin moments, we get ordinary matrix multiplication of moments
\begin{equation}
  \mathbf{D}(n, M, {\rm FS}, {\rm RS}) = \mathbf{A}(n, M, {\rm FS}, {\rm RS})
  \mathbf{\widehat{D}}(n)  \label{scdefpdf}
\end{equation}
and the expansion of $\mathbf{A}(n, M, {\rm FS}, {\rm RS})$ is expressed as follows
\begin{equation}
  \mathbf{A}(n, M, {\rm FS}, {\rm RS}) = \sum_{k=0}^{\infty} a^k(M, {\rm RS})
  \mathbf{A}\!^{(k)} (n, {\rm FS}, {\rm RS}), \quad  \mathbf{A}\!^{(0)}(n, {\rm FS},
  {\rm RS}) = \mathbf{1} .
\end{equation}
The form of relation (\ref{scdefpdf}) allows us to obtain expression for
the bare parton distribution functions $\mathbf{\widehat{D}}(n)$:
\begin{equation}
  \mathbf{\widehat{D}}(n) = \mathbf{B}(n, M, {\rm FS}, {\rm RS}) \mathbf{D}(n, M,
  {\rm FS}, {\rm RS})   \label{scbarepdfmm}
\end{equation}
where $\mathbf{B}(n, M, {\rm FS}, {\rm RS})$ is the matrix inversion of
$\mathbf{A}(n, M, {\rm FS}, {\rm RS})$, which can be expanded in powers of $a(M,{\rm RS})$:
\begin{equation}
  \mathbf{B}(n, M, {\rm FS}, {\rm RS}) = \sum_{k=0}^{\infty} a^k(M, {\rm RS})
  \mathbf{B}^{(k)} (n, {\rm FS}, {\rm RS})
\end{equation}
where the coefficients $\mathbf{B}^{(k)} (n, {\rm FS}, {\rm RS})$ are determined by
the following recurrence relation
\begin{equation}
  \mathbf{B}^{(k)}(n, {\rm FS}, {\rm RS}) = -\sum_{l=0}^{k-1} \mathbf{A}\!^{(k-l)}(n,
  {\rm FS}, {\rm RS}) \mathbf{B}^{(l)}(n, {\rm FS}, {\rm RS}), \quad
  \mathbf{B}^{(0)}(n, {\rm FS}, {\rm RS}) = \mathbf{1} ,   \label{screkrelbkoef}
\end{equation}
which implies that $\mathbf{B}^{(k)}(n, {\rm FS}, {\rm RS})$ is given as a sum of
terms of the form
\begin{equation}
   c\mathbf{A}\! ^{(l_1)}(n, {\rm FS}, {\rm RS}) \mathbf{A}\! ^{(l_2)}(n, {\rm FS}, {\rm RS})
   \dotsm \mathbf{A}\! ^{(l_m)}(n, {\rm FS}, {\rm RS})
\end{equation}
where the coefficient $c$ is an integer and $\sum\limits_{i=1}^m l_i = k$. This means that
$\mathbf{B}^{(k)}(n, {\rm FS}, {\rm RS})$ depends on $\mathbf{A}^{(l)}(n, {\rm FS},
{\rm RS})$ for $1 \leq l \leq k$ and is independent of $\mathbf{A}^{(l)}(n, {\rm FS},
{\rm RS})$ for $l > k$. Relation (\ref{screkrelbkoef}) and its consequence are
useful for analyzing formulae (\ref{scsplitfunccoeff}) and (\ref{scdeftkoef}).

The following three subsections will be devoted to changing the unphysical quantities
on which the dressed parton distribution functions depend. The obtained results will
then allow us to investigate the dependence of the splitting functions and the hard
scattering cross-sections on the unphysical quantities associated with
the factorization procedure.

\subsection{Changing the renormalization scheme}

If we want to change the renormalization scheme used for the factorization procedure
from ${\rm RS}_1$ to ${\rm RS}_2$, then we should convert the expansion parameter
$a(M, {\rm RS}_1)$ to $a(M, {\rm RS}_2)$. Hence, let us express $A_{ij} (x, M,
{\rm FS}, {\rm RS}_1)$ as an expansion in powers of $a(M, {\rm RS}_2)$:
\begin{multline}
  A_{ij}(x, M, {\rm FS}, {\rm RS}_1) = \delta_{ij}\delta(1-x) + {} \\ {} +
  \sum_{k=1}^{\infty} a^k(M, {\rm RS}_2) \sum_{l=1}^k h^{(l)}_{k-l}
  ({\rm RS}_1, {\rm RS}_2) \, A_{ij}^{(l)} (x, {\rm FS}, {\rm RS}_1) ,  \label{scchangersexpan}
\end{multline}
where we have used formula (\ref{abgenchangescheme}). Seeing that the coefficients
$h^{(k)}_{l} ({\rm RS}_1, {\rm RS}_2)$ are independent of $\varepsilon$, the preceding
expansion has the form that is consistent with formulae (\ref{scdefamatrix}) and
(\ref{scdefakmatrix}), and therefore there must exist such a factorization scheme
$\mathcal{FS}({\rm RS}_1, {\rm RS}_2, {\rm FS})$ that
\begin{equation}
  A_{ij}(x, M, {\rm FS}, {\rm RS}_1) = A_{ij}(x, M,
  \mathcal{FS}({\rm RS}_1, {\rm RS}_2, {\rm FS}), {\rm RS}_2) .  \label{scchangerseqv}
\end{equation}
The pair of $\{ {\rm FS}, {\rm RS}_1 \} $ thus defines, independently of the
factorization scale $M$, the same singular factors that are absorbed into parton
distribution functions as the pair of $\{ \mathcal{FS}({\rm RS}_1, {\rm RS}_2,
{\rm FS}), {\rm RS}_2 \} $, and therefore both pairs are equivalent in the sense of
the definition of dressed parton distribution functions. Changing the renormalization
scheme from ${\rm RS}_0$ to RS for the fixed factorization scheme FS and
factorization scale $M$ can thus be converted to changing the
factorization scheme FS to the factorization scheme $\mathcal{FS}({\rm RS}, {\rm RS}_0,
{\rm FS})$ for the fixed renormalization scheme ${\rm RS}_0$ and factorization scale $M$.
The change of the factorization scheme will be discussed in Subsection \ref{prtchangingfs}.

In order to be able to change the renormalization scheme in the above described manner,
we have to know some prescription for the determination of the factorization scheme
$\mathcal{FS}({\rm RS}_1, {\rm RS}_2, {\rm FS})$. Such a prescription can be obtained
as follows. Comparing the expansion (\ref{scdefamatrix}) of the right hand side of
(\ref{scchangerseqv}) with the expansion (\ref{scchangersexpan}) of the left hand
side of (\ref{scchangerseqv}), we find
\begin{equation}
  A_{ij}^{(k)}(x, \mathcal{FS}({\rm RS}_1, {\rm RS}_2, {\rm FS}), {\rm RS}_2) =
  \sum_{l=1}^k h^{(l)}_{k-l} ({\rm RS}_1, {\rm RS}_2) \, A^{(l)}_{ij} (x, {\rm FS},
  {\rm RS}_1 ), \quad k \geq 1 .
\end{equation}
Using formula (\ref{scdefakmatrix}) and exploiting the fact that the coefficients
$h^{(k)}_{l} ({\rm RS}_1, {\rm RS}_2)$ are independent of $\varepsilon$, we get
\begin{equation}
  A_{ij}^{(km)}(x, \mathcal{FS}({\rm RS}_1, {\rm RS}_2, {\rm FS}), {\rm RS}_2) =
  \sum_{l=1}^k h^{(l)}_{k-l} ({\rm RS}_1, {\rm RS}_2) \, A^{(lm)}_{ij} (x, {\rm FS},
  {\rm RS}_1 )    \label{scdeffsfunc}
\end{equation}
where $k \geq 1$ and $m \geq 0$. The desired prescription is then represented by
the preceding equation (\ref{scdeffsfunc}) for $m = 0$.

\subsection{Evolution equations}

The evolution equations describe the dependence of dressed parton distribution
functions on the factorization scale. From formula (\ref{scdefpdf}), we get
\begin{equation}
  \frac{{\rm d} \mathbf{D}(n, M, {\rm FS}, {\rm RS})}{{\rm d}\ln M} =
  \frac{{\rm d} \mathbf{A}(n, M, {\rm FS},
  {\rm RS})}{{\rm d}\ln M} \mathbf{B}(n, M, {\rm FS}, {\rm RS})
  \mathbf{D}(n, M, {\rm FS}, {\rm RS}),
\end{equation}
where we have used formula (\ref{scbarepdfmm}) to express the bare parton
distribution functions $\mathbf{\widehat{D}}(n)$ in terms of the dressed
parton distribution functions $\mathbf{D}(n, M, {\rm FS}, {\rm RS})$.
Expanding in powers of $a(M, {\rm RS})$, using equation (\ref{abbeta})
and after some manipulations, we obtain
\begin{equation}
  \frac{{\rm d} \mathbf{A}(n, M, {\rm FS}, {\rm RS})}{{\rm d}\ln M}
  \mathbf{B}(n, M, {\rm FS}, {\rm RS}) = a(M, {\rm RS})
  \mathbf{P}(n, M, {\rm FS}, {\rm RS})  \label{scdefsplitfcenonper}
\end{equation}
where the splitting functions $\mathbf{P}(n, M, {\rm FS}, {\rm RS})$
have an expansion in powers of $a(M, {\rm RS})$
\begin{equation}
  \mathbf{P}(n, M, {\rm FS}, {\rm RS}) = \sum_{k=0}^{\infty} a^{k}
  (M, {\rm RS})\, \mathbf{P}^{(k)}(n, {\rm FS}, {\rm RS})
\end{equation}
with the coefficients given by the formula
\begin{equation}
  \mathbf{P}^{(k)}(n, {\rm FS}, {\rm RS}) = \sum_{l=0}^k \left( \sum_{p=0}^l
  (p+1) \beta_{l-p}({\rm RS}) \mathbf{A}\! ^{(p+1)}(n, {\rm FS}, {\rm RS}) \right)
  \mathbf{B}^{(k-l)} (n, {\rm FS}, {\rm RS}) .  \label{scsplitfunccoeff}
\end{equation}
The evolution equations expressed in terms of Mellin moments thus read
\begin{equation}
  \frac{{\rm d} \mathbf{D}(n, M, {\rm FS}, {\rm RS})}{{\rm d}\ln M} =
  a(M, {\rm RS}) \mathbf{P}(n, M, {\rm FS}, {\rm RS})
  \mathbf{D}(n, M, {\rm FS}, {\rm RS}) .
\end{equation}
Converting the preceding equation into $x$-space, we obtain the evolution
equations in the form of (\ref{sbxevolrov}).

The splitting functions $\mathbf{P}^{(k)}(n, {\rm FS}, {\rm RS})$, determined
by (\ref{scsplitfunccoeff}), must be finite in the limit $\varepsilon\to 0$.
A detailed analysis of this condition shows that
\begin{equation}
 \begin{array}{l}
   \mathbf{A}\! ^{(kl)}(n, {\rm FS}, {\rm RS}) = 0 \text{ for $l > k \geq 1$,}\\[6pt]
   \mathbf{A}\! ^{(kl)}(n, {\rm FS}, {\rm RS})
     \text{ for $k \geq l \geq 2$ is uniquely determined by}\\
     \mathbf{A}\! ^{(m0)}(n, {\rm FS}) \text{ and } \mathbf{A}\! ^{(m1)}(n, {\rm FS},
     {\rm RS}) \text{ for $1 \leq m \leq k-1$.}
 \end{array}  \label{scconseqpfinite}
\end{equation}
Since the functions $A^{(k0)}_{ij}(x)$ can be choosen at will, all nontrivial
information concerning the properties of the theory has to be contained
in the functions $A^{(k1)}_{ij}(x, {\rm FS}, {\rm RS})$.

\subsection{Changing the factorization scheme}
\label{prtchangingfs}

The formula for changing the factorization scheme can be obtained from relations
(\ref{scdefpdf}) and (\ref{scbarepdfmm}):
\begin{equation}
  \mathbf{D}(n, M, {\rm FS}_1, {\rm RS}) = \mathbf{T}(n, M, {\rm FS}_1, {\rm FS}_2,
  {\rm RS}) \mathbf{D}(n, M, {\rm FS}_2, {\rm RS})
\end{equation}
where
\begin{equation}
  \mathbf{T}(n, M, {\rm FS}_1, {\rm FS}_2, {\rm RS}) = \mathbf{A}(n, M, {\rm FS}_1,
  {\rm RS}) \mathbf{B}(n, M, {\rm FS}_2, {\rm RS}) .   \label{scdeftmatrixfunc}
\end{equation}
Expanding the preceding relation in powers of $a(M, {\rm RS})$, we get
\begin{equation}
  \mathbf{T}(n, M, {\rm FS}_1, {\rm FS}_2, {\rm RS}) = \sum_{k=0}^{\infty}
  a^k(M, {\rm RS}) \, \mathbf{T}^{(k)}(n, {\rm FS}_1, {\rm FS}_2, {\rm RS})
\end{equation}
where the coefficients of the expansion are determined by the formula
\begin{equation}
  \mathbf{T}^{(k)}(n, {\rm FS}_1, {\rm FS}_2, {\rm RS}) = \sum_{l=0}^k
  \mathbf{A}\! ^{(k-l)}(n, {\rm FS}_1, {\rm RS}) \mathbf{B}^{(l)}(n, {\rm FS}_2,
  {\rm RS}).  \label{scdeftkoef}
\end{equation}
Using relation (\ref{screkrelbkoef}), we find that $\mathbf{T}^{(0)}(n, {\rm FS}_1,
{\rm FS}_2, {\rm RS}) = \mathbf{1}$ and $\mathbf{T}^{(k)}(n, {\rm FS}_1, {\rm FS}_2,
{\rm RS})$ for $k \geq 1$ is given as a polynomial expression in $\mathbf{A}\! ^{(l)}
(n, {\rm FS}_1, {\rm RS})$ and $\mathbf{A}\! ^{(l)} (n, {\rm FS}_2, {\rm RS})$ for
$1 \leq l \leq k$ with coefficients independent of $\varepsilon$.
Seeing that $\mathbf{T}^{(k)}(n, {\rm FS}_1, {\rm FS}_2, {\rm RS})$ must be
finite in the limit $\varepsilon\to 0$, we can do the following
replacement\footnote{The functions $A^{(l0)}_{ij}(x)$, which define the factorization
scheme, do not depend on the renormalization scheme.}
\begin{equation}
  \mathbf{A}\!^{(l)}(n, {\rm FS}_1, {\rm RS}) \to \mathbf{A}\!^{(l0)}(n, {\rm FS}_1),
  \quad \mathbf{A}\!^{(l)}(n, {\rm FS}_2, {\rm RS}) \to \mathbf{A}\!^{(l0)}(n, {\rm FS}_2)
\end{equation}
in the expression for $\mathbf{T}^{(k)}(n, {\rm FS}_1, {\rm FS}_2, {\rm RS})$
because the singular parts of $\mathbf{A}\!^{(l)}(n, {\rm FS}_1, {\rm RS})$ and
$\mathbf{A}\!^{(l)}(n, {\rm FS}_2, {\rm RS})$ cannot contribute to the finite
part of $\mathbf{T}^{(k)}(n, {\rm FS}_1, {\rm FS}_2, {\rm RS})$ (see formula
(\ref{scdefakmatrix})). Hence, $\mathbf{T}^{(k)}(n, {\rm FS}_1, {\rm FS}_2,
{\rm RS})$ does not depend on the renormalization scheme RS.

The condition that $\mathbf{T}^{(k)}(n, {\rm FS}_1, {\rm FS}_2)$ is finite
in the limit $\varepsilon\to 0$ allows us to determine the dependence
of $\mathbf{A}\! ^{(l1)}(n, {\rm FS}, {\rm RS})$ on the factorization scheme.
Consider the so called minimal subtraction (MS) factorization scheme, which
is defined by setting the functions $A^{(k0)}_{ij}(x)$ equal to zero.
From relations (\ref{scdefakmatrix}) and (\ref{screkrelbkoef}), we get
\begin{equation}
  \mathbf{B}^{(k)}(n, {\rm MS}, {\rm RS}) = -\frac{1}{\varepsilon} \mathbf{A}\! ^{(k1)}
  (n, {\rm MS}, {\rm RS}) + \text{terms containing at least $\frac{1}{\varepsilon^2}$}.
  \label{scbformsfs}
\end{equation}
The requirement of the absence of a term proportional to $1/ \varepsilon$ on
the right hand side of (\ref{scdeftkoef}) then implies
\begin{equation}
  \mathbf{A}\! ^{(k1)} (n, {\rm FS}, {\rm RS}) = \mathbf{A}\! ^{(k1)} (n, {\rm MS},
  {\rm RS}) + \sum_{l=1}^{k-1} \mathbf{A}\! ^{(k-l,0)} (n, {\rm FS})
  \mathbf{A}\! ^{(l1)} (n, {\rm MS}, {\rm RS}) .  \label{scconseqtfinite}
\end{equation}
This formula determines the dependence of the functions $A^{(k1)}_{ij}(x, {\rm FS}, {\rm RS})$
on the arbitrary functions $A^{(k0)}_{ij}(x)$, which specify the factorization
scheme. The dependence of the functions $A^{(k1)}_{ij}(x, {\rm MS}, {\rm RS})$ on
the renormalization scheme is then described by the following formula resulting
from (\ref{scdeffsfunc}):
\begin{equation}
  A_{ij}^{(k1)}(x, {\rm MS}, {\rm RS}) = \sum_{l=1}^k h^{(l)}_{k-l} ({\rm RS}_0, {\rm RS}) \,
  A^{(l1)}_{ij} (x, {\rm MS}, {\rm RS}_0),  \label{scchangersforms}
\end{equation}
where we have used the fact that $\mathcal{FS}({\rm RS}_0, {\rm RS}, {\rm MS}) = {\rm MS}$,
which follows from equation (\ref{scdeffsfunc}) for $m = 0$. The obtained relations
(\ref{scconseqpfinite}), (\ref{scconseqtfinite}) and (\ref{scchangersforms}) thus imply that
if we know the functions $A_{ij}^{(k1)}(x, {\rm FS}, {\rm RS})$ for some factorization scheme
and some renormalization scheme, then we are able to determine the complete singular factors
$A_{ij}(x, M, {\rm FS}, {\rm RS})$ for an arbitrary factorization scale, factorization scheme
and renormalization scheme.

Since the pair of $\{ {\rm FS}, {\rm RS} \} $ is equivalent to the pair of
$\{ \mathcal{FS}({\rm RS}, {\rm RS}_0, {\rm FS}), {\rm RS}_0 \} $, we can convert
the simultaneous change of the factorization scheme and the renormalization scheme
from $\{ {\rm FS}_0, {\rm RS}_0 \}$ to $\{ {\rm FS}, {\rm RS} \}$ to changing the factorization
scheme from ${\rm FS}_0$ to $\mathcal{FS}({\rm RS}, {\rm RS}_0, {\rm FS})$ for the
fixed renormalization scheme ${\rm RS}_0$.

\subsection{Specification of factorization scheme via splitting functions}
\label{prtspecfsviasplitfunc}

The relations presented in the preceding subsections allow us to investigate
the dependence of the splitting functions on the factorization scheme, which
is one of the main aims of this text. This dependence interests us only in
the limit $\varepsilon\to 0$ (in four space-time dimensions).

From formula (\ref{scsplitfunccoeff}), we get that the LO splitting functions
for $\varepsilon = 0$ are given as
\begin{equation}
  \mathbf{P}^{(0)}(x, {\rm FS}, {\rm RS}) = \lim_{\varepsilon\to 0} \left(
  -2\varepsilon \mathbf{A}\! ^{(1)}(x, {\rm FS}, {\rm RS}) \right) =
  -2\mathbf{A}\! ^{(11)}(x, {\rm FS}, {\rm RS}) . \label{scp0xformula}
\end{equation}
Relations (\ref{scconseqtfinite}), (\ref{scchangersforms}) and (\ref{abchangescheme})
then imply that $\mathbf{A}\! ^{(11)}(x, {\rm FS}, {\rm RS})$ is independent of
the factorization scheme and the renormalization scheme, and therefore the LO
splitting functions $\mathbf{P}^{(0)}(x)$ are unique (for $\varepsilon = 0$). However,
this is not true for the higher order splitting functions. Using (\ref{screkrelbkoef}),
(\ref{scsplitfunccoeff}), (\ref{scconseqtfinite}), (\ref{scp0xformula}) and the properties
of the coefficients $\beta_k ({\rm RS})$ mentioned in Appendix \ref{prtqcdcouplparameps},
we find that the higher order splitting functions $\mathbf{P}^{(k)}(n, {\rm FS}, {\rm RS})$,
$k \geq 1$ for $\varepsilon = 0$ are expressed as
\begin{equation}
  \mathbf{P}^{(k)}(n, {\rm FS}, {\rm RS}) = \left[ \mathbf{A}\! ^{(k0)}(n, {\rm FS}),
  \mathbf{P}^{(0)}(n) \right] - kb\mathbf{A}\! ^{(k0)}(n, {\rm FS}) +
  \mathbf{R}_k (n, {\rm FS}, {\rm RS})  \label{schigherpfsdependence}
\end{equation}
where $\mathbf{R}_k (n, {\rm FS}, {\rm RS})$ is a polynomial expression in
$\mathbf{A}\! ^{(l1)}(n, {\rm MS}, {\rm RS})$ for $1 \leq l \leq k+1$ and
$\mathbf{A}\! ^{(l0)}(n, {\rm FS})$ for $1 \leq l \leq k-1$ with the coefficients
depending on the renormalization scheme RS. The preceding relation (\ref{schigherpfsdependence})
and the relations presented in Appendix \ref{prtsolutionmatrixeq} imply
that for any set of arbitrarily chosen splitting functions $\mathbf{P}^{(l)}(n)$,
$1 \leq l \leq k$ where $k$ is arbitrary, there exists exactly one set of
the corresponding $\mathbf{A}\! ^{(l0)}(n)$, $1 \leq l \leq k$. The higher order
splitting functions $\mathbf{P}^{(k)}(x, {\rm FS}, {\rm RS})$, $k \geq 1$
for $\varepsilon = 0$ can thus be chosen at will and can be used for labeling
factorization schemes.\footnote{If we use the splitting functions for labeling
factorization schemes, then the complete specification of the factorization scheme
requires also the specification of the corresponding renormalization scheme because
the relation between the splitting functions and the functions $A_{ij}^{(k0)}(x)$,
which define the factorization scheme, depends on the renormalization scheme.}

The specification of factorization schemes via splitting functions can be useful
for phenomenology. In order to simplify the application of this kind of specification,
it is necessary to derive formulae for expressing $\mathbf{T}^{(k)}(x, {\rm FS}_1,
{\rm FS}_2)$ and $\mathcal{FS}({\rm RS}_1, {\rm RS}_2, {\rm FS})$ in terms of
splitting functions. The relation between $\mathbf{T}^{(k)}(x, {\rm FS}_1,
{\rm FS}_2)$ and the corresponding splitting functions can be obtained as follows.
Let us start with the relation
\begin{equation}
  \mathbf{A}(n, M, {\rm FS}_1, {\rm RS}) = \mathbf{T}(n, M, {\rm FS}_1, {\rm FS}_2,
  {\rm RS}) \mathbf{A}(n, M, {\rm FS}_2, {\rm RS}),
\end{equation}
which follows from (\ref{scdeftmatrixfunc}). Differentiating the preceding
relation with respect to $\ln M$ and using equation (\ref{scdefsplitfcenonper}),
we find
\begin{multline}
  a(M, {\rm RS}) \mathbf{P}(n, M, {\rm FS}_1, {\rm RS}) \mathbf{T}(n, M, {\rm FS}_1,
  {\rm FS}_2, {\rm RS}) = \frac{{\rm d} \mathbf{T}(n, M, {\rm FS}_1, {\rm FS}_2,
  {\rm RS})}{{\rm d}\ln M} + {} \\ {} + a(M, {\rm RS}) \mathbf{T}(n, M, {\rm FS}_1,
  {\rm FS}_2, {\rm RS}) \mathbf{P}(n, M, {\rm FS}_2, {\rm RS}).
\end{multline}
Expanding the preceding equation for $\varepsilon = 0$ in powers of $a(M, {\rm RS})$,
using equation (\ref{sbbeta}) in the form
\begin{equation}
  \frac{{\rm d} a(\mu, {\rm RS})}{{\rm d} \ln\mu} = -b\sum_{k=0}^{\infty}
  c_{k}({\rm RS}) a^{k+2}(\mu, {\rm RS}), \quad c_0({\rm RS}) = 1,
  \quad c_1({\rm RS}) = c
\end{equation}
and after some manipulations, we get the desired formula
\begin{multline}
  \left[ \mathbf{T}^{(k)}(n, {\rm FS}_1, {\rm FS}_2), \mathbf{P}^{(0)}(n) \right]
  - kb \mathbf{T}^{(k)}(n, {\rm FS}_1, {\rm FS}_2) =  \mathbf{P}^{(k)}(n, {\rm FS}_1,
  {\rm RS}) - {} \\ {} - \mathbf{P}^{(k)}(n, {\rm FS}_2, {\rm RS})
  + \sum_{l=1}^{k-1} \left\{ \mathbf{P}^{(k-l)}(n, {\rm FS}_1, {\rm RS})
  \mathbf{T}^{(l)}(n, {\rm FS}_1, {\rm FS}_2)  -  {} \right. \\ \left. {} -
  \mathbf{T}^{(l)}(n, {\rm FS}_1, {\rm FS}_2) \mathbf{P}^{(k-l)}(n, {\rm FS}_2, {\rm RS})
  + lbc_{k-l}({\rm RS}) \mathbf{T}^{(l)}(n, {\rm FS}_1, {\rm FS}_2) \right\}  \label{sctcoeffviasplitfce}
\end{multline}
where $k \geq 1$. The preceding formula forms a set of equations for
$\mathbf{T}^{(k)}(n, {\rm FS}_1, {\rm FS}_2)$, which can be solved iteratively.

The formula for $\mathcal{FS}({\rm RS}_1, {\rm RS}_2, {\rm FS})$ follows from
\begin{equation}
  a(M, {\rm RS}_1) \mathbf{P}(x, M, {\rm FS}, {\rm RS}_1) = a(M, {\rm RS}_2)
  \mathbf{P}(x, M, \mathcal{FS}({\rm RS}_1, {\rm RS}_2, {\rm FS}), {\rm RS}_2),
\end{equation}
which is a consequence of (\ref{scchangerseqv}) and (\ref{scdefsplitfcenonper}).
Expanding the preceding equation in powers of $a(M, {\rm RS}_2)$ (the left hand
side is expanded using (\ref{abgenchangescheme})), we obtain
\begin{equation}
  \mathbf{P}^{(k)}(x, \mathcal{FS}({\rm RS}_1, {\rm RS}_2, {\rm FS}), {\rm RS}_2)
  = \sum_{l=0}^k h^{(l+1)}_{k-l}({\rm RS}_1, {\rm RS}_2)\, \mathbf{P}^{(l)}
  (x, {\rm FS}, {\rm RS}_1).
\end{equation}
Contrary to formula (\ref{sctcoeffviasplitfce}), the preceding relation holds
for general $\varepsilon$.

\subsection{Transformation of coefficient functions}

The subject of this subsection is the determination of the dependence of
coefficient functions on the unphysical quantities associated with factorization.
An arbitrary structure function $F\!\left(x, Q^2\right)$ is given as
\begin{equation}
  F\!\left(n, Q^2\right) = \mathbf{\widehat{C}}\!\left(n, Q^2 \right)
  \mathbf{\widehat{D}}(n)  \label{scdefstructfuncbare}
\end{equation}
where $\mathbf{\widehat{C}}\!\left(n, Q^2 \right)$ represents the corresponding
bare coefficient functions.\footnote{The coefficient functions form a row vector
whereas the parton distribution functions are represented by a column vector,
and therefore the multiplication on the right hand side of (\ref{scdefstructfuncbare})
yields a number (a matrix $1\times 1$).} Using relation (\ref{scdefpdf}), we get
\begin{equation}
  F\!\left(n, Q^2\right) = \mathbf{C}\!\left(n, Q^2, M, {\rm FS}, {\rm RS} \right)
  \mathbf{D}(n, M, {\rm FS}, {\rm RS})
\end{equation}
where the (finite) coefficient functions $\mathbf{C}\!\left(n, Q^2, M, {\rm FS},
{\rm RS} \right)$ are given by
\begin{equation}
  \mathbf{C}\!\left(n, Q^2, M, {\rm FS}, {\rm RS} \right) =
  \mathbf{\widehat{C}}\!\left(n, Q^2 \right) \mathbf{B}(n, M, {\rm FS}, {\rm RS}) .
  \label{scfincoefffuncdef}
\end{equation}
The dependence of coefficient functions on the unphysical quantities associated with
the factorization procedure then follows from the preceding relation.

From relation (\ref{scfincoefffuncdef}), we find
\begin{multline}
  \frac{{\rm d} \mathbf{C}\!\left(n, Q^2, M, {\rm FS}, {\rm RS} \right)}{{\rm d}\ln M} =
  \mathbf{\widehat{C}}\!\left(n, Q^2 \right) \frac{{\rm d} \mathbf{B}(n, M,
  {\rm FS}, {\rm RS})}{{\rm d}\ln M} = {} \\ {} = \mathbf{C}\!\left(n, Q^2, M,
  {\rm FS}, {\rm RS} \right) \mathbf{A}(n, M, {\rm FS}, {\rm RS})
  \frac{{\rm d} \mathbf{B}(n, M, {\rm FS}, {\rm RS})}{{\rm d}\ln M} . \label{scfincoefffuncmdepbezp}
\end{multline}
Equation (\ref{scdefsplitfcenonper}) and $\mathbf{A}(n, M, {\rm FS}, {\rm RS})
\mathbf{B}(n, M, {\rm FS}, {\rm RS}) = \mathbf{1}$ then imply that
\begin{multline}
  \mathbf{A}(n, M, {\rm FS}, {\rm RS}) \frac{{\rm d}\mathbf{B}(n, M, {\rm FS},
  {\rm RS})}{{\rm d}\ln M} = {} \\ {} = - \frac{{\rm d}\mathbf{A}(n, M, {\rm FS},
  {\rm RS})}{{\rm d}\ln M} \mathbf{B}(n, M, {\rm FS}, {\rm RS}) =
  - a(M, {\rm RS}) \mathbf{P}(n, M, {\rm FS}, {\rm RS}) ,  \label{scabderexpression}
\end{multline}
which allows us to rewrite formula (\ref{scfincoefffuncmdepbezp}) in the form
\begin{equation}
  \frac{{\rm d} \mathbf{C}\!\left(n, Q^2, M, {\rm FS}, {\rm RS} \right)}{{\rm d}\ln M} =
  -a(M, {\rm RS})\, \mathbf{C}\!\left(n, Q^2, M, {\rm FS}, {\rm RS} \right)
  \mathbf{P}(n, M, {\rm FS}, {\rm RS}) .  \label{sccoefffuncmdepnonper}
\end{equation}
Changing the factorization scheme from ${\rm FS}_0$ to FS is described by the following formula
\begin{equation}
  \mathbf{C}\!\left(n, Q^2, M, {\rm FS}, {\rm RS} \right) = \mathbf{C}\!\left(n, Q^2,
  M, {\rm FS}_0, {\rm RS} \right) \mathbf{T}(n, M, {\rm FS}_0, {\rm FS}, {\rm RS}),
  \label{sccoefffuncfsdepnonper}
\end{equation}
which follows from relations (\ref{scdeftmatrixfunc}) and (\ref{scfincoefffuncdef}).

The coefficient functions $\mathbf{C}\!\left(n, Q^2, M, {\rm FS}, {\rm RS} \right)$
are (at least in principle) fully calculable within the framework of perturbative QCD
and therefore can be expanded in powers of $a(\mu, {\rm RS})$:
\begin{equation}
  \mathbf{C}\!\left(n, Q^2, M, {\rm FS}, {\rm RS} \right) = \sum_{k=0}^{\infty}
  a^k (\mu, {\rm RS})\, \mathbf{C}^{(k)}\!\!\left(n, Q^2, \mu, M, {\rm FS},
  {\rm RS} \right) \! .
\end{equation}
Substituting the preceding expansion in formulae (\ref{sccoefffuncmdepnonper})
and (\ref{sccoefffuncfsdepnonper}), we obtain the following formulae describing
the dependence of the coefficient functions $\mathbf{C}^{(k)}\!\!\left(n, Q^2,
\mu, M, {\rm FS}, {\rm RS} \right)$ on the factorization scale
\begin{align}
  \frac{{\rm d}\mathbf{C}^{(k)}\!\!\left(n, Q^2, \mu, M, {\rm FS}, {\rm RS}
  \right)}{{\rm d}\ln M} & = - \sum_{l=0}^{k-1} \mathbf{C}^{(l)}\!\!\left(n, Q^2,
  \mu, M, {\rm FS}, {\rm RS} \right) \times {} \nonumber\\ & {} \times\sum_{m=0}^{k-l-1}
  g^{(m+1)}_{k-l-m-1} (M, \mu, {\rm RS}) \mathbf{P}^{(m)}(n, {\rm FS}, {\rm RS})
  \label{sccoefffuncchangescale}
\end{align}
and scheme
\begin{align}
  \mathbf{C}^{(k)}\!\!\left(n, Q^2, \mu, M, {\rm FS}, {\rm RS} \right) & =
  \sum_{l=0}^k \mathbf{C}^{(l)}\!\!\left(n, Q^2, \mu, M, {\rm FS}_0, {\rm RS} \right)
  \times {} \nonumber\\ & {} \times \sum_{m=0}^{k-l} g^{(m)}_{k-l-m} (M, \mu, {\rm RS})
  \mathbf{T}^{(m)}(n, {\rm FS}_0, {\rm FS}) . \label{sccoefffuncchangescheme}
\end{align}

Relations (\ref{scchangerseqv}) and (\ref{scfincoefffuncdef}) imply that
\begin{equation}
  \mathbf{C}\!\left(n, Q^2, M, {\rm FS}, {\rm RS}_1 \right) =
  \mathbf{C}\!\left(n, Q^2, M, \mathcal{FS}({\rm RS}_1, {\rm RS}_2, {\rm FS}),
  {\rm RS}_2 \right) .
\end{equation}
Hence, the simultaneous change of the factorization scheme and the renormalization
scheme from $\{ {\rm FS}_0, {\rm RS}_0 \}$ to $\{ {\rm FS}, {\rm RS} \}$ can be
converted to changing the factorization scheme from ${\rm FS}_0$ to
$\mathcal{FS}({\rm RS}, {\rm RS}_0, {\rm FS})$ for the fixed renormalization
scheme ${\rm RS}_0$, similarly as in the case of parton distribution functions.

Formulae (\ref{sccoefffuncchangescale}), (\ref{sccoefffuncchangescheme}) and
(\ref{abkoefchangeep}) are sufficient for changing all unphysical parameters
associated with the renormalization and factorization procedure (even in the case
if the renormalization scheme of the coupling parameter that is employed for
expanding the coefficient functions is different from the renormalization scheme
used for the factorization procedure).

\subsection{Transformation of hard scattering cross-sections for lepton-hadron collisions}

According to the parton model, any inclusive cross-section $\sigma (P)$ (in general
differential) depending on observables $P$ and describing a lepton-hadron collision
is given as\footnote{The formula (\ref{scdefstructfuncbare}) for structure
functions is a particular case of this formula.}
\begin{equation}
  \sigma (P) = \sum_i \int_0^1 \!{\rm d}x\, \widehat{\sigma}_i (x, P) \widehat{D}_i (x) .
\end{equation}
The preceding relation expresses the cross-section $\sigma (P)$ in terms of
bare quantities. Using relations (\ref{scbarepdfmm}), (\ref{abmeltransconvul}),
(\ref{abconeqncrosssec}) and (\ref{abdefotimes}), we get
\begin{equation}
  \sigma (P) = \sum_i \int_0^1 \!{\rm d}x\, \sigma_i (x, P, M, {\rm FS}, {\rm RS})
  D_i (x, M, {\rm FS}, {\rm RS})
\end{equation}
where the (finite) hard scattering cross-section $\sigma_i (x, P, M, {\rm FS}, {\rm RS})$
is given by
\begin{equation}
  \sigma_i (x, P, M, {\rm FS}, {\rm RS}) = \sum_j \int_0^1 \!{\rm d}y\,
  \widehat{\sigma}_j (xy, P) B_{ji}(y, M, {\rm FS}, {\rm RS}) .  \label{sclhhscsdeffinite}
\end{equation}
The inversion of the preceding relation has the form
\begin{equation}
  \widehat{\sigma}_i (x, P) = \sum_j \int_0^1 \! {\rm d}y\,
  \sigma_j (xy, P, M, {\rm FS}, {\rm RS}) A_{ji}(y, M, {\rm FS}, {\rm RS}) , \label{sclhhscsrelbare}
\end{equation}
which follows from formulae (\ref{sclhhscsdeffinite}), (\ref{abmeltransconvul}),
(\ref{abdefotimes}), (\ref{abconrelcrosssec}) and (\ref{abdeltafcejeident}).

The dependence of the hard scattering cross-section $\sigma_i (x, P, M, {\rm FS}, {\rm RS})$
on the unphysical quantities associated with the factorization procedure can be investigated
in a similar way as in the case of the coefficient functions in the preceding subsection.
From equations (\ref{sclhhscsdeffinite}), (\ref{sclhhscsrelbare}) and (\ref{abconrelcrosssec}),
we find
\begin{multline}
  \frac{{\rm d} \sigma_i(x, P, M, {\rm FS}, {\rm RS})}{{\rm d}\ln M} = \sum_j \left[
  \widehat{\sigma}_j (P) \otimes \frac{{\rm d}B_{ji}(M, {\rm FS}, {\rm RS})}{{\rm d}\ln M}
  \right] (x) = {} \\ {} = \sum_{jk} \left[ \sigma_k (P, M, {\rm FS}, {\rm RS}) \otimes
  \left( A_{kj} (M, {\rm FS}, {\rm RS}) \ast \frac{{\rm d} B_{ji}(M, {\rm FS}, {\rm RS})}
  {{\rm d}\ln M} \right) \right] (x) .
\end{multline}
Applying equation (\ref{scabderexpression}) then gives
\begin{multline}
  \frac{{\rm d}\sigma_i (x, P, M, {\rm FS}, {\rm RS})}{{\rm d}\ln M} = {} \\ {} =
  - a(M, {\rm RS}) \sum_j \int_0^1 \!{\rm d}y\, \sigma_j(xy, P, M, {\rm FS}, {\rm RS})
  P_{ji}(y, M, {\rm FS}, {\rm RS}) .  \label{schscsnonperdepfscale}
\end{multline}
The formula for changing the factorization scheme from ${\rm FS}_0$ to FS follows from
relations (\ref{sclhhscsdeffinite}), (\ref{sclhhscsrelbare}) and (\ref{abconrelcrosssec}):
\begin{multline}
  \sigma_i (x, P, M, {\rm FS}, {\rm RS}) = {} \\ {} = \sum_{jk} \left[
  \sigma_j (P, M, {\rm FS}_0, {\rm RS}) \otimes \Bigl( A_{jk} (M, {\rm FS}_0, {\rm RS})
  \ast B_{ki}(M, {\rm FS}, {\rm RS}) \Bigr) \right] (x) .
\end{multline}
Taking into account relation (\ref{scdeftmatrixfunc}), we can write the preceding
equation as
\begin{equation}
  \sigma_i (x, P, M, {\rm FS}, {\rm RS}) = \sum_j \int_0^1 \!{\rm d}y\,
  \sigma_j (xy, P, M, {\rm FS}_0, {\rm RS}) T_{ji} (y, M, {\rm FS}_0, {\rm FS},
  {\rm RS}) . \label{schscsnonperdepfscheme}
\end{equation}

Expanding the hard scattering cross-section $\sigma_i (x, P, M, {\rm FS}, {\rm RS})$
in powers of $a(\mu, {\rm RS})$:
\begin{equation}
  \sigma_i (x, P, M, {\rm FS}, {\rm RS}) = \sum_{k=0}^{\infty} a^{k+k_0}(\mu, {\rm RS})
  \, \sigma_i^{(k)} (x, P, \mu, M, {\rm FS}, {\rm RS}), \quad k_0 \geq 0
\end{equation}
and inserting the expansion into equations (\ref{schscsnonperdepfscale}) and
(\ref{schscsnonperdepfscheme}), we obtain the formulae describing the dependence
of the hard scattering cross-sections $\sigma_i^{(k)} (x, P, \mu, M, {\rm FS}, {\rm RS})$
on the factorization scale
\begin{multline}
  \frac{{\rm d}\sigma_i^{(k)}(x, P, \mu, M, {\rm FS}, {\rm RS})}{{\rm d}\ln M} =
  -\sum_j \int_0^1 \!{\rm d}y \left\{ \sum_{l=0}^{k-1} \sigma_j^{(l)} (xy, P, \mu,
  M, {\rm FS}, {\rm RS}) \times {} \right. \\ \left. {} \times \sum^{k-l-1}_{m=0}
  g^{(m+1)}_{k-l-m-1} (M, \mu, {\rm RS}) P^{(m)}_{ji}(y, {\rm FS}, {\rm RS}) \right\}
  \label{schscsfscaledep}
\end{multline}
and scheme
\begin{multline}
  \sigma_i^{(k)} (x, P, \mu, M, {\rm FS}, {\rm RS}) = \sum_j \int_0^1 \!{\rm d}y
  \left\{ \sum_{l=0}^k \sigma_j^{(l)} (xy, P, \mu, M, {\rm FS}_0, {\rm RS})
  \times {} \right. \\ \left. {} \times \sum_{m=0}^{k-l} g^{(m)}_{k-l-m}(M, \mu,
  {\rm RS}) T^{(m)}_{ji}(y, {\rm FS}_0, {\rm FS}) \right\} . \label{schscsfschemedep}
\end{multline}
The preceding formulae (\ref{schscsfscaledep}) and (\ref{schscsfschemedep})
represent an analogy of formulae (\ref{sccoefffuncchangescale}) and
(\ref{sccoefffuncchangescheme}) and together with formula (\ref{abkoefchangeep})
are sufficient for changing all unphysical quantities associated with the
renormalization and factorization procedure.

\subsection{Transformation of hard scattering cross-sections for hadron-hadron collisions}

In the case of a hadron-hadron collision, any inclusive cross-section $\sigma (P)$
depending on observables $P$ can be expressed as
\begin{equation}
  \sigma (P) = \sum_{ij} \int_0^1 \!\!\int_0^1 \!{\rm d}x_1{\rm d}x_2 \,
  \widehat{\sigma}_{ij} (x_1, x_2, P) \widehat{D}_{i/H_1}(x_1) \widehat{D}_{j/H_2}(x_2) .
\end{equation}
Using relations (\ref{scbarepdfmm}), (\ref{abmeltransconvul}), (\ref{abconeqncrosssec})
and (\ref{abdefotimes}), we can rewrite the preceding formula for $\sigma(P)$ in the form
\begin{multline}
  \sigma (P) = \sum_{ij} \int_0^1 \!\!\int_0^1 \!{\rm d}x_1{\rm d}x_2 \,
  \sigma_{ij}(x_1, x_2, P, M_1, {\rm FS}_1, {\rm RS}_1, M_2, {\rm FS}_2, {\rm RS}_2)
  \times {} \\ {} \times D_{i/H_1}(x_1, M_1, {\rm FS}_1, {\rm RS}_1)
  D_{j/H_2}(x_2, M_2, {\rm FS}_2, {\rm RS}_2)
\end{multline}
where the hard scattering cross-section $\sigma_{ij}(x_1, x_2, P, M_1, {\rm FS}_1,
{\rm RS}_1, M_2, {\rm FS}_2, {\rm RS}_2)$ is given as
\begin{multline}
  \!\!\!\sigma_{ij}(x_1, x_2, P, M_1, {\rm FS}_1, {\rm RS}_1, M_2, {\rm FS}_2, {\rm RS}_2) =
  \sum_{kl} \int_0^1 \!\!\! \int_0^1 \!{\rm d}y_1 {\rm d}y_2\, \widehat{\sigma}_{kl}
  (x_1 y_1, x_2 y_2, P) \times {} \\ {} \times B_{ki}(y_1, M_1, {\rm FS}_1, {\rm RS}_1)
  B_{lj}(y_2, M_2, {\rm FS}_2, {\rm RS}_2) .  \label{schscshhfindef}
\end{multline}

The preceding relation (\ref{schscshhfindef}) can be written as
\begin{multline}
  \sigma_{ij}(x_1, x_2, P, M_1, {\rm FS}_1, {\rm RS}_1, M_2, {\rm FS}_2, {\rm RS}_2)
  = {} \\ {} = \sum_k \int_0^1 \!{\rm d}y\, \widetilde{\sigma}_{kj} (x_1 y, x_2, P, M_2,
  {\rm FS}_2, {\rm RS}_2) B_{ki}(y, M_1, {\rm FS}_1, {\rm RS}_1) \label{schscshhfinmodf}
\end{multline}
where
\begin{equation}
  \widetilde{\sigma}_{kj} (x_1, x_2, P, M_2, {\rm FS}_2, {\rm RS}_2) = \sum_l
  \int_0^1 \!{\rm d}y\, \widehat{\sigma}_{kl} (x_1, x_2 y, P) B_{lj} (y, M_2,
  {\rm FS}_2, {\rm RS}_2) .
\end{equation}
Relation (\ref{schscshhfinmodf}) has the same form as relation (\ref{sclhhscsdeffinite}),
and therefore we can immediately write the formulae for changing the factorization scale $M_1$
\begin{multline}
  \frac{{\rm d} \sigma_{ij}^{(k)}(x_1, x_2, P, \mu, M_1, {\rm FS}_1, {\rm RS}_1, M_2,
  {\rm FS}_2, {\rm RS}_2)}{{\rm d}\ln M_1} = {} \displaybreak[0]\\ {} = - \sum_{r} \int_0^1
  \!{\rm d}y \left\{ \sum_{l=0}^{k-1} \sigma_{rj}^{(l)} (x_1 y, x_2, P, \mu, M_1, {\rm FS}_1,
  {\rm RS}_1, M_2, {\rm FS}_2, {\rm RS}_2) \times {} \right. \\ \left. {} \times
  \sum_{m=0}^{k-l-1} g^{(m+1)}_{k-l-m-1} (M_1, \mu, {\rm RS}_1) P^{(m)}_{ri} (y,
  {\rm FS}_1, {\rm RS}_1) \right\}  \label{schscshhfscalef}
\end{multline}
and the factorization scheme associated with hadron $H_1$ (from ${\rm FS}_1^{(0)}$ to ${\rm FS}_1$)
\begin{multline}
  \sigma_{ij}^{(k)}(x_1, x_2, P, \mu, M_1, {\rm FS}_1, {\rm RS}_1, M_2, {\rm FS}_2,
  {\rm RS}_2) = {} \displaybreak[0]\\ {} = \sum_{r} \int_0^1 \!{\rm d}y \left\{
  \sum_{l=0}^k \sigma^{(l)}_{rj}(x_1 y, x_2, P, \mu, M_1, {\rm FS}_1^{(0)}, {\rm RS}_1,
  M_2, {\rm FS}_2, {\rm RS}_2) \times {} \right. \\ \left. {} \times \sum_{m=0}^{k-l}
  g^{(m)}_{k-l-m}(M_1, \mu, {\rm RS}_1) T^{(m)}_{ri} (y, {\rm FS}_1^{(0)}, {\rm FS}_1)
  \right\} .  \label{schscshhfschemef}
\end{multline}

Similarly, rewriting formula (\ref{schscshhfindef}) in the form
\begin{multline}
  \sigma_{ij}(x_1, x_2, P, M_1, {\rm FS}_1, {\rm RS}_1, M_2, {\rm FS}_2, {\rm RS}_2)
  = {} \\ {} = \sum_l \int_0^1 \!{\rm d}y\, \widetilde{\sigma}_{il} (x_1, x_2 y, P, M_1,
  {\rm FS}_1, {\rm RS}_1) B_{lj}(y, M_2, {\rm FS}_2, {\rm RS}_2)
\end{multline}
where
\begin{equation}
  \widetilde{\sigma}_{il} (x_1, x_2, P, M_1, {\rm FS}_1, {\rm RS}_1) = \sum_k
  \int_0^1 \!{\rm d}y\, \widehat{\sigma}_{kl} (x_1 y, x_2, P) B_{ki} (y, M_1,
  {\rm FS}_1, {\rm RS}_1) ,
\end{equation}
we obtain the formulae describing the change of the factorization scale $M_2$
\begin{multline}
  \frac{{\rm d} \sigma_{ij}^{(k)}(x_1, x_2, P, \mu, M_1, {\rm FS}_1, {\rm RS}_1, M_2,
  {\rm FS}_2, {\rm RS}_2)}{{\rm d}\ln M_2} = {} \displaybreak[0]\\ {} = - \sum_{r} \int_0^1
  \!{\rm d}y \left\{ \sum_{l=0}^{k-1} \sigma_{ir}^{(l)} (x_1, x_2 y, P, \mu, M_1, {\rm FS}_1,
  {\rm RS}_1, M_2, {\rm FS}_2, {\rm RS}_2) \times {} \right. \\ \left. {} \times
  \sum_{m=0}^{k-l-1} g^{(m+1)}_{k-l-m-1} (M_2, \mu, {\rm RS}_2) P^{(m)}_{rj} (y,
  {\rm FS}_2, {\rm RS}_2) \right\}  \label{schscshhfscales}
\end{multline}
and the factorization scheme associated with hadron $H_2$ (from ${\rm FS}_2^{(0)}$ to ${\rm FS}_2$)
\begin{multline}
  \sigma_{ij}^{(k)}(x_1, x_2, P, \mu, M_1, {\rm FS}_1, {\rm RS}_1, M_2, {\rm FS}_2,
  {\rm RS}_2) = {} \displaybreak[0]\\ {} = \sum_{r} \int_0^1 \!{\rm d}y \left\{
  \sum_{l=0}^k \sigma^{(l)}_{ir}(x_1, x_2 y, P, \mu, M_1, {\rm FS}_1, {\rm RS}_1,
  M_2, {\rm FS}_2^{(0)}, {\rm RS}_2) \times {} \right. \\ \left. {} \times \sum_{m=0}^{k-l}
  g^{(m)}_{k-l-m}(M_2, \mu, {\rm RS}_2) T^{(m)}_{rj} (y, {\rm FS}_2^{(0)}, {\rm FS}_2)
  \right\} .  \label{schscshhfschemes}
\end{multline}

It is important to point out that formulae (\ref{schscshhfscalef}) and
(\ref{schscshhfschemef}) correspond to the expansion in powers of $a(\mu, {\rm RS}_1)$
whereas formulae (\ref{schscshhfscales}) and (\ref{schscshhfschemes}) correspond
to the expansion in powers of $a(\mu, {\rm RS}_2)$. As well as in the case of
hard scattering cross-sections for lepton-hadron collisions, formulae (\ref{schscshhfscalef}),
(\ref{schscshhfschemef}), (\ref{schscshhfscales}), (\ref{schscshhfschemes}) and
(\ref{abkoefchangeep}) are sufficient for changing all unphysical parameters
associated with the renormalization and factorization procedure.

\acknowledgments{The author would like to thank J. Ch\'yla for careful reading of
the manuscript and valuable suggestions. This work was supported by the projects LC527
of Ministry of Education and AVOZ10100502 of the Academy of Sciences of the Czech Republic.}

\end{document}